\documentclass[fleqn,usenatbib]{mnras}

\usepackage{newtxtext,newtxmath}

\usepackage[T1]{fontenc}
\usepackage{ae,aecompl}
\usepackage{comment}
 \usepackage{booktabs}
 \usepackage{multi row}
 
\usepackage{epsfig}
\usepackage{amssymb}
\usepackage{natbib}

\usepackage{graphicx}	
\usepackage{amsmath}	
\usepackage{amssymb}	
\usepackage{nccmath}

\title[catalog B-D]{A catalog of polychromatic bulge-disk decompositions of $\sim 17.600$ galaxies in CANDELS}

\author[Dimauro et al.]{
Paola Dimauro$^{1}$,\thanks{E-mail: paola.dimauro@obspm.fr}
Marc Huertas-Company$^{1,2,3}$,
Emanuele Daddi$^{4}$,
\newauthor
Pablo G. P\'erez-Gonz\'alez$^{5}$, Mariangela Bernardi$^{2}$, Guillermo Barro$^{6}$, 
\newauthor
Fernando Buitrago$^{7}$, Fernando Caro$^{1,3,9}$, Andrea Cattaneo$^{8}$,  
\newauthor
Helena Dominguez-S\'{a}nchez$^{1}$,Sandra M. Faber$^{9}$, Boris H\"au\ss ler$^{10}$, Dale D. Kocevski$^{11}$, 
\newauthor
Anton M. Koekemoer$^{12}$, David C. Koo$^{9}$, Christoph T. Lee$^{9}$, Simona Mei$^{1,3}$, 
 \newauthor
Berta Margalef-Bentabol$^{1}$, Joel Primack$^{9}$, Aldo Rodriguez-Puebla$^{13}$, Mara Salvato$^{14}$,
 \newauthor
 Francesco Shankar$^{15}$, Diego Tuccillo$^{1,16}$
\\
$^{1}$Sorbonne Universit\'e, Observatoire de Paris, Universit\'e PSL, CNRS, LERMA, F-75014, Paris, France\\
$^{2}$ Department of Physics and Astronomy, University of Pennsylvania, Philadelphia, PA 19104, USA\\
$^{3}$ Universit\'e Paris Diderot, 5 Rue Thomas Mann, 75013, France\\
$^{4}$ CEA Saclay, Laboratoire AIM-CNRS-Universit\'e Paris Diderot, Irfu/SAp, Orme des Merisiers, 91191, Gif-sur-Yvette, France\\
$^{5}$ Departamento de Astrof\'isica, Facultad de CC. F\'isicas, Universidad Complutense de Madrid, E-28040 Madrid, Spain\\
$^{6}$ University of California, Berkeley, CA 94720, USA 0000-0001-6813-875X\\
$^{7}$ Instituto de Astrofisica e Ciencias do Espaco, Universidade de Lisboa, OAL, Tapada da Ajuda, P-1349-018 Lisbon, Portugal \\
$^{8}$ GEPI, Observatoire de Paris, 61 Avenue de l'Observatoire, 75014, Paris, France\\
$^{9}$ University of California, Santa Cruz, CA 95064, USA\\
$^{10}$ European Southern Observatory, Alonso de Cordova 3107, Vitacura, Casilla 19001, Santiago, Chile\\
$^{11}$Department of Physics and Astronomy, Colby College, Waterville, ME 04961, USA \\
$^{12}$Space Telescope Science Institute, 3700 San Martin Drive, Baltimore, MD 21218, USA \\
$^{13}$ Instituto de Astronom\'ia, Universidad Nacional Aut\'onoma de M\'exico, A.P. 70-264, 04510 M\'exico, D.F., Mexico\\
$^{14}$ Max Planck Institut fur Plasma Physik and Excellence Cluster, 85748 Garching, Germany \\
$^{15}$ Department of Physics and Astronomy, University of Southampton, Highfield, SO17 1BJ, UK\\
$^{16}$MINES Paristech, PSL Research University, Centre for Mathematical Morphology, Fontainebleau, France
}

\date{Accepted XXX. Received YYY; in original form ZZZ}

\pubyear{2018}

\begin{document}
\label{firstpage}
\pagerange{\pageref{firstpage}--\pageref{lastpage}}
\maketitle

\newcommand{\Galfitm}{\textsc{GalfitM}\:}
\newcommand{\GalfitM}{\textsc{GalfitM}\:}
\newcommand{\galfit}{\textsc{Galfit}\:}
\newcommand{\Sersic}{S\'{e}rsic\:}

\begin{abstract}
Understanding how bulges grow in galaxies is critical step towards unveiling the link between galaxy morphology and star-formation. To do so, it is necessary to decompose large sample of galaxies at different epochs into their main components (bulges and disks). This is particularly challenging, especially at high redshifts, where galaxies are poorly resolved. 
This work presents a catalog of bulge-disk decompositions of the surface brightness profiles of $\sim17.600$ H-band selected galaxies in the CANDELS fields ($F160W<23$, $0<z<2$) in 4 to 7 filters covering a spectral range of $430-1600nm$. This is the largest available catalog of this kind up to $z = 2$. By using a novel approach based on deep-learning to select the best model to fit, we manage to control systematics arising from wrong model selection and obtain less contaminated samples than previous works. We show that the derived structural properties are within $\sim10-20\%$ of random uncertainties.  We then fit stellar population models to the decomposed SEDs (Spectral Energy Distribution) of bulges and disks and derive stellar masses (and stellar mass bulge-to-total ratios) as well as rest-frame colors (U,V,J) for bulges and disks separately. All data products are publicly released with this paper and through the web page  \url{https://lerma.obspm.fr/huertas/form_CANDELS} and will be used for scientific analysis in forthcoming works.
\end{abstract}





\section{Introduction}
Most galaxies have two main components: disks and bulges. These two components are believed to have very different formation mechanisms. Disks are generally rotationally supported and confined to a thin plane. They are believed to be the consequence of gas infall into halos, which transfers angular momentum to the baryons. Bulges generally have a 3D shape and larger stellar velocity dispersions. Their formation requires dissipative processes and a loss of angular momentum. Mergers of two disks is the classical channel to grow bulges (e.g.\citealp{Toomre1977}). However numerical models show that disks, especially at high redshift when they are more unstable and gas rich, can also self generate a bulge through instabilities (e.g.~\citealp{Bournaud2016}) and/or inflow of cold gas towards the center (e.g.~\citealp{Zolotov2015}). Properly understanding how all these different processes come together to assemble galaxies into their main components requires identifying bulges and disks in galaxies and studying their evolution across cosmic time. Since disks and bulges have different projected surface brightness distributions, the decomposition of the light by fitting analytic \Sersic models \citealp{Sersic1968} to the 1D or 2D light profiles has been widely used in the literature. Extending this approach to large datasets arising from deep surveys, where objects cannot be checked individually, is particularly challenging. Not only because of the computing time, but also due to the large amount of systematics that need to be controlled. 
Many works have obtained bulge-disk decompositions of several hundreds of thousands of galaxies at low redshift, where galaxies are reasonably well resolved \citep{Simard2011,Meert2015,Lange2016,Allen2006,Mendel2014,Lackner2012}. A significant amount of post processing is required to assess the quality of the fits and eventually identify unphysical solutions. One key issue, for instance, is deciding whether two components are really needed to model the light profile or if one unique component is better suited. This is usually addressed by performing a-posteriori statistical tests to measure if the addition of an extra component improves the fit (e.g. \citealp{Vika2014,Meert2015,Allen2006}). 

At high redshift, the situation is even more dramatic, both because of lower S/N and because galaxies start to be less well resolved even with space based imaging. This is why most of the works involving surface brightness fitting of large samples of distant galaxies tend to use one single $S\'{e}rsic\:$ component, reducing the amount of free parameters (e.g. \citealp{haeussler2007,vanderWel2012}). Two component fitting is generally done on smaller datasets (e.g. \citealp{Bruce2014a,Lang2014,Margalef2016,Margalef2018}). Even there, degeneracies are reduced by adding more constraints on the parameters. For example, \cite{Bruce2014a} forced the \Sersic index of the bulge to be $4$. However, many works have shown that bulges have a wider distribution of the \Sersic index (e.g \citealp{Meert2015}) so this might not be the ideal solution. 

An additional issue of bulge-disk decompositions is that they are performed on the light profiles, while models predict stellar mass distributions. Deriving stellar masses from light distributions requires assuming an mass/light (thereafter M/L) ratio, which can be different for bulges and disks and also from galaxy to galaxy. Assuming that the light profile directly trace the stellar mass (a unique M/L for all galaxies/components) is clearly an oversimplification which can introduce additional systematics. This is especially true for high redshift studies, where very different cosmic epochs are probed. 
 
In this paper we present a catalog of bulge-disk decompositions of $\sim 17.600$ galaxies in the CANDELS fields. This is the largest catalog of this kind for objects spanning such a wide redshift range ($z < 2$).  This work also introduces several novel techniques to address some of the issues discussed above. Firstly, we develop a method based on deep-learning to estimate the optimal model that should be used to fit the light profile (namely one or two components). While other techniques existing in the literature which rely on the fitting residuals, our method acts before the modeling phase, at the pixel level. Additionally, our fits are done simultaneously in 4 to 7 (depending on the fields) high resolution filters using the modified version of \textsc{Galfit}, \textsc{GalfitM} ~\citep{haeussler2013,Vika2014}. This multi-wavelength fitting allows us to increase the Signal/Noise ($S/N$) and to reduce the random uncertainties but also to estimate Spectral Energy Distributions (SEDs) of bulges and disks and a M/L for every component by fitting stellar populations models with the FAST code \citep{Kriek2009}. We  thus provide stellar population properties (stellar masses, SFRs) and rest-frame colors (U,V,J) for bulges and disks. This should enable a less biased comparison with predictions of galaxy formation models. The catalog will be made public upon publication of this work\footnote{\url{https://lerma.obspm.fr/huertas/form_CANDELS}}. 

The paper proceeds as follows. We describe the dataset in section~\ref{sec:data}. The methodology used for profile fitting is discussed in sections~\ref{sec:mod_sel} and sections~\ref{sec:galfitM}. The accuracy of the catalog is quantified in section~\ref{sec:errors}. The  stellar population properties are described in section~\ref{sec:SEDs}. All magnitudes are measured in the AB system. 

\section{Data}
\label{sec:data}
Our starting point for the selection are the official CANDELS \citep{Grogin2011,Koekemoer2011} H-band (F160W) selected catalogs (\citealp{Galametz2013} for UDS, \citealp{Guo2013} for GOODS-S, Barro et al. (2017)  for GOODS-N and \citealp{Stefanon2017} for COSMOS and AEGIS). For this study, we only consider galaxies brighter than $F_{160W}=23$. This magnitude selection is applied to ensure reliable two component decompositions as detailed in section~\ref{sec:errors}. In addition to the three NIR images (F105, F125, F160), observed as part of the CANDELS survey, we use ancillary data in four additional bands for GOODS-N and GOODS-S (F435W, F606W, F775W, F850L) and two in  the AEGIS, UDS and COSMOS fields (F606W, F814W). All images are resampled to a common pixel scale of $0.06$ $arcsec/pixel$. This is required to perform simultaneous multi-wavelength fits to the surface brightness profiles as described in section~\ref{sec:galfitM}. 

We also use the 2D single \Sersic fits published in \cite{vanderWel2012} in three NIR filters (F105W, F125W, F160W) and the deep-learning based \emph{visual} morphologies published in \cite{Huertas2015}. The official CANDELS redshifts are used. More details can be found in \cite{Dahlen2013}. Spectroscopic redshifts are used when available. If not, we use photometric redshifts derived through SED fitting by combining different available codes.  Although we derive stellar masses of bulges of disks (described in section~\ref{sec:SEDs}) we also use total stellar masses from previous works for comparison~\citep{Huertas2016}. Namely, the best available redshift is used to estimate stellar masses based on the PEGASE01 stellar population models (Fioc \& Rocca-Volmerange 1999). We assume solar metallicity, exponentially declining star formation histories, Salpeter (1995) IMF and a Calzetti et al. (2000) extinction law. The $M_*/L$ ratios are converted to a Chabrier (2000) IMF by applying a constant 0.22 dex offset. Rest-frame magnitudes (U,V,J) are also derived as part of the SED fitting procedure and are used to divide galaxies between star-forming and quiescent systems with the now standard way (see \citealp{Whitaker2012}).

\begin{figure}
$\begin{array}{c}
\includegraphics[width=0.45\textwidth]{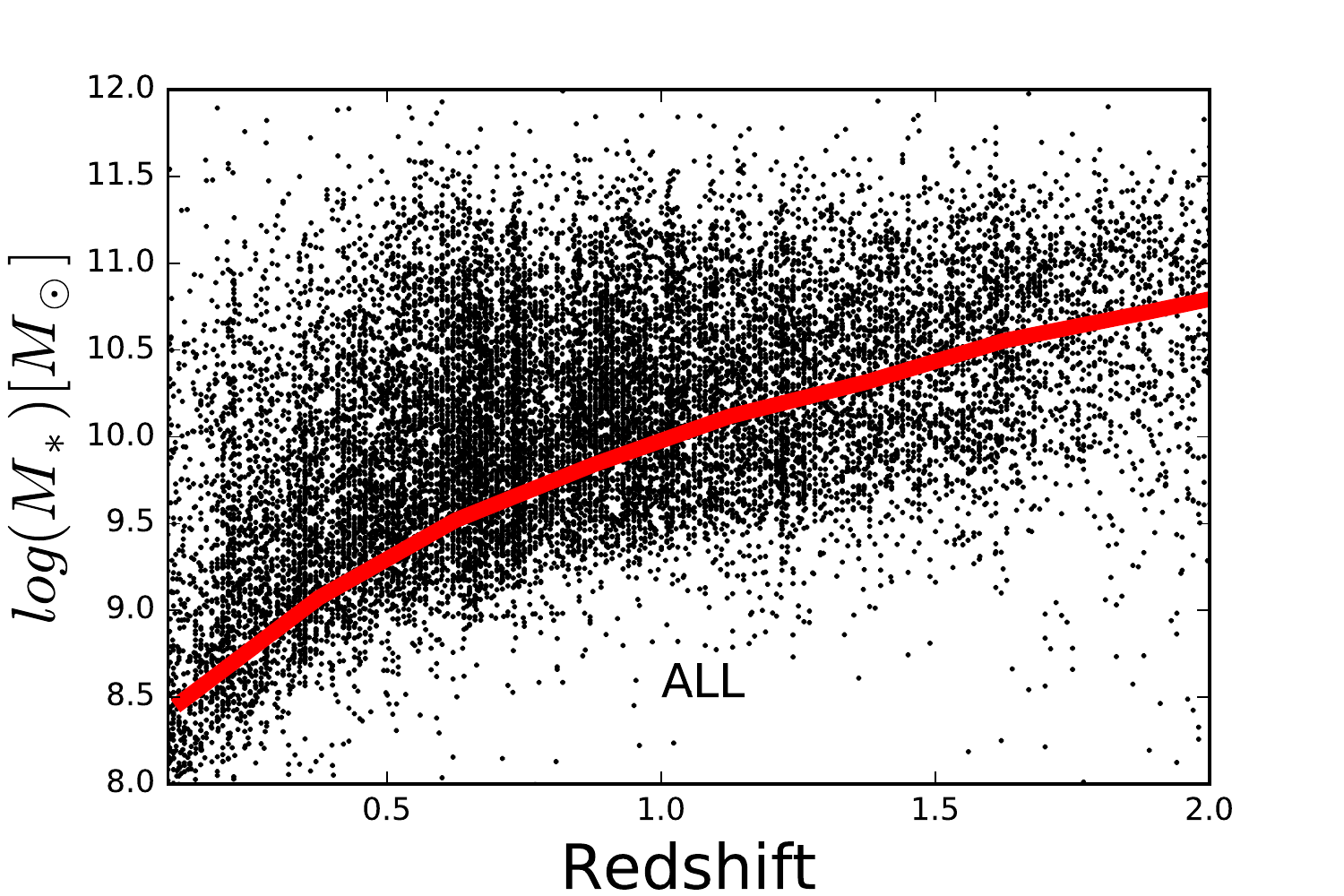} \\ 
\includegraphics[width=0.45\textwidth]{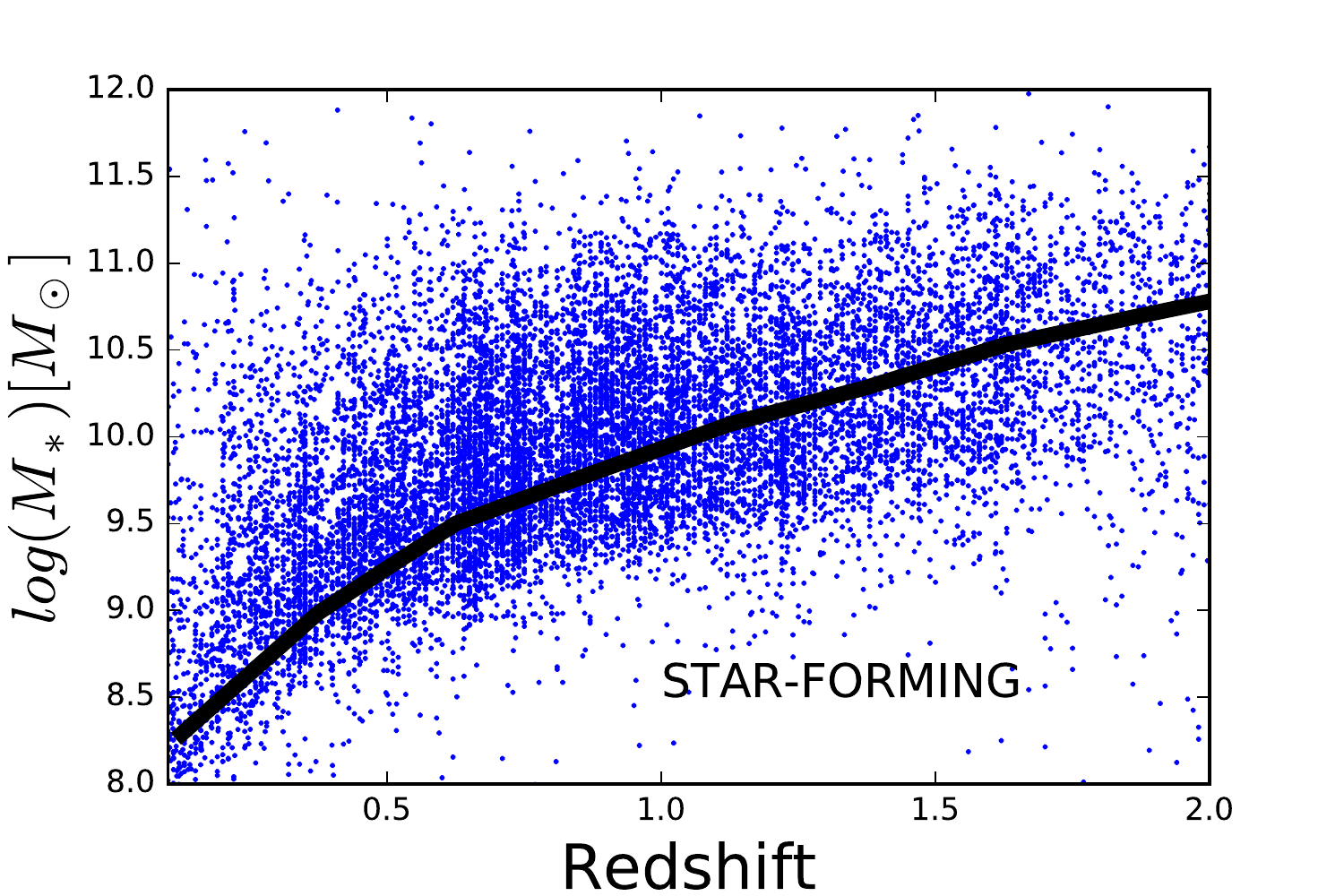}\\ 
\includegraphics[width=0.45\textwidth]{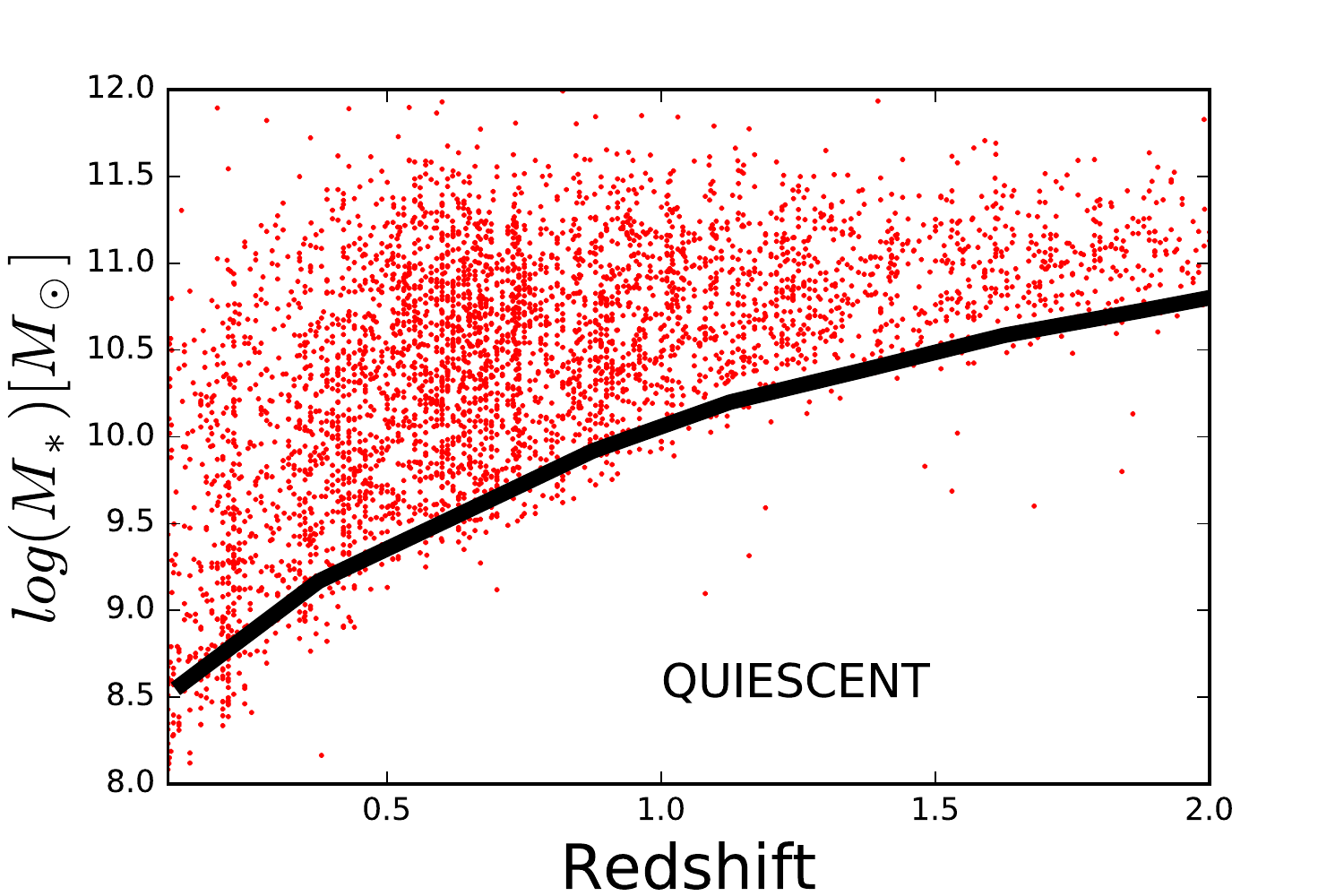} \\
\includegraphics[width=0.45\textwidth]{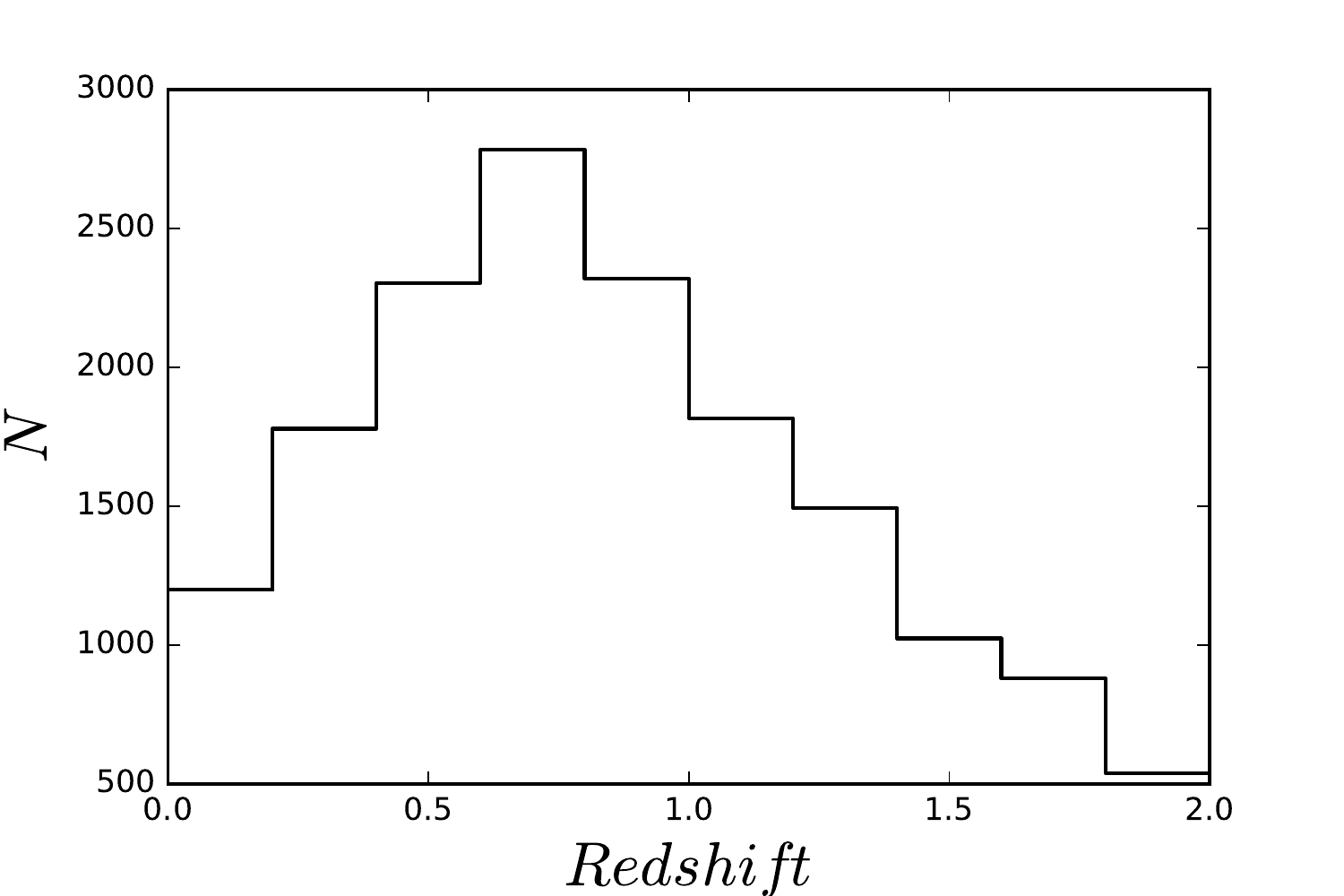}
\end{array}$
\caption{Stellar mass completeness of the selected sample. The panels show the relation between redshift and stellar mass for galaxies in our sample ($F160W<23$). Top: all galaxies, middle: star-forming, bottom: quiescent. The shaded regions indicate the stellar mass completeness estimated using the method described in the text. The bottom panel shows the redshift distribution.} 
\label{fig:completeness}
\end{figure}

Our final sample consists of $\sim17.600$ galaxies out of $\sim19.000$ with $F160W<23$ in the CANDELS survey. The difference comes from the fact that there is not a perfect overlap between the fields in the different filters. We restricted our analysis to galaxies observed with at least 4 filters. Figure~\ref{fig:completeness} shows the distribution of the selected galaxies in the $M_*-z$ plane. We estimate the stellar mass completeness using the method described in~\cite{Pozzetti2010}. The lower limit stellar mass ($M^*_{lim}$) is computed as: $log(M^*_{lim})=log(M_*)+0.4\times(m_{f160W}-23)$. $m_{f160W}$ is the apparent magnitude in the $F160W$ filter where the selection is done. The completeness is then estimated as the 90th percentile of the distribution of $M^*_{lim}$. We repeat the same procedure for all galaxies as well as for passive and star-forming galaxies (selected according to their UVJ colors). The obtained values are plotted in figure~\ref{fig:completeness} and summarized in table~\ref{tbl:comp}. 
We estimate a mass completeness limit of $\sim 10.7$ at $z\sim2$. We notice that for the passive population, very few galaxies lie below the completeness limit, despite the bright magnitude cut. This is expected given the evolution of the stellar mass function of passive galaxies (e.g. \citealp{Ilbert2013}), which at $z>1.5$, is dominated by massive galaxies. As a sanity check, since the CANDELS catalogs are several magnitudes deeper than our selected catalog, the stellar mass completeness is also estimated by computing the stellar mass above which at least $80\%$ of galaxies from the deep catalog at a given mass and redshift are selected (see also~\citealp{Huertas2016}). We obtain similar results. We also provide in table~\ref{tbl:comp} an estimate of the completeness for embedded components (bulges and disks). 

\begin{table}
\centering
\begin{tabular}{|c |c |c| c| }

\hline
z        &All&  $Q $& $SF$ \\
\hline
\hline
0-0.5   &   9.0    &  9.16 &  8.98 \\
\hline
0.5-1.0&  9.75   &  9.91 &  9.79  \\
\hline
1.0-1.4 &  10.3   & 10.38 & 10.28 \\
\hline
1.4-2.0 &  10.7  & 10.72 & 10.69\\
\hline
\end{tabular}
\caption{Stellar mass completeness of the sample used in this work. We show the values for galaxies (All), quiescent (Q), star-forming (SF).}
\label{tbl:comp}
\end{table}

The key measurement added by this work is a one/two component (bulge/disk) decomposition of the surface brightness profiles in different filters. In the following we describe the details of the methodology as well as the procedure used to estimate the uncertainties. 

\section{Best-model selection with deep learning}
\label{sec:mod_sel}

A fundamental problem with surface brightness profile fitting is to decide how many components are needed to model the galaxy light. By using two component models, we force the fitting algorithm to find a solution with 2 components within a given set of constraints even if the galaxy might be better fitted with a single \Sersic profile or with another combination of profiles. This can lead to unphysical solutions, introducing a systematic error in our subsequent analysis of bulges and disks properties. The reason is that some light might be associated with a bulge and/or a disk even if there is not such a component in the galaxy. This systematic uncertainty can potentially dominate over random uncertainties when performing a scientific analysis (e.g. \citealp{Meert2015}).

Several works have used a statistical approach to tackle this problem. By looking at the residuals of the resulting fits it is possible to establish a probability that adding a profile actually improves the fit (e.g. \citealp{Simard2011, Meert2015}). This is sometimes combined with a visual inspection \citep{Margalef2016}. This approach still has the problem that a better fit does not necessarily mean a physically meaningful result and that the light is actually properly associated to bulges and disks (\citealp{Allen2006,Mendel2014,Head2014,Mendez2017,Lackner2012}). 

Here we introduce a novel alternative technique based on unsupervised feature learning (deep learning). The main novelty is that the best model to fit a galaxy is set a-priori, instead of by looking at the residuals maps a-posteriori. The objective is then to measure, given a galaxy image, which analytic model among a finite set of possibilities is preferred to describe the surface brightness distribution. Recall that this is different from a morphological classification. We are not aiming at obtaining the \emph{true} morphology but to assess if a given analytical model is appropriate to describe the galaxy.

We proceed in two main steps described in the following. 

\subsection{Training on simulated analytic galaxies}

We first simulate a set of $100.000$ synthetic galaxies reasonably spanning all the range of structural parameters expected (see table \ref{tab:table_sim}) using the \textsc{GalSim} code\footnote{\url{http://galsim-developers.github.io/GalSim/index.html}}. Images are convolved with a real PSF and realistic noise from CANDELS images is added as explained in section~\ref{sec:sims}. For this particular application, we only simulate one filter (F160W) that will be used to define the model to be fitted. The H-band filter is chosen as a reference since it is the detection band and also the deepest. \\

\begin{table}
     \centering
     \begin{tabular}{|c|c|c|c|c|c|c|}
      \hline
    $magT_{F160}$ & $B/T$ &  $n_b$ & $n_d $& $re_b (")$ & r$e_d(") $& $b/a (b,d$)\\
      \hline
       \hline
     18-24 & 0-1 &   0-6 & 1 & 0-1.5 & 0-3 & nc\\
      \hline
      \end{tabular} 
      \caption[Range of values used in the simulations]{Range of values used in the simulation for the bulge and the disk components [See text for details]. $n$ is the \Sersic index, $r_e$ the effective radius, and $b/a$ the axis ratio. Magnitude on bulge and disk are assigned depending on the B/T and the total magnitude values. No constraints are applied on the axis ratio. Sizes are taken from the real catalog in multi-bands simulations since we want to kept the wavelength behavior, while a log-uniform distribution is used for mock galaxies of the training sample. }
       \label{tab:table_sim}
 \end{table}

We then define 4 types of profiles among the simulated galaxies:

\begin{itemize}
\item Pure \Sersic: $B/T>0.8$ and $n_{bulge}>2.5$. These are galaxies for which the surface brightness profile should be  well described with a Single \Sersic model.
\item Pure Exponential: $B/T<0.2$ or $B/T>0.8$ and $0.5<n_{bulge}<1.5$. Objects that are disk dominated thus for the surface brightness profile is well captured with a single exponential profile or a one component \Sersic profile with a \Sersic index <2.

\item Bulge + Exponential: $0.2<B/T<0.8$ and $n_{bulge}>2.5$. Systems that clearly require two components, one with an exponential profile and another with a large \Sersic index.
\item Pseudo-bulge + Exponential: $0.2<B/T<0.8$ and $n_{bulge}<2$. Systems that still require two \Sersic components, but both with low values of the \Sersic index. 
\end{itemize}

Independent Convolutional Neural Networks (CNN) are trained in a binary classification mode to isolate the given type of profile from the others in the simulated galaxies. An introduction to CNNs is out of the scope of this work. For more information, we refer the reader to \citealp{Dominguez2017} and \citealp{Tuccillo2017} where more details are given. In this work we train 4 different machines with the same architecture. The input of the network is  a simulated 2D image (with noise and PSF) centered on the galaxy (64$\times$64 pixels) and the output is a probability that the image is described by the model it was trained to identify. The model has 4 convolutional layers of increasing depth (from 16 to 64) and 2 fully connected layers. A $3\times3$ max pooling is performed after each convolutional layer to reduce the number of parameters and a $10\%$ dropout is applied during training to avoid over-fitting. Additionally, a $1\%$ gaussian noise is added in the first layer to avoid that the network learns features on the noise pattern. The model configuration was established after testing different architectures. Slight modifications do not change the main results. The model is trained until convergence and evaluated on the validation dataset. 

\begin{figure*}
\begin{center}
$\begin{array}{c c}
\includegraphics[width=0.46\textwidth]{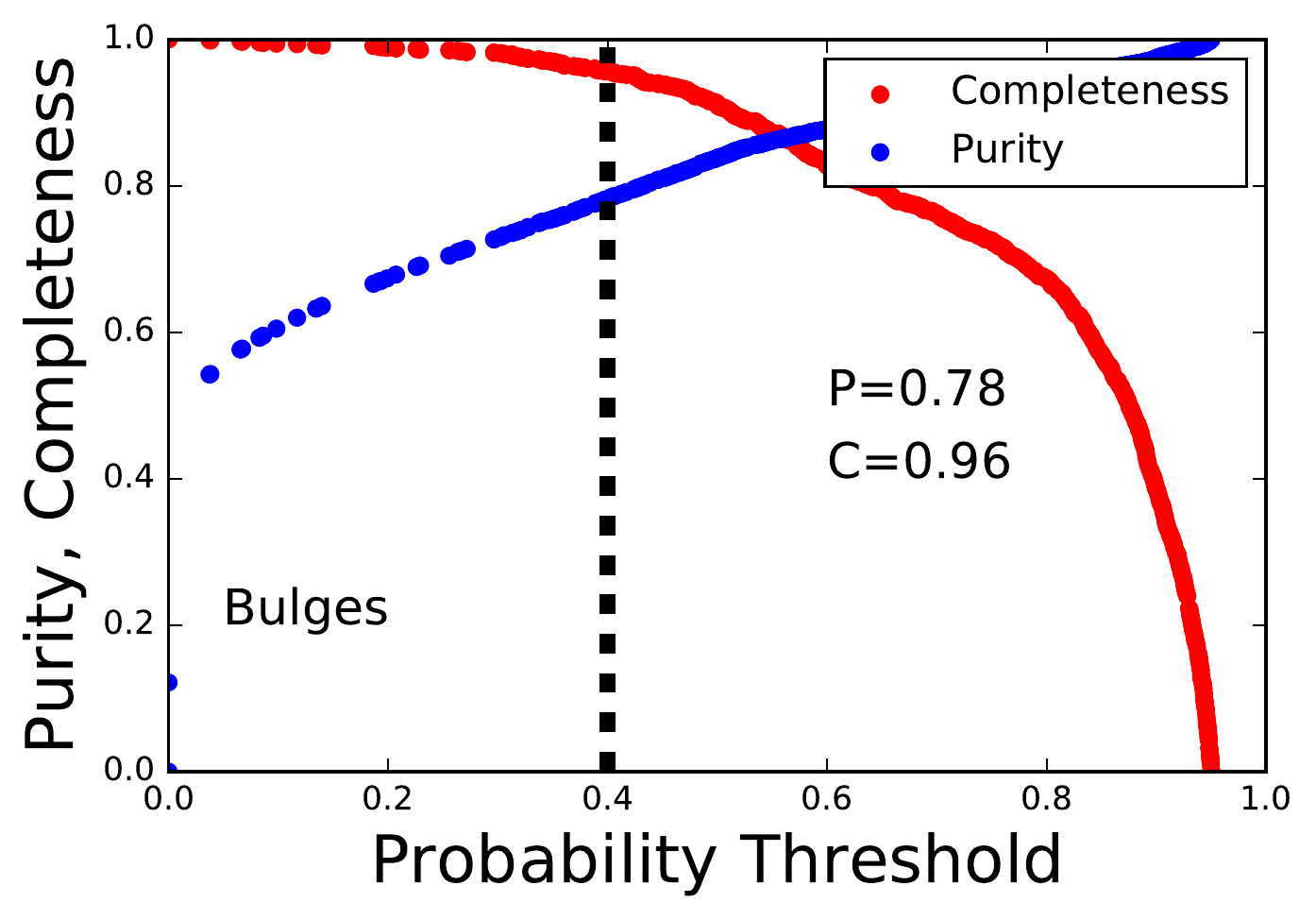} & \includegraphics[width=0.46\textwidth]{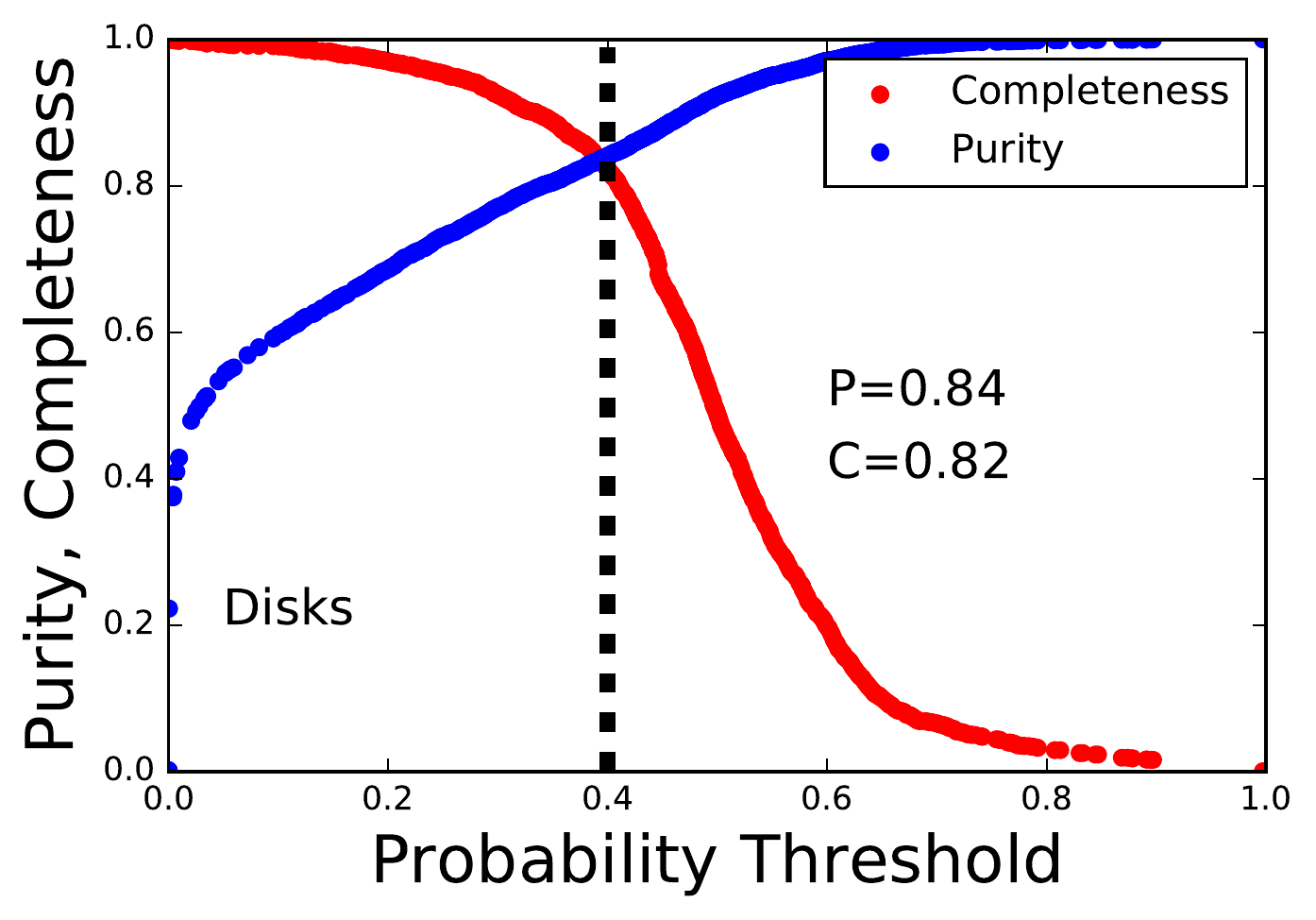} \\
\includegraphics[width=0.46\textwidth]{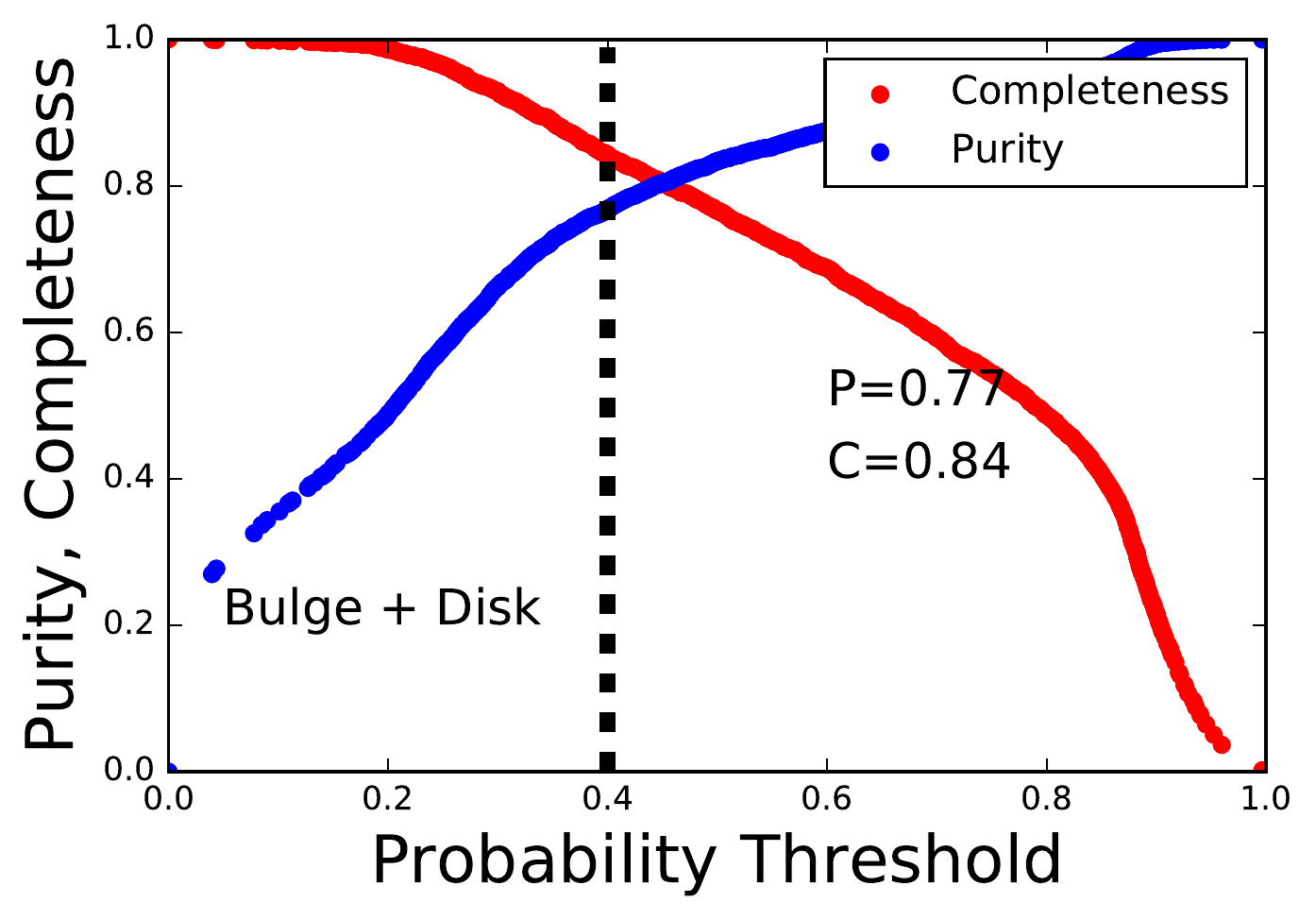} & \includegraphics[width=0.46\textwidth]{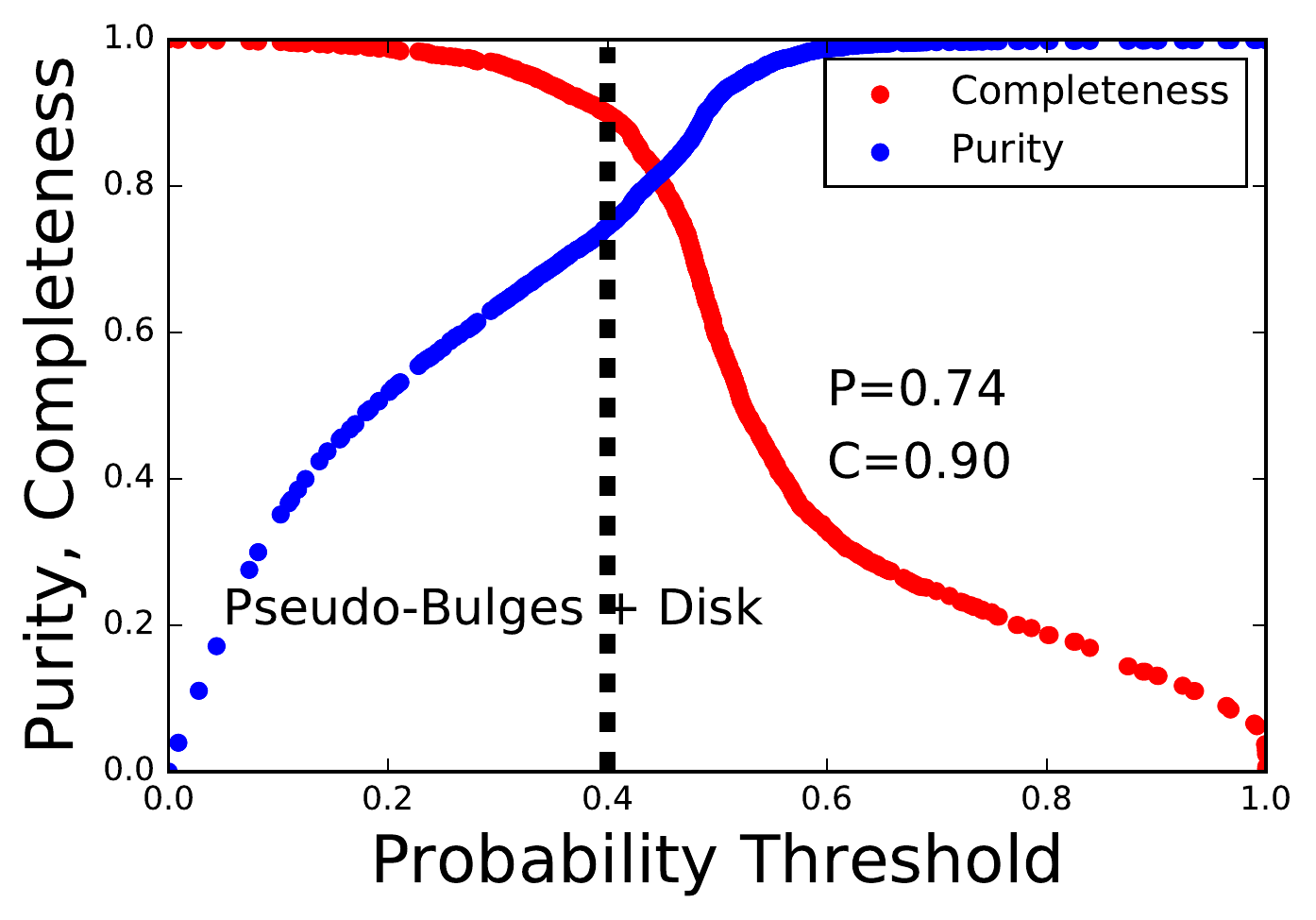} \\

\end{array}$
\caption{Purity and completeness as a function of the probability threshold for the 4 types of models considered in this work. The labels in each plot show the values for the adopted threshold of P=0.4 (black dashed line).} \label{fig:model_sel}
\end{center}
\end{figure*}

\begin{figure*}
\begin{center}
$\begin{array}{c c}
\includegraphics[width=0.43\textwidth]{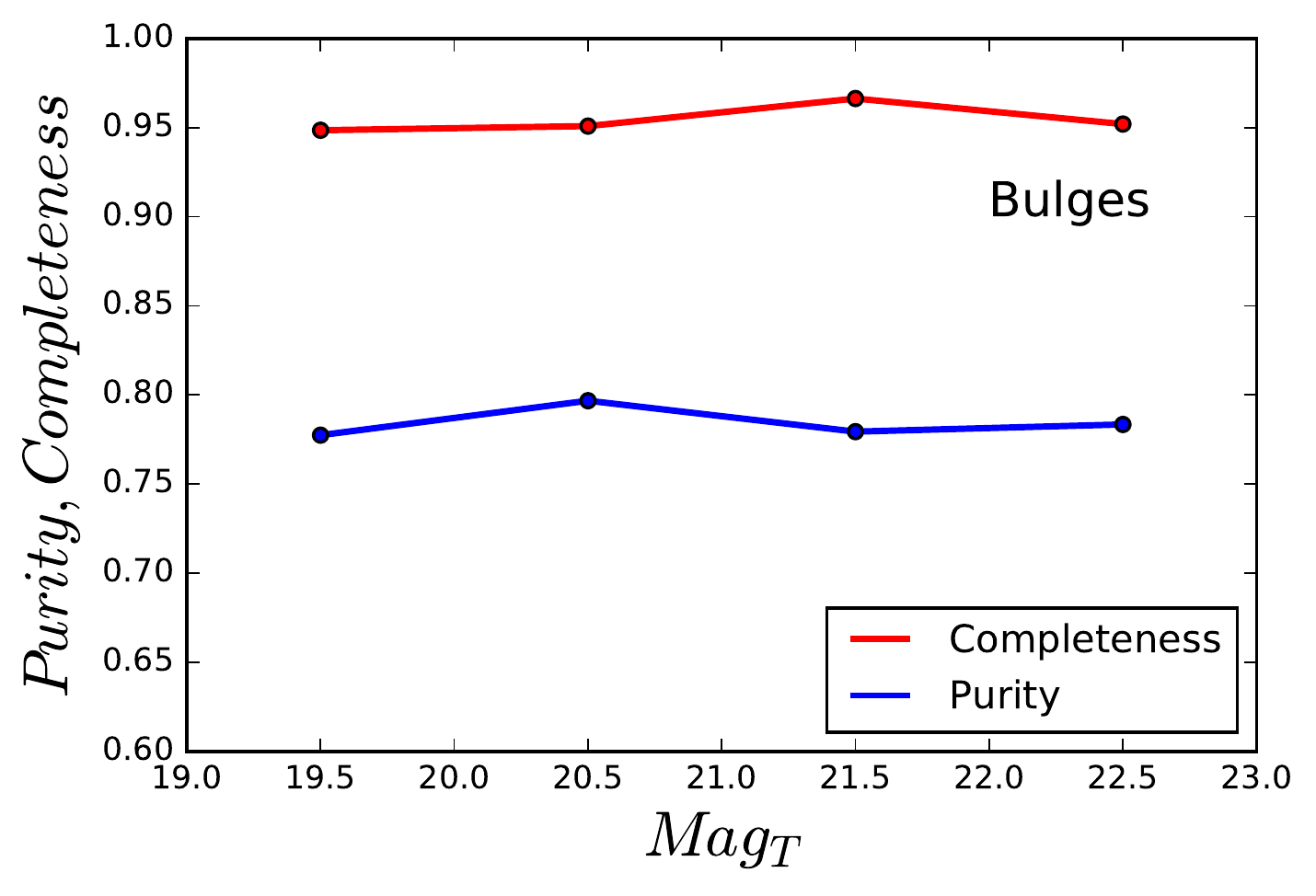} &
 \includegraphics[width=0.43\textwidth]{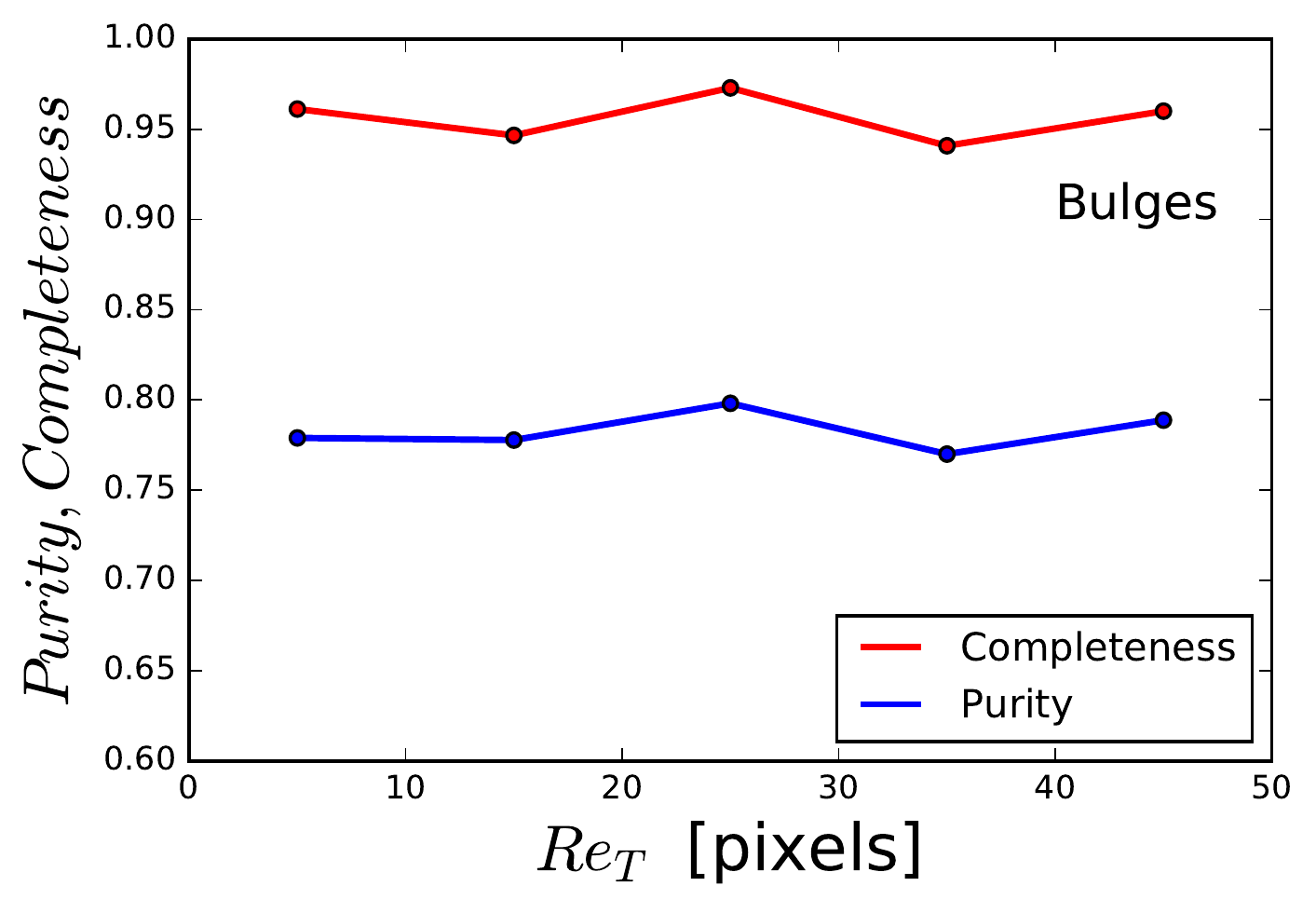} \\
\includegraphics[width=0.43\textwidth]{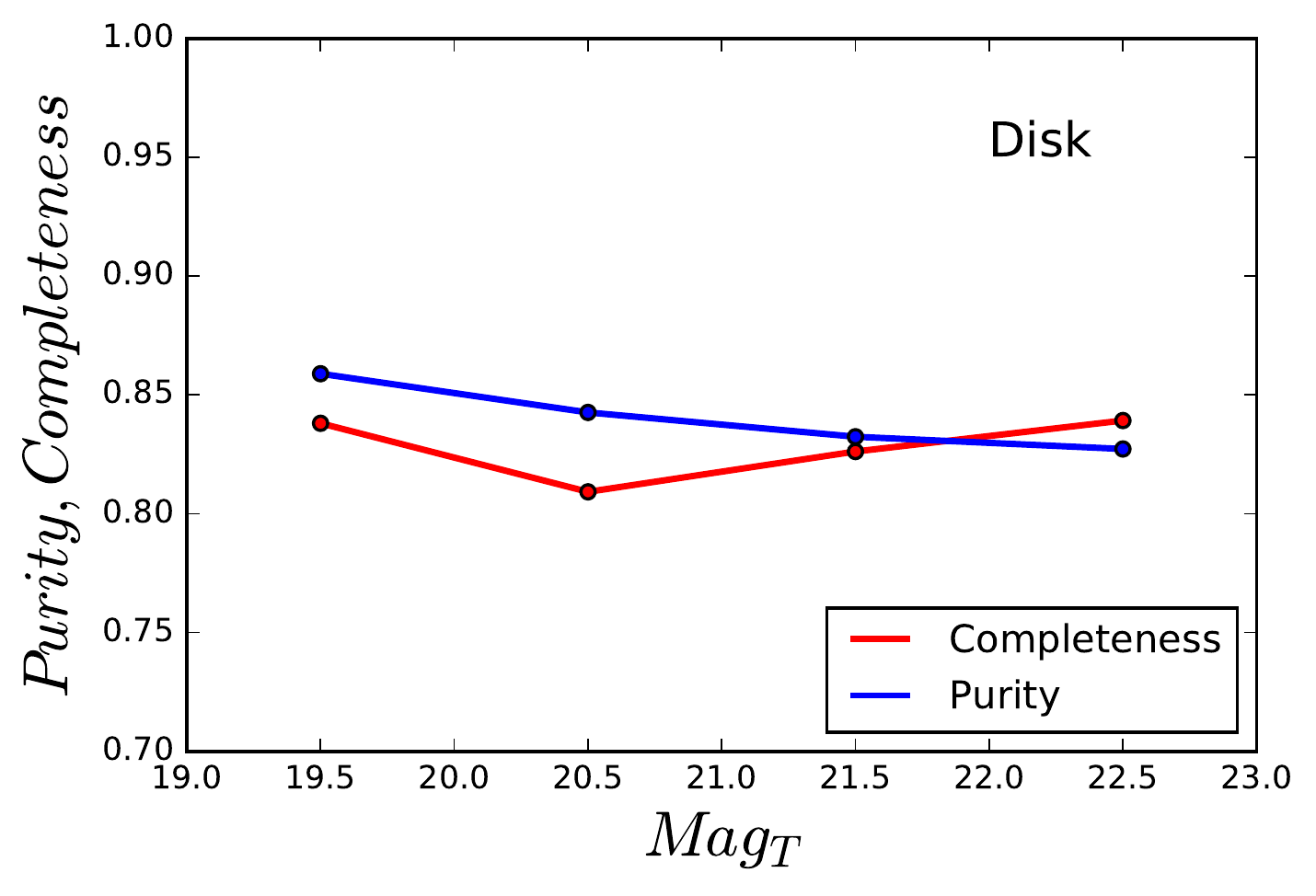} &
 \includegraphics[width=0.43\textwidth]{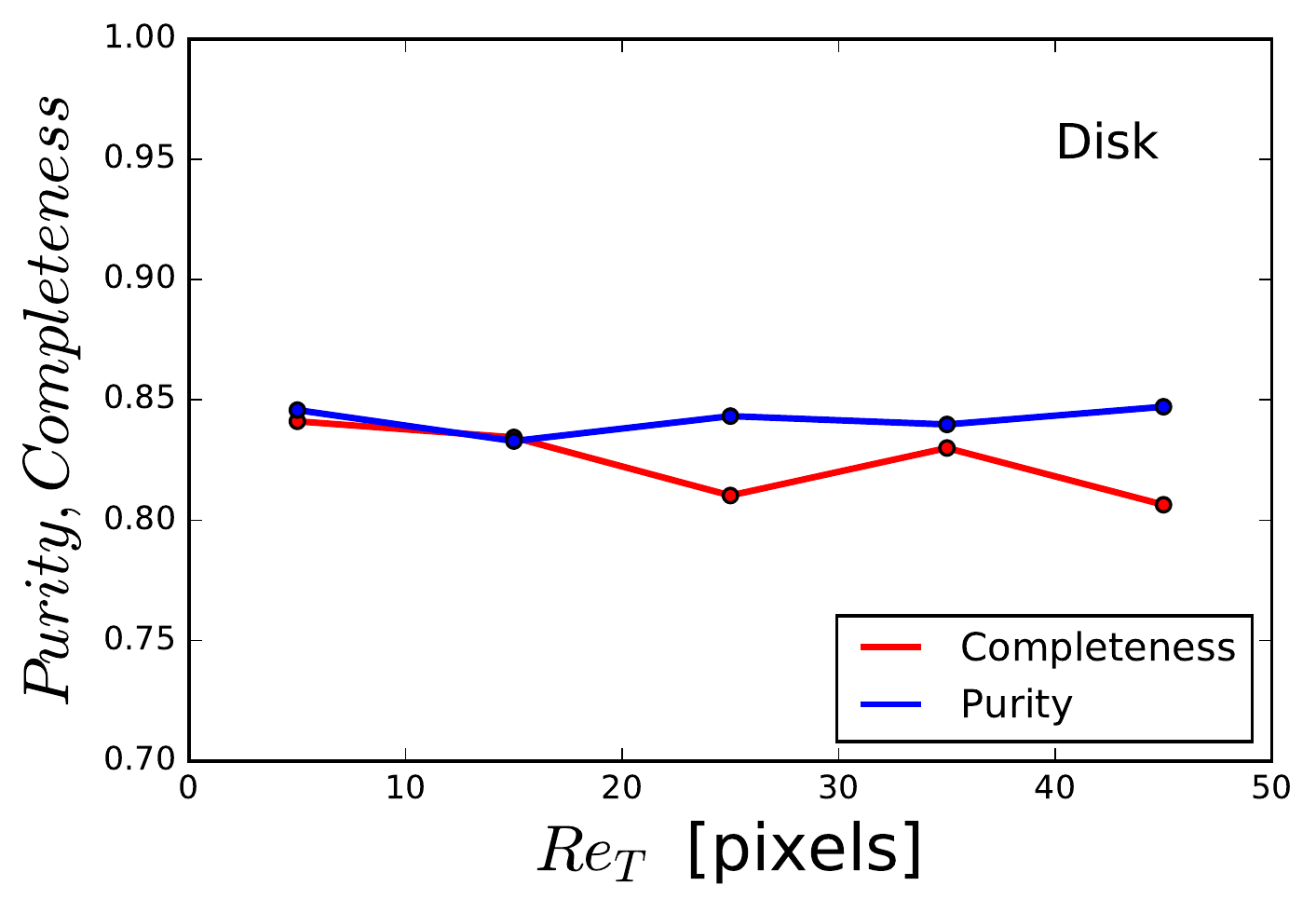} \\ 
\includegraphics[width=0.43\textwidth]{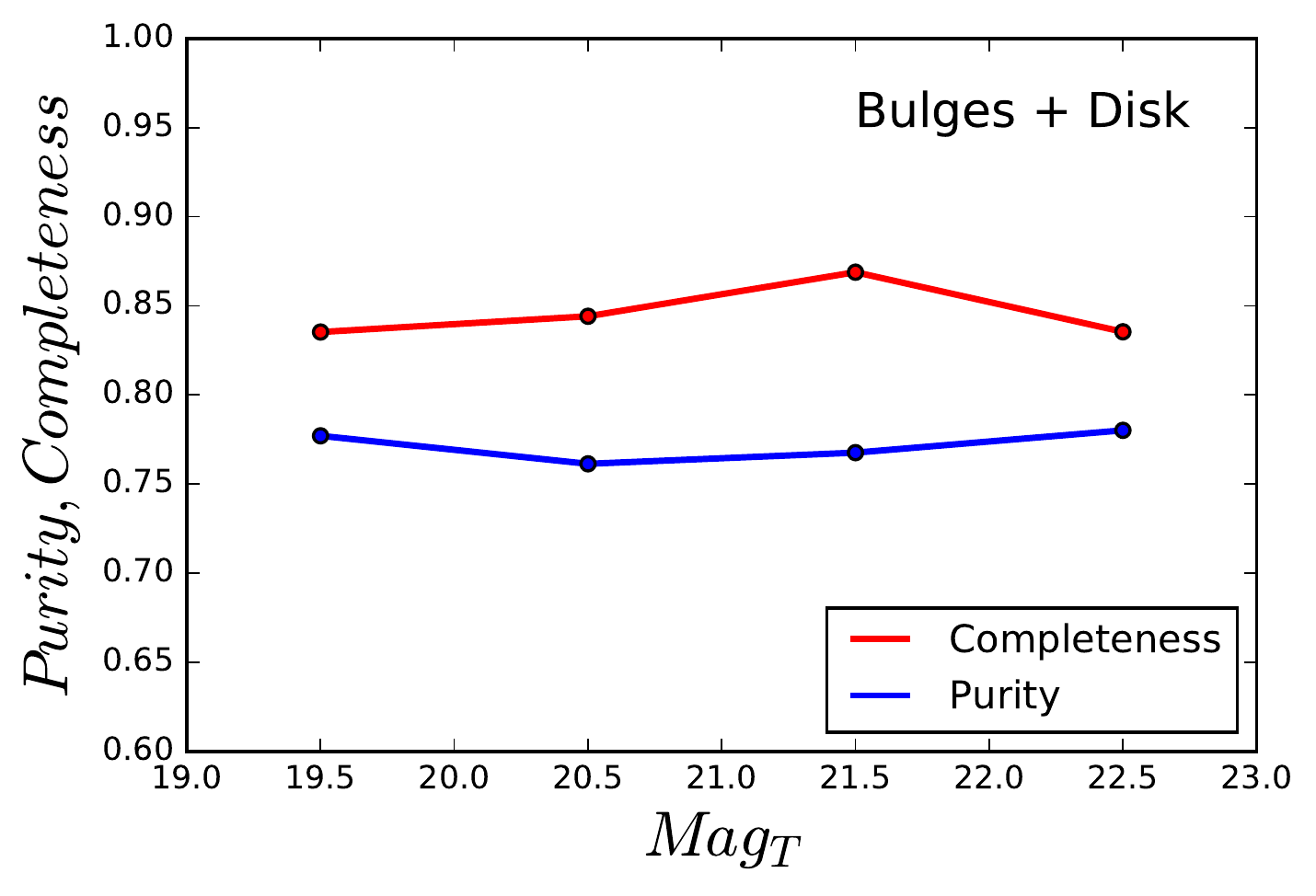} &
 \includegraphics[width=0.43\textwidth]{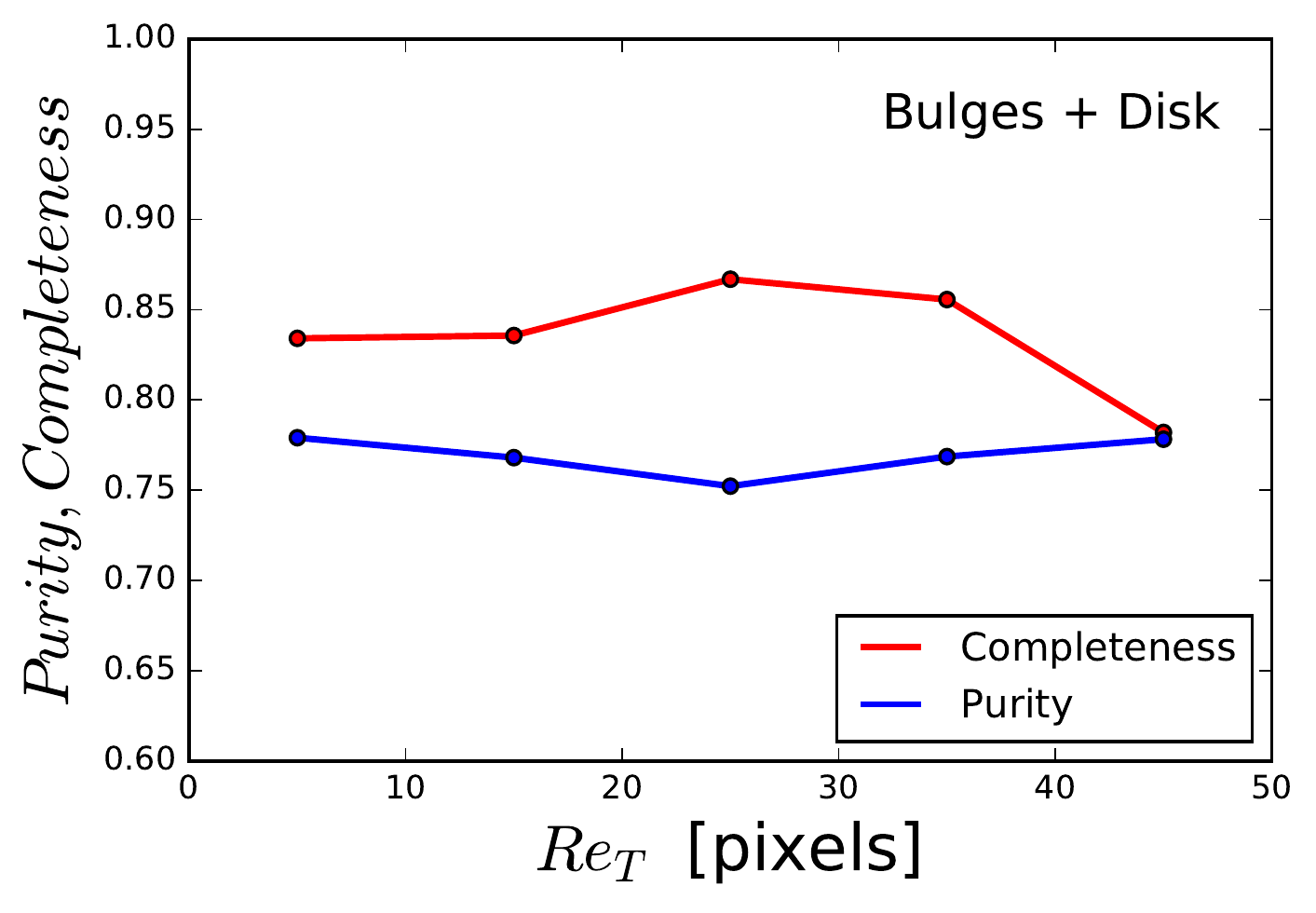} \\
 \includegraphics[width=0.43\textwidth]{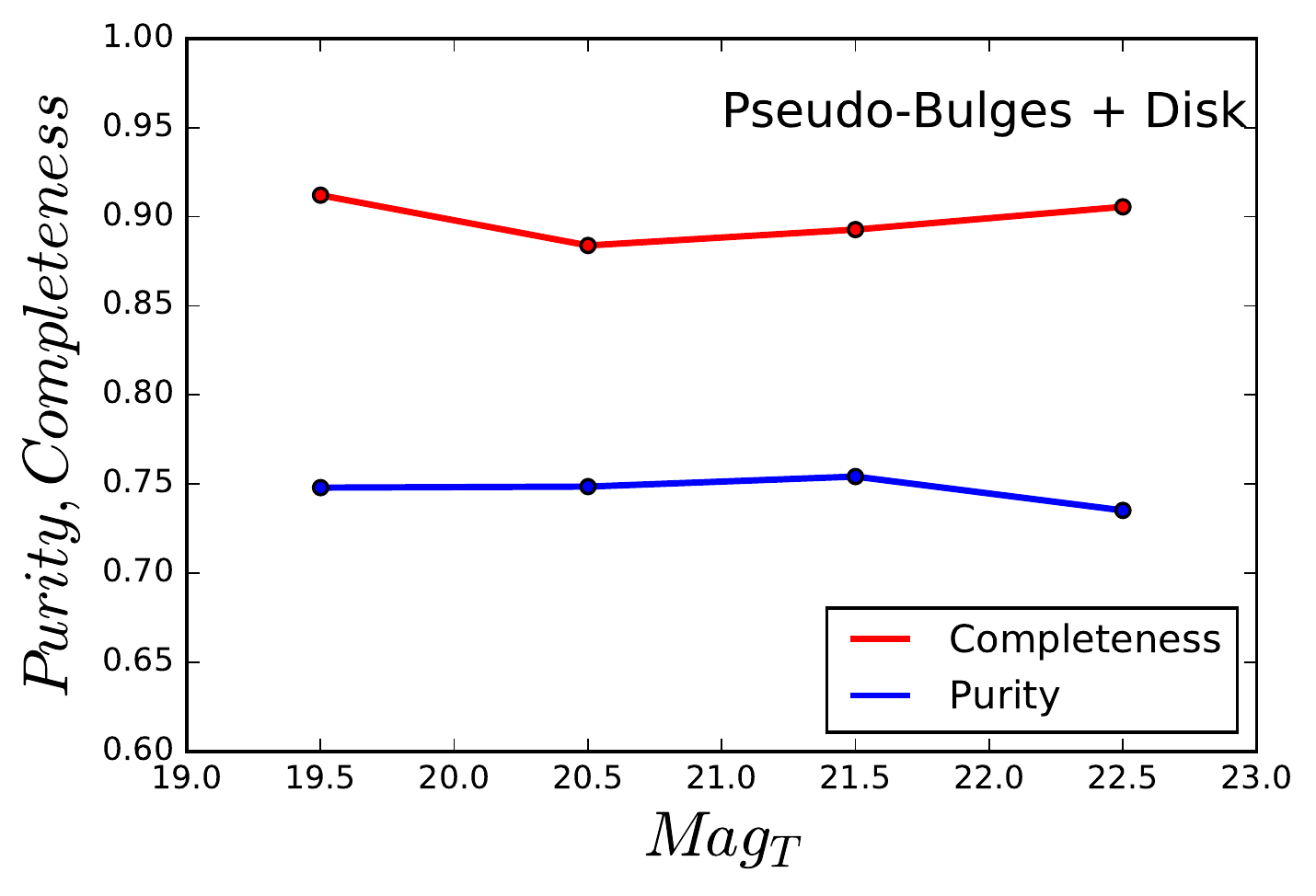} &
 \includegraphics[width=0.43\textwidth]{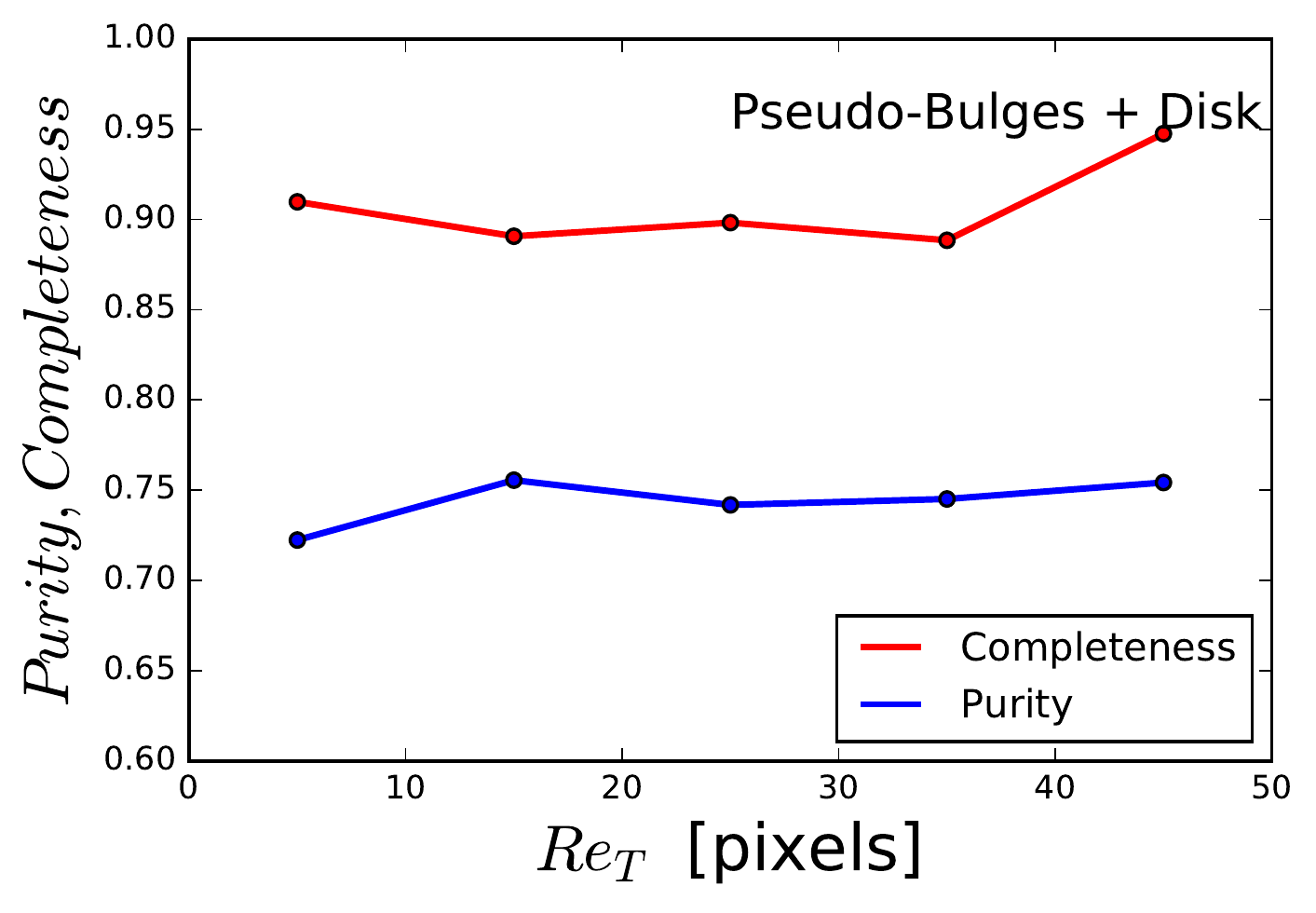} \\
\end{array}$
\caption{Purity and Completeness at fixed threshold (0.4) as a function of the total magnitude on the left panels and total half light size.}
\label{fig:model_sel2}
\end{center}
\end{figure*}

At the end of the training process, each simulated galaxy has 4 associated probabilities. Recall that the probabilities do not add to 1 since they were estimated with four independent CNNs. Since for the simulated galaxies, we know the model that was generated, we can quantify the ability of the CNN to distinguish between different profiles on an independent test dataset which was not used during the training phase. Following a standard procedure, we use the area under the ROC curve. The ROC curve is defined as the relation between the True Positive Rate (TPR) and False Positive Rate (FPR) for different probability thresholds. Ideally one would like to have large TPR and small FPR values. So the largest is the area under the ROC curve, the better. The classification is always a trade off between both. In order to define the optimal threshold needed for the classification of our sample, we computed the following two parameters: Purity ($P$) and Completeness ($C$):
$$C=TPR=\frac{TP}{TP+FN}$$
$$P=1-FPR=\frac{TN}{TN+FP}$$
TP and FP stand for true and false positives respectively, while TN and FN are true and false negatives. Specificity is a measurement of how contaminated a selection of a given class is by galaxies not belonging to that class. Sensitivity is a measurement of how well the machine recovers all galaxies belonging to a given class. In figure~\ref{fig:model_sel} we show how these two quantities change depending on the applied probability threshold. As expected, higher probabilities indicate more pure but less complete samples.
The plots confirm that the CNN models are able to distinguish between the 4 different types of profiles. We notice that a probability threshold of $p=0.4$ results in a reasonable trade-off between purity and completeness, around $80\%-90\%$. In order to test the robustness of this choice, figure \ref{fig:model_sel2} shows how C and P depend on galaxy properties (total magnitude and half light size) for each class. Both quantities are stable, confirming that the chosen threshold can be used over the full parameter space covered by the catalog.\\

\subsection{Knowledge transfer to real galaxies}

The above results are based on simulations. The critical step is to use the 4 machines to classify our real galaxies. Our aim is to provide, for each object, a probability that a given model (i.e. pure bulge, pure disk, bulge+exponential, pseudo-bulge+exponential) is preferred to describe its surface brightness profile. We insist that this is not a morphological classification and it is fundamentally different than the visual morphological catalog presented in \cite{Huertas2015}. It measures, given a set of available analytic profiles, which one is better suited to fit the surface brightness distribution of a given galaxy. This information is intended to be used as input for fitting the light profiles as explained in the following sections.

In transferring the trained machine to real data, there is a risk that the CNN does not provide reliable results since simulations and real galaxies are obviously different. For example, even though our simulations include realistic instrumental effects and noise, we did not include close companions in the training set nor irregular galaxies or bars. The validity of the knowledge transfer between simulations and real data is difficult to evaluate since, as opposed to the simulations, the \emph{true} profile is not known in the real sample 

\begin{figure*}
\begin{center}
\includegraphics[width=0.9\textwidth]{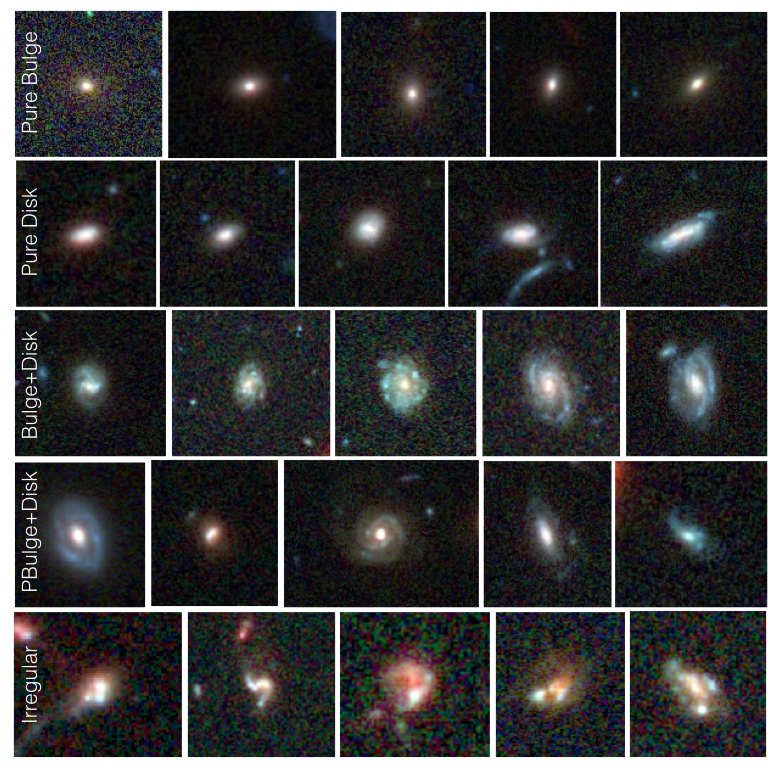} 
\caption{Color images of galaxies of the 4 types of models, and unclassified. From top to bottom: galaxies for which a pure bulge model is fitted, galaxies for which a pure disk model is preferred, galaxies for which a 2 component model with $n_b>2.5$ is preferred, objects for which a low \Sersic index bulge is the best solution, irregular/unclassified galaxies.} 
\label{fig:ex_stamps}
\end{center}
\end{figure*}

We perform some posterior sanity checks In order to verify that the trained model properly represents reality:

\begin{itemize}

\item  First, we visually inspect a significant number of color images of different classes provided by the CNN. Although, this is not a quantitative measurement, it allows to identify obvious errors. Examples of the galaxies classified according to the different types of profiles are shown in figure~\ref{fig:ex_stamps}. Generally speaking, when the algorithm indicates that 2 components are required, a bulge and a disk are observed in the color image.

 \item In terms of types of models, we find that for the majority of the galaxies ($\sim 55\%$) a bulge+exponential model is preferred.  $\sim20\%$ of the remaining galaxies are well fitted with an exponential model, $\sim15\%$ are preferentially fitted with 2 low \Sersic index components and $\sim10\%$ are well described with one \Sersic profile with $n>2$. The fraction of classical double component galaxies strongly depend on the stellar mass and  on the morphology. Since we are studying massive galaxies ($M _*>10.3 M_{\odot}$), we expect that most of them are characterized by two components. Our results is in rather good agreement with the predictions (\citealp{Huertas2016}). The majority of galaxies are described by a \emph{classical} two component model with a central bulge with large \Sersic index and an exponential disk.  However, it is worth noticing that if no model pre-selection is done and a 2 component fit is blindly applied to the entire sample, the systematic error reaches $\sim30\%$ and thus most probably dominates the error budget. 
 
  \item We measure that $<10\%$ of the galaxies in the real sample have all 4 probabilities below $P=0.4$, i.e. their surface brightness profile are not properly described by any of the 4 considered models. All other galaxies have at least one of the probabilities above this threshold. This is a good indication that the network did not find drastic differences between the training/test simulated samples and the real data. Otherwise the probability distributions would have been very low for all real galaxies.
 
 \item For galaxies for which no model is preferred ($P<0.4$), the visual classification taken from \cite{Huertas2015} indicates that the vast majority of them ($>80\%$) are classified as irregulars with high probability (see also figure~\ref{fig:ex_stamps}). It is expected that their surface brightness profile is not properly described by any of the considered models. We notice additionally that for the irregulars for which an optimal profile was actually found a \emph{pure} disk like profile is the best solution.
 \end{itemize}

We include in the catalog these probabilities, which indicate how accurately each of the four light profiles fits the light distribution of each galaxy.
This allows to select a model to fit a-priori and that way reduce systematic uncertainties as described in section~\ref{sec:errors}. 

\section{Multi-$\lambda$ fits with galfitM}
\label{sec:galfitM}
The main tools we used to perform the fits are \textsc{GalfitM} and \textsc{Galapagos-2} from the \textsc{MEGAMORPH} project \citep{haeussler2013,Vika2014}. They are based on \textsc{Galapagos} and \textsc{Galfit} \citep{Barden2012,Peng2002}. The main difference is that they can simultaneously fit all images at different wavelengths (as opposed to an independent fit for each band). As shown in the aforementioned works, the advantage of such an approach is that, by combining data from all filters, we effectively increase the $S/N$ and naturally use the color information. Therefore the fit is better constrained down to fainter magnitudes than when considering all bands independently. To accomplish this, the wavelength dependence of the structural parameters of galaxies is parametrized with a family of Chebyshev polynomials. The order of the polynomial for each quantity is a user-configurable parameter which sets the degree of freedom. The fitting algorithm then minimizes the coefficients of the function for each structural parameter. If the degree of freedom is equal to the number of filters, then the parameter is effectively independent in each band, as it is obviously the case for the fluxes. For the other parameters, the choice (of the degrees of freedom allowed) is a trade-off between allowing total independence or setting no variation with wavelength (thus reducing the number of free parameters). More details can be found in~\cite{haeussler2013} and \cite{Vika2014}.

There is no obvious way of selecting the optimal configuration. The wavelength dependence of the structural parameters will certainly vary from galaxy to galaxy. Our approach has been to empirically test different configurations and use them to estimate random uncertainties as discussed in section~\ref{sec:errors}. For each galaxy we fit 2 types of models: a 1 component \Sersic model and a 2 component \Sersic+Exponential model. Then, for each of the models we adopt three different setups for \textsc{GalfitM} as shown in table~\ref{tbl:table1}. In all setups, the fluxes of both components are left free, the centroids of galaxies are held constant over wavelength (we assume that the images were properly aligned). The position angles of the galaxy and the axis ratios are also kept constant since these quantities are not expected to present strong wavelength dependence. The most critical parameters are the \Sersic index and the effective radius. We explore the effect of the wavelength dependence of the size by allowing a quadratic variation in setups 1 and 4 and restricting to constant in setups 2 and 5. Additionally, the maximum degree of freedom is reduced to the number of bands used in each field. For the \Sersic index of the bulge (\Sersic+Exponential model) we only allow a linear variation (given that the bulge is normally dominated by old stellar populations, we do not expect a strong wavelength dependence of the \Sersic index). However, we changed the range from 0-8 in setups 4 and 5 and  to 2.5-8 in setup 6. This is used, as explained in section~\ref{sec:errors}, to evaluate our procedure for model selection based on CNNs (see section~\ref{sec:mod_sel}). The properties of all runs performed are summarized in table~\ref{tbl:table1}.

\begin{table*}
     \begin{center}
     \begin{tabular}{c|c|c|c|c|c|c|c|c|c|c|c|}
     \hline
       & & & x & y & mag & $r_e$ & n & q & pa \\
     \hline
      \multirow{2}*{\Sersic} & setup 1 & & 0 & 0 & 6   & 2 & 1 & 0 & 0 \\
      &setup 2 & & 0 & 0 & 6   & 0 & 1 & 0 & 0 \\
      &setup 3 & & 0 & 0 & 6   & 1 & 1 & 0 & 0 \\
     \hline
     \hline
      \multirow{4}*{\Sersic+Exp} & \multirow{2}*{setup 4} &BULGE & 0 & 0 & 6   & 2 & 1 & 0 & 0 \\
     & & DISK  & 0 & 0 & 6   & 2 & fix & 0 & 0 \\
     \cline{3-10}
     & \multirow{2}*{setup 5} & BULGE & 0 & 0 & 6   & 0 & 1 & 0 & 0 \\
     & & DISK  & 0 & 0 & 6   & 0 & fix & 0 & 0 \\
    \cline{3-10}
     & \multirow{2}*{setup 6} & BULGE & 0 & 0 & 6   & 1 & 1* & 0 & 0 \\
     & & DISK  & 0 & 0 & 6   & 1 & fix & 0 & 0 \\
      \hline
     \hline
     \end{tabular}
     \end{center}
     \caption{Orders of the polynomial functions used in the \textsc{GalfitM} run for each parameter. Each galaxy was fitted with 2 models (\Sersic / \Sersic + Exp) and three different setups. 0=constant over all wavelengths, 1=linear, 2=quadratic function, 6 =free. The main difference between the setups resides on the degree of freedom allowed in the size wavelength dependence. For setup 6, the \Sersic index of the bulge component is only allowed to vary in the range $2.5-8$.  }
     \label{tbl:table1}
 \end{table*}

  \section{Quantification of uncertainties}
 \label{sec:errors}
 
 In the following sections, we evaluate the overall accuracy of our final bulge/disk catalog using different methods.
 
 \subsection{Accuracy of model selection}
\begin{figure*}
\begin{center}
\includegraphics[width=\textwidth]{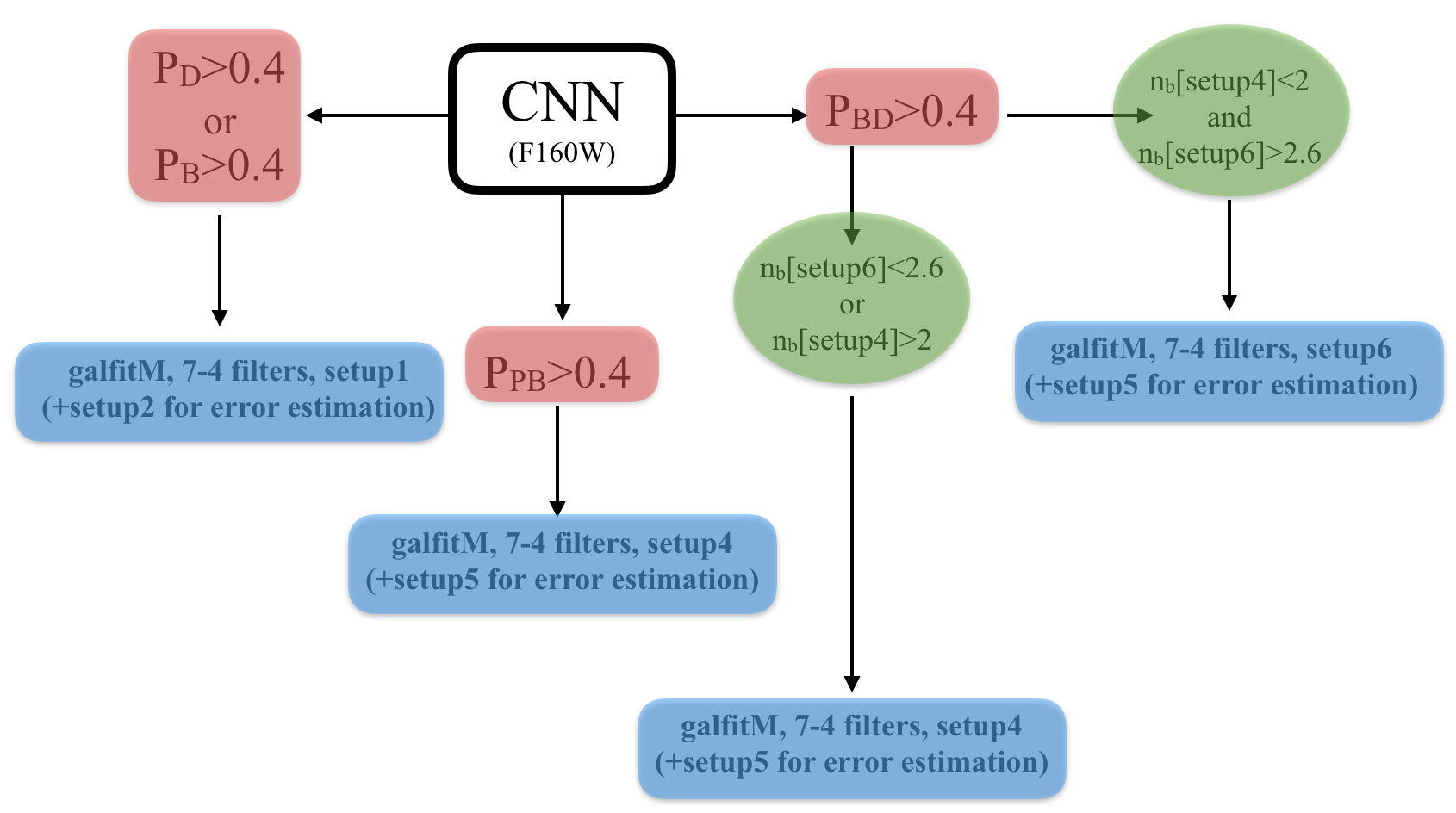} 
\end{center}
\caption{ Flow chart showing the use of the different setups depending on the CNN based classification. This is the selection used to build the released \emph{Gold catalog}. }
\label{fig:flow_chart}
\end{figure*}

\begin{figure}
\begin{center}
\includegraphics[width=0.44\textwidth]{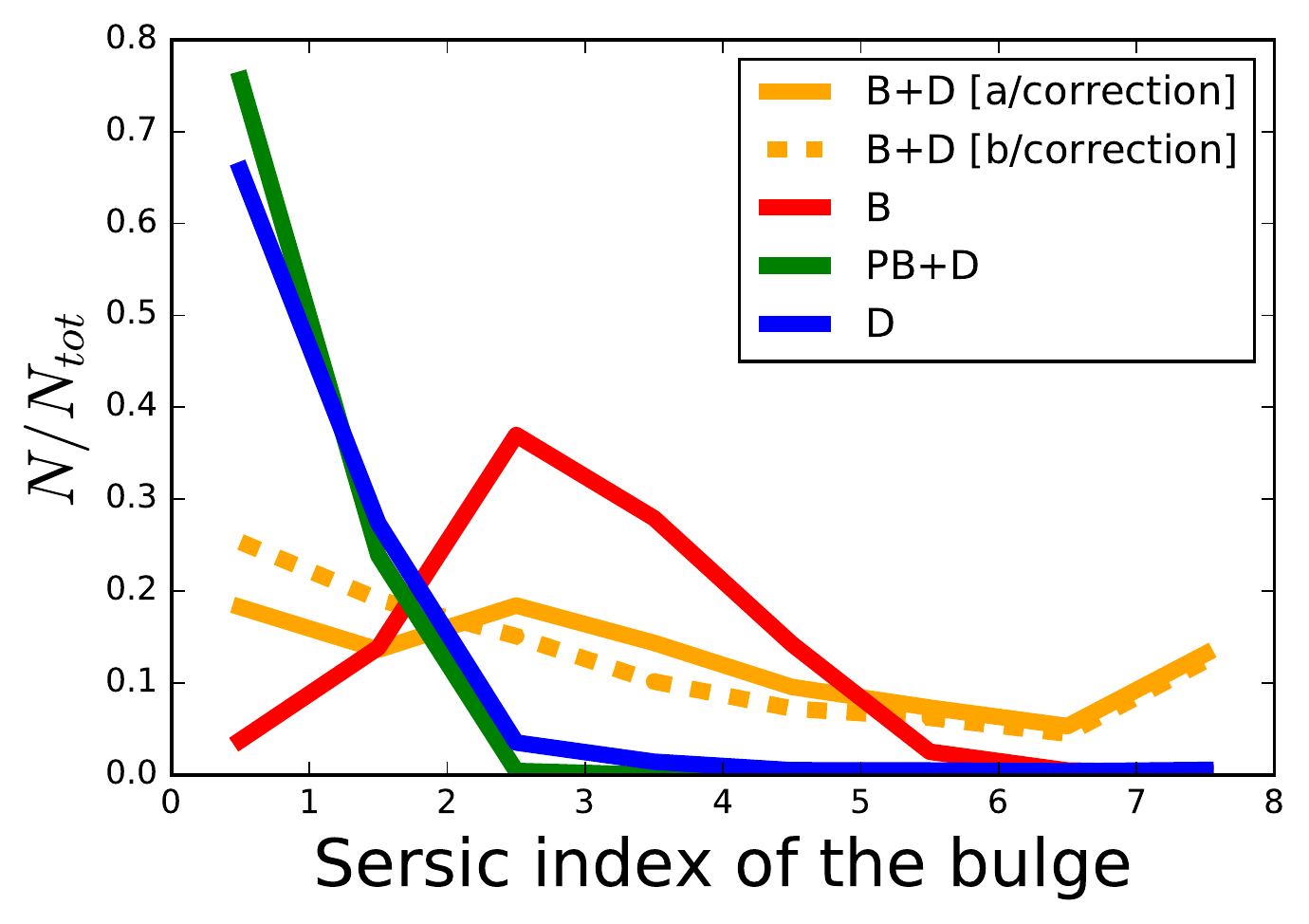} 
\caption{H band bulge \Sersic index distribution for the different CNN derived models (see text for details). For bulges (B) and disks (D), the global \Sersic index is plotted. For the B+D and PB+D we plot the \Sersic index of the bulge component only. Moreover, for the B+D systems, we show the distribution of the \Sersic index before (yellow dotted line) and after the correction.}
 \label{fig:NB_dist}
\end{center}
\end{figure}

\begin{figure}
\begin{center}
\includegraphics[width=0.46\textwidth]{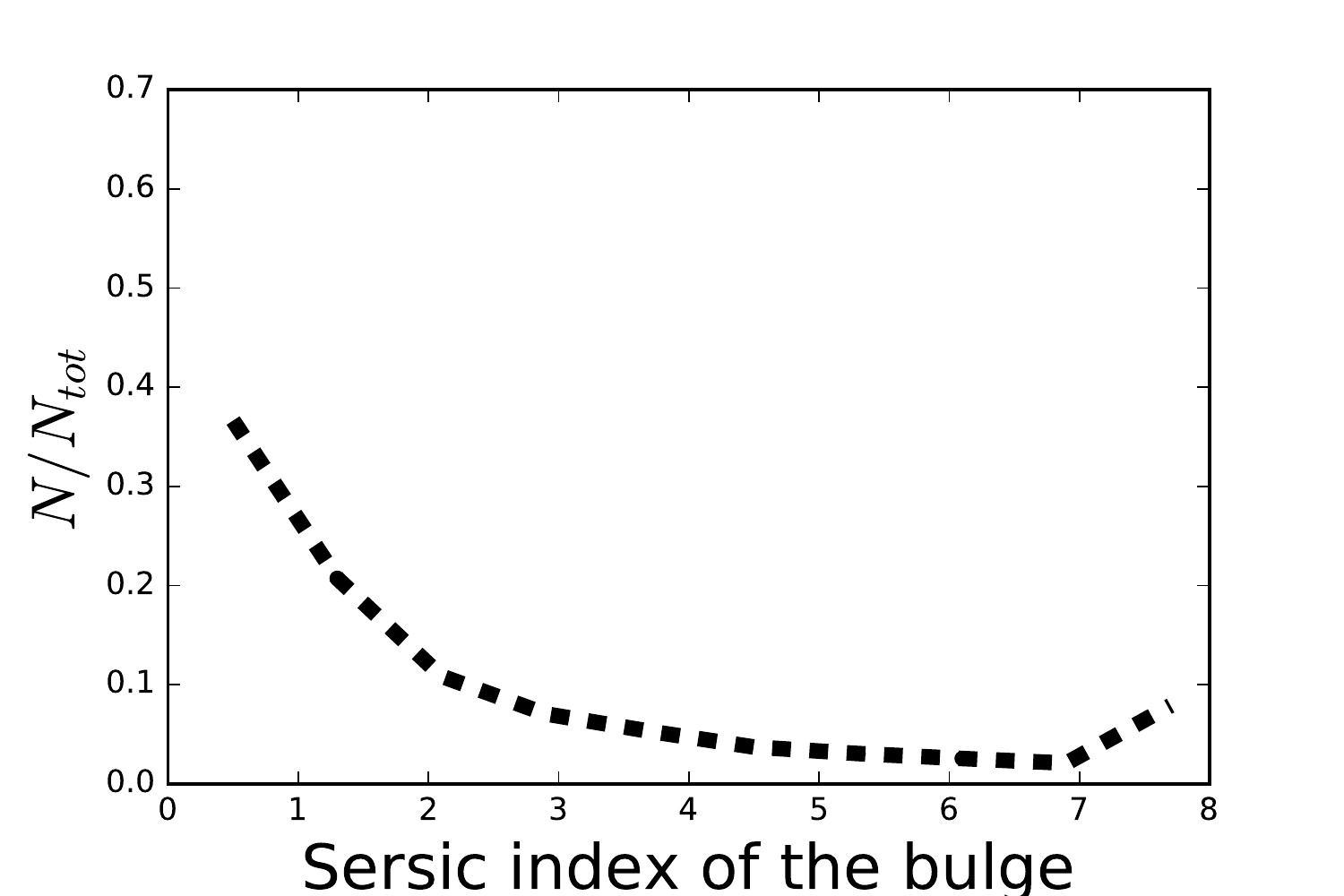} 
\caption{H band bulge \Sersic index distribution of all galaxies with no model pre selection. If no model selection is applied, $\sim50\%$ of the bulges have \Sersic indices lower than 2. } \label{fig:NB_dist_all}
\end{center}
\end{figure}

 In order to test the validity of our methodology to select the best model, we compare the outputs delivered by \textsc{galfitM} with the expectations according to the CNN based classes. If both the best model \emph{class} and the fitting procedure work as expected, one would expect that the best fit model converges towards the expected \emph{best} profile. In figure~\ref{fig:NB_dist}, we show the H-band \Sersic index distributions of the bulge component for galaxies classified in the 4 profile classes detailed previously (using a probability threshold of 0.4). For obvious reasons, for objects for which a single \Sersic model is preferred, we plot the global \Sersic index as well as for pure exponential profiles. Figure~\ref{fig:flow_chart} summarizes the criteria used for the selection. We clearly see that the distributions are different for every type of model and follow the expected trends. Pure disks and pseudo-bulges have almost all \Sersic indices lower than 2. Pure bulges peak at values of $n\sim 3-4$. The distribution for objects that require a 2 component fit extends to large values as well. However, there is a fraction of objects for which our CNN based model selection technique would have preferred a model with a high \Sersic index bulge while the fitting procedure converges to a solution with a lower value (dashed orange line in figure~\ref{fig:NB_dist}). Given that we expect a contamination of $\sim15\%$ (see fig.~\ref{fig:model_sel}), our results is higher than anticipated. In order to test if this is a problem of \textsc{galfit} converging to a local minimum, we use the results of setup 6 (table~\ref{tbl:table1}) in which the \Sersic index of the bulge was forced to be larger than 2.5 at all wavelengths. For approximately $50\%$ of the objects, the fitting procedure converged to a new solution with $n_b$ exactly equal to 2.5, i.e. the boundary condition. We considered therefore that for these objects a low \Sersic index bulge is the best solution. However, for the remaining $50\%$, the new setup provided a solution with $n_b>2.5$ in agreement with the CNN classification. The corrected distribution is shown in figure~\ref{fig:NB_dist}. As expected, the number of bulges with $n_b<2$, has been reduced and is now compatible with the expected contamination level. We notice that, if no model selection is applied, half of the bulges have low \Sersic index values ($<2$) as shown in figure~\ref{fig:NB_dist_all}.

\begin{figure}
\begin{center}
\includegraphics[width=0.45\textwidth]{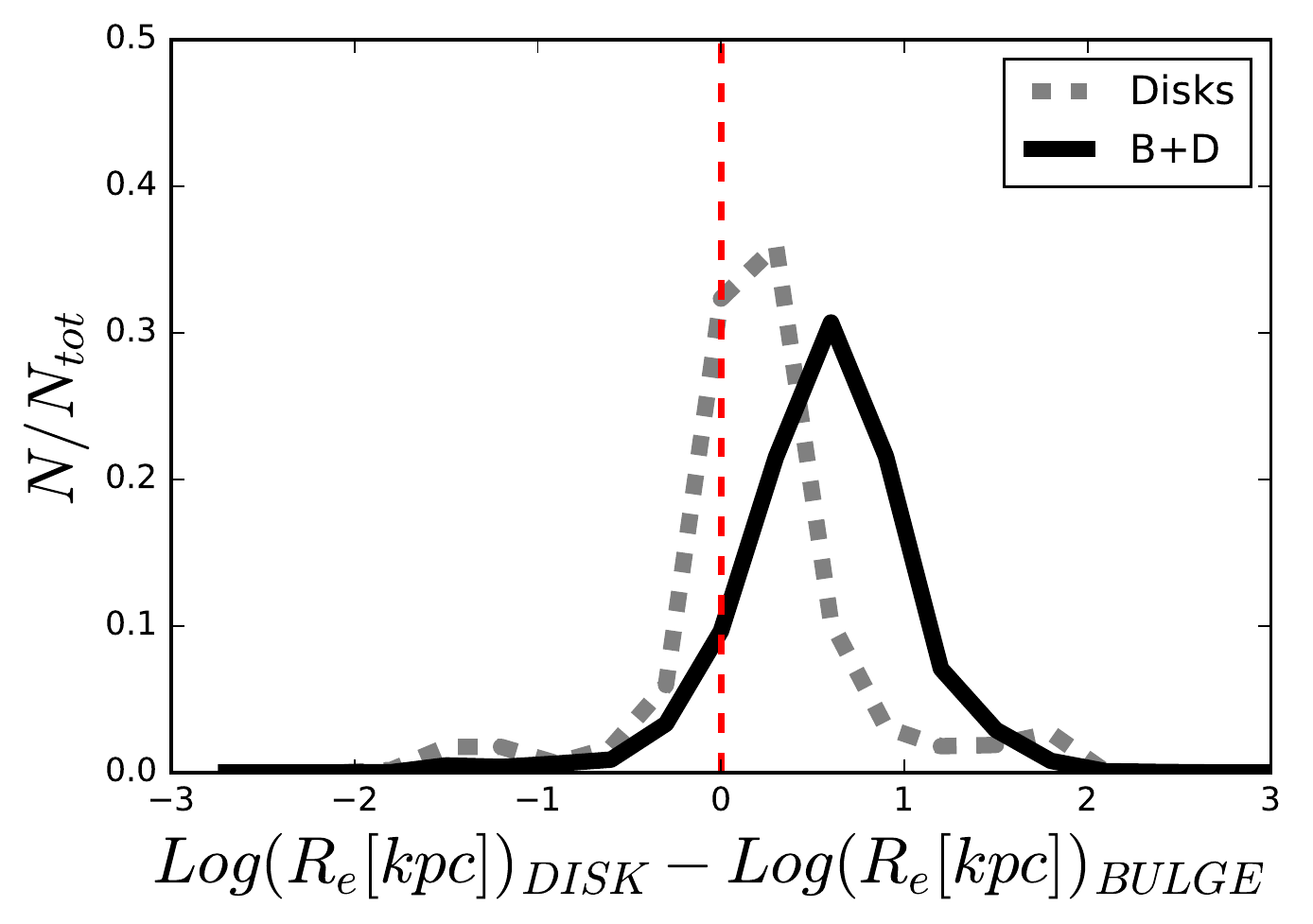} 
\caption{Distribution of the difference between the sizes of disks and bulges for our 2 component systems (black line). The vast majority of the objects have larger disks as expected. The dashed line shows the same distribution for one component disks according to the CNN classification (see text for details). For roughly half of the objects, the fit converges to a solution in which the bulge is larger than the disk. This confirms that 2 components are probably not needed for these systems.} \label{fig:rb_red}
\end{center}
\end{figure}

In figure~\ref{fig:rb_red}, we show the distribution of the difference between the sizes of disks and bulges for objects identified as requiring two components. As expected, for the vast majority of the objects, the disk component has a larger effective radius than the bulge. For comparison, we also show the same distribution for objects that were classified as pure disks by the CNN model. In that case, roughly half of the population has a bulge larger than the disk. This confirms that a 2 component model is not suited for these galaxies. Including the bulges of these objects in any scientific analysis would definitely introduce a systematic error that could potentially bias the results.

 \subsection{Statistic systematic uncertainties with simulations}
\label{sec:sims}

We use simulations to estimate the global accuracy of the fits as is commonly done in the literature. In particular, one key quantity to calibrate is the limiting magnitude (or $S/N$) below which measurements remain unbiased. From \cite{vanderWel2012}, we know that for galaxies fainter than $F160W=24.5$ (at the \emph{wide} depth), statistical errors on the structural parameters derived from 1 component \Sersic fits exceed $\sim 20\%$. This threshold needs to be calibrated for 2 component fits. 

 In order to do so, we follow a standard procedure based on simulations of analytic profiles as done in several previous works (e.g. \citealp{haeussler2007, vanderWel2012,Delaye2014}). Namely, we generate mock galaxies with 2 analytic profiles (a \Sersic profile for the bulge component and an exponential for the disk). The structural parameters of each component as well as the total magnitudes and $B/T$ ratios are randomly distributed to span the range of values found in real data. We then convolve each profile with a real PSF and embed the galaxies in a real background. 

\begin{figure}
\centering
\includegraphics[width=86mm]{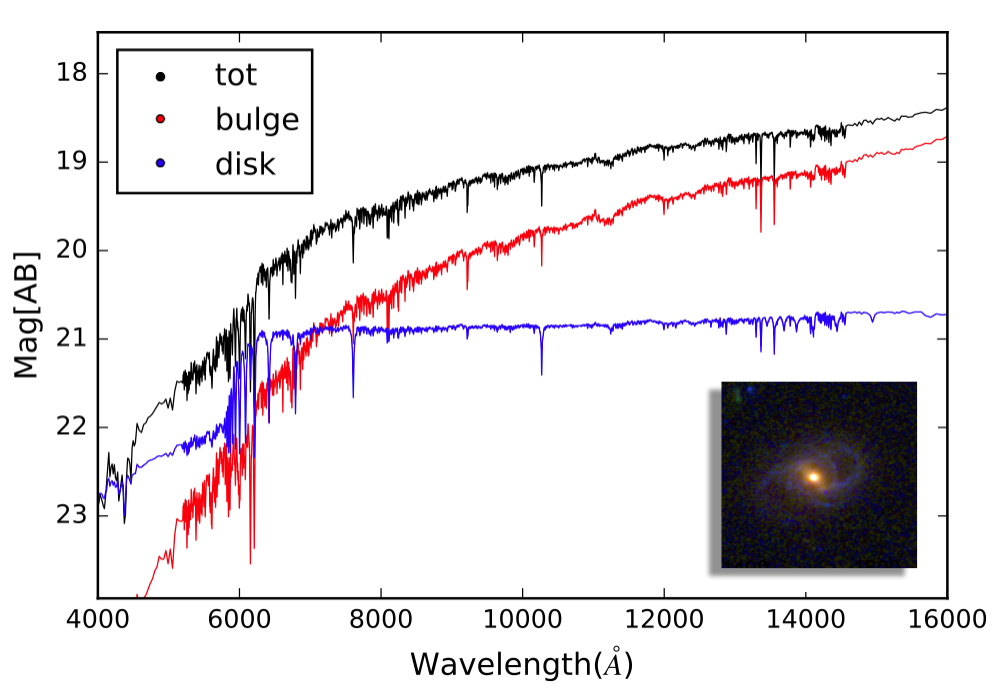}
\caption{Template galaxy used for the simulation. The bulge SED is shown in red and the disk in blue. The black line shows the global SED. Random variations of this template are used to build a simulated sample (see text for details).}
\label{fig:simulation}
\end{figure}

A key difference compared to previous approaches is that our fitting procedure takes into account all wavelengths simultaneously. This needs to be captured in the simulations too, i.e. the profile needs to be simulated in every band. A random distribution of all parameters as a function of wavelength is not a good approximation to reality. As done in \cite{Vika2013}, we simplified the problem by selecting a real low redshift galaxy ($z\sim0.5$) that clearly has two components (from visual inspection and our CNN classification) from which we obtained a bulge/disk decomposition (see figure~\ref{fig:simulation}). We then use the best SED models for the bulge and the disk as templates for our simulations (see section~\ref{sec:SEDs} for details on the SED fitting). 

We first assign a random $B/T$ (in the i-band rest-frame) and a total magnitude (in the H band) in the range of $[18-25]$ to every simulated galaxy. This allows us to fix the bulge and disk fluxes in the H band. We also associate a random redshift in the range $[0.01,3]$ and shift the SED template described above accordingly.  
We can then associate a realistic magnitude in all other bands with a typical SED of a bulge and a disk while also accounting for the real redshift distribution. The surface brightness profile of the disk is rendered using an exponential profile. For the bulge  component, we draw a random value of the \Sersic index between 2 and 4, and we assume no wavelength dependence. In order to have a realistic wavelength dependence for the radii, we used random examples from the real data, requiring that the bulge always be smaller than the disk. The final simulated sample contains  $\sim 4000$ galaxies.

We then run \textsc{Galapagos-2} with exactly the same settings used for the real data and compare the input and recovered parameters to assess statistical errors. 

\begin{figure*}
\begin{center}
$\begin{array}{c c}
\includegraphics[width=0.5\textwidth]{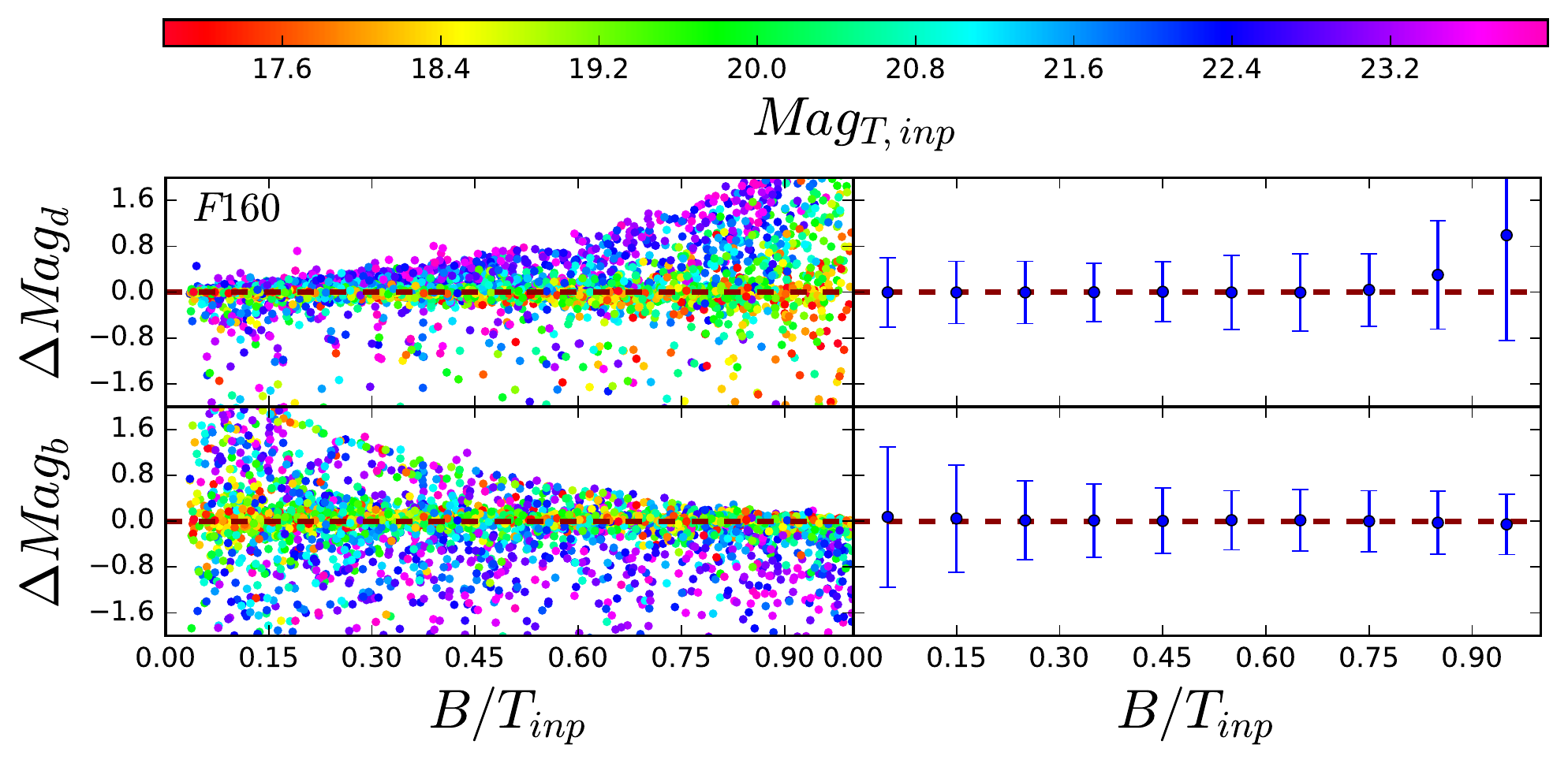} &
\includegraphics[width=0.5\textwidth]{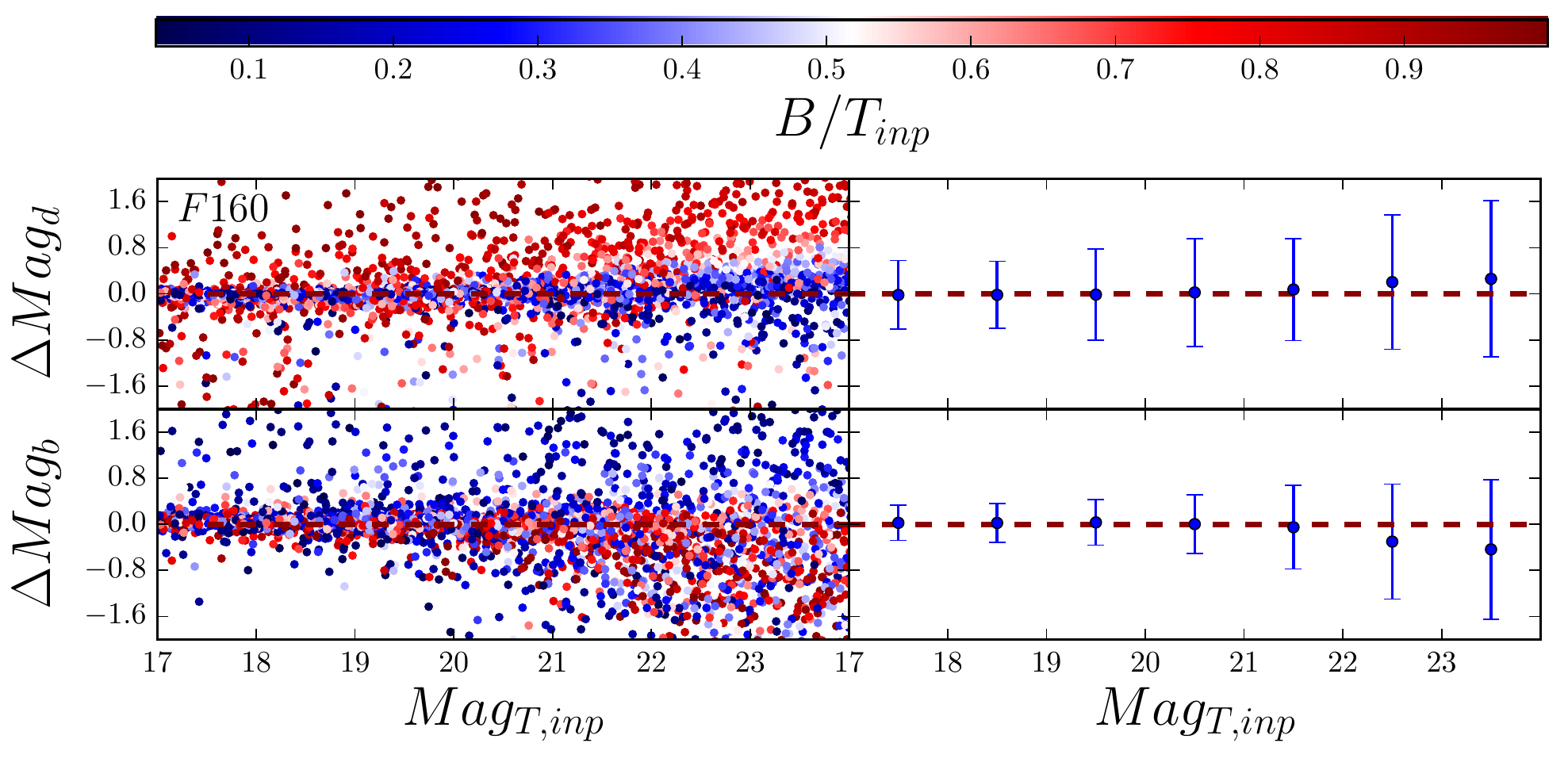} \\
\includegraphics[width=0.5\textwidth]{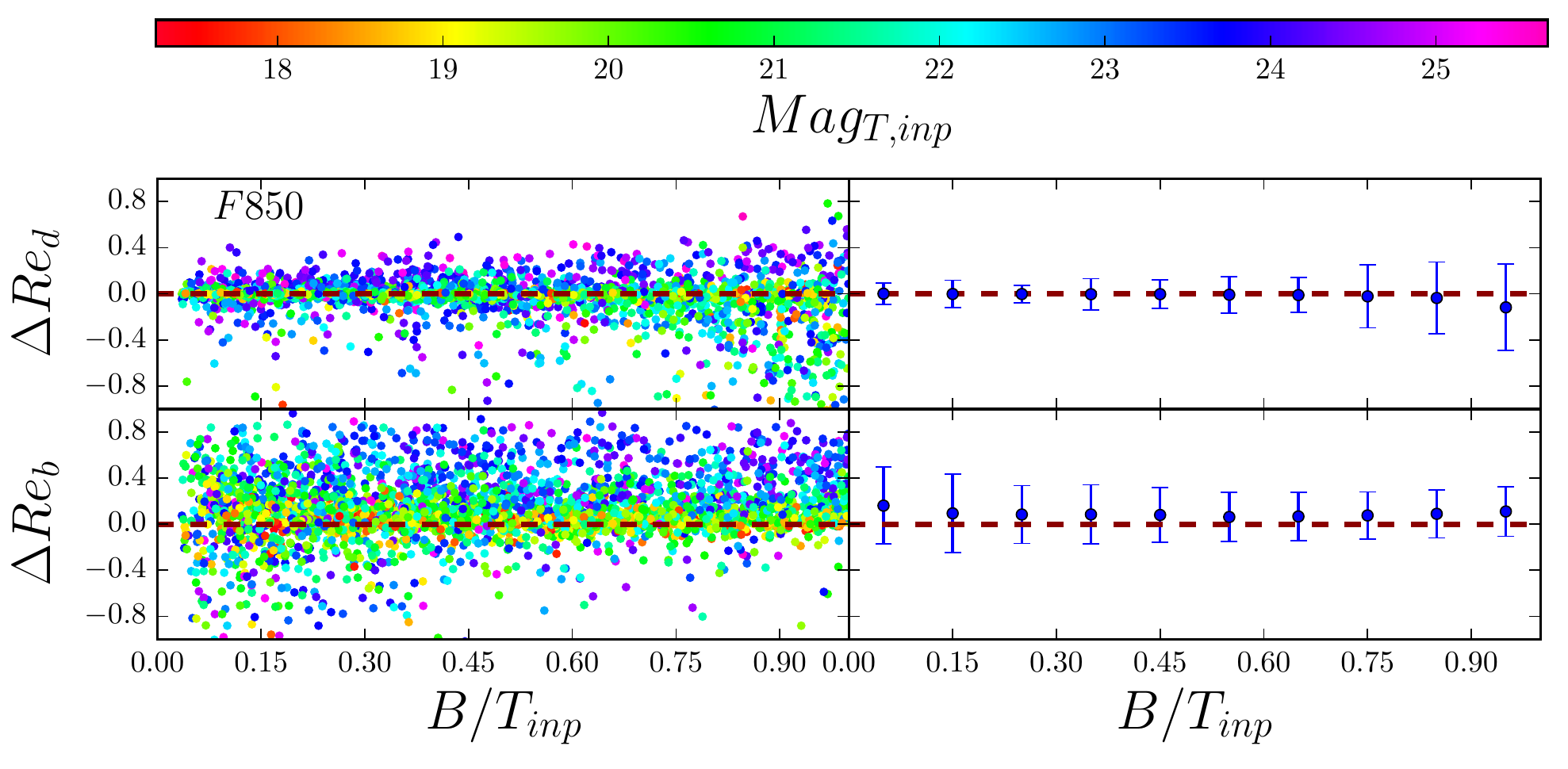} &
\includegraphics[width=0.5\textwidth]{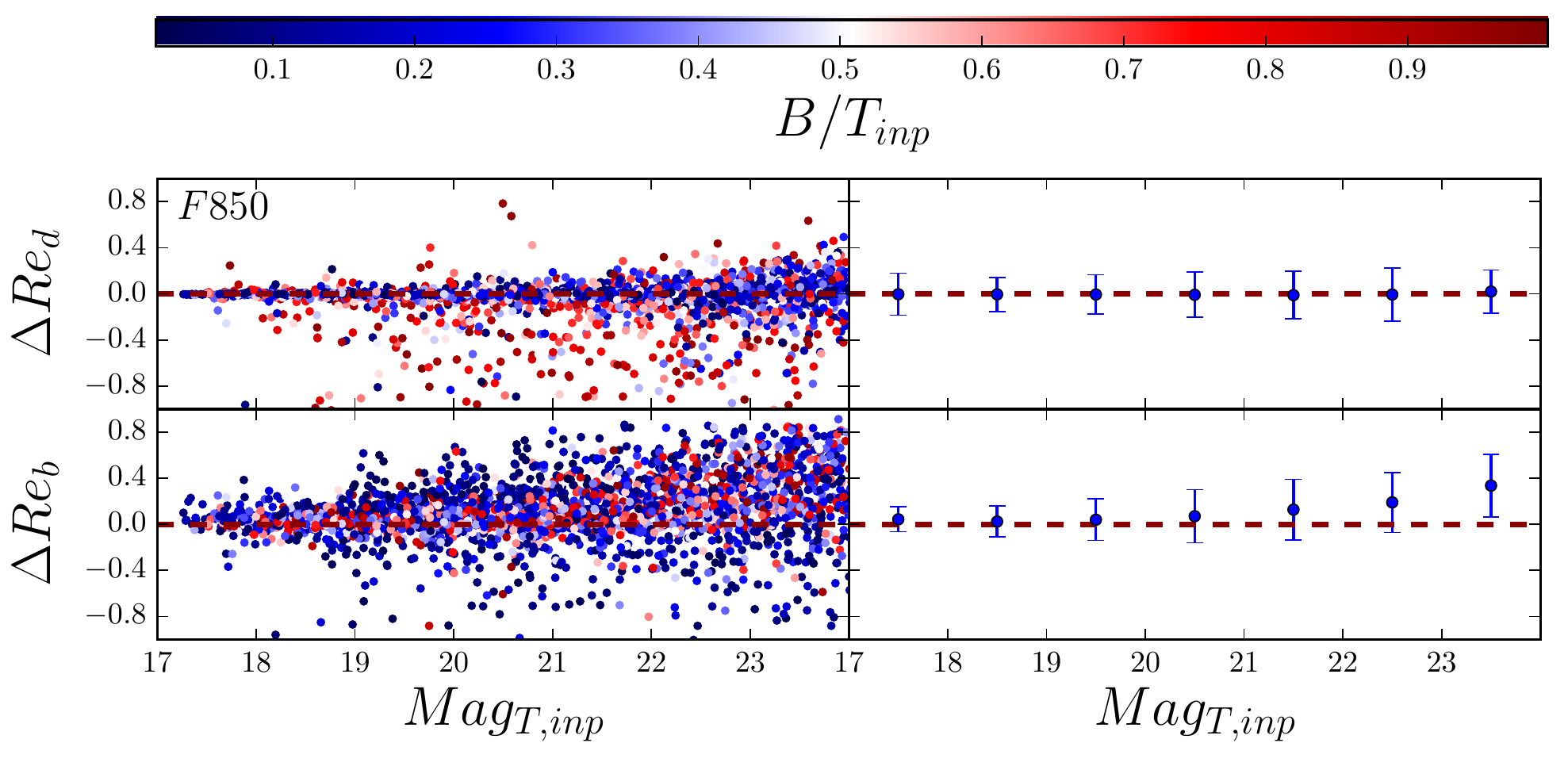} \\
\includegraphics[width=0.5\textwidth]{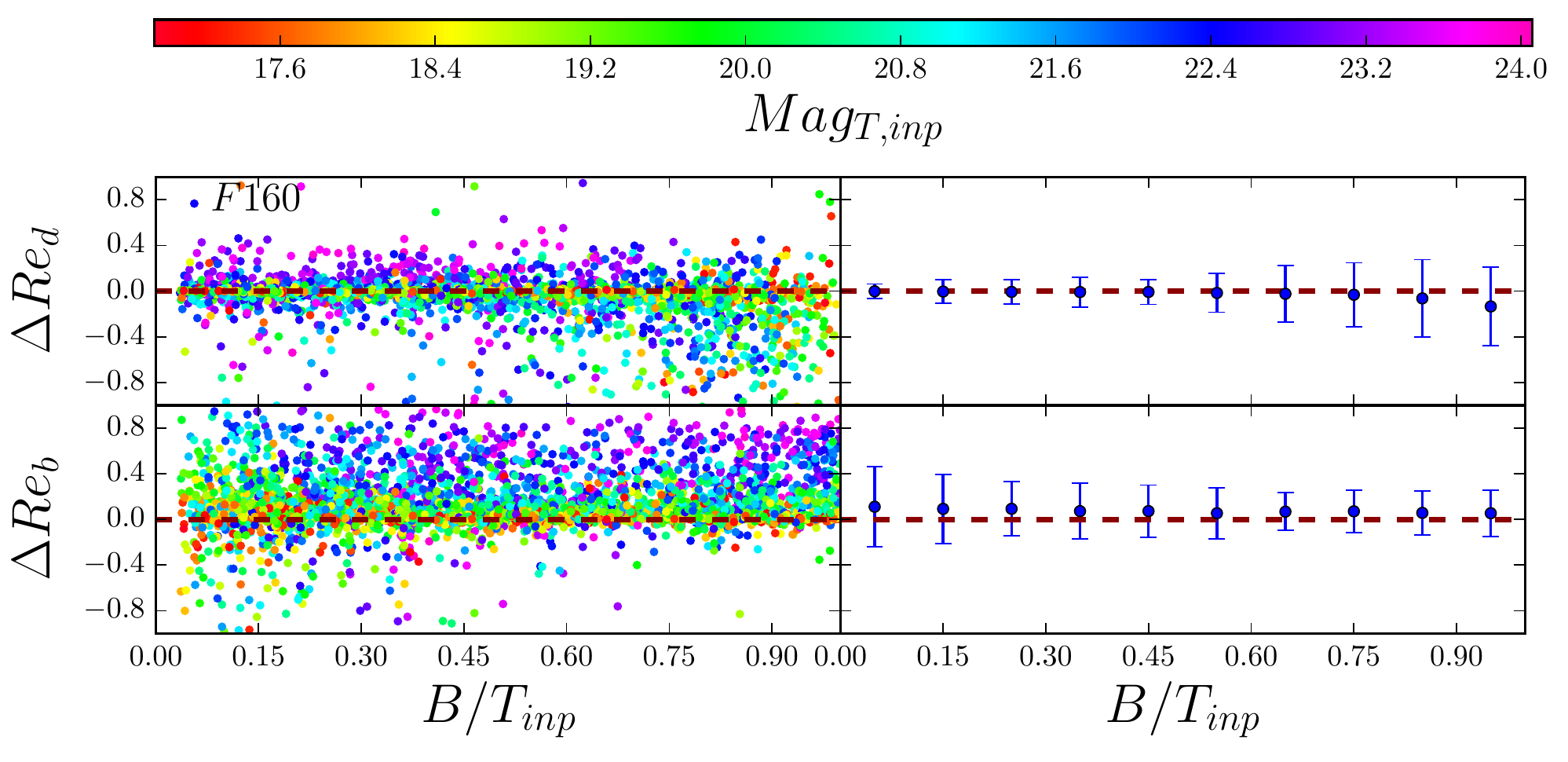} &
\includegraphics[width=0.5\textwidth]{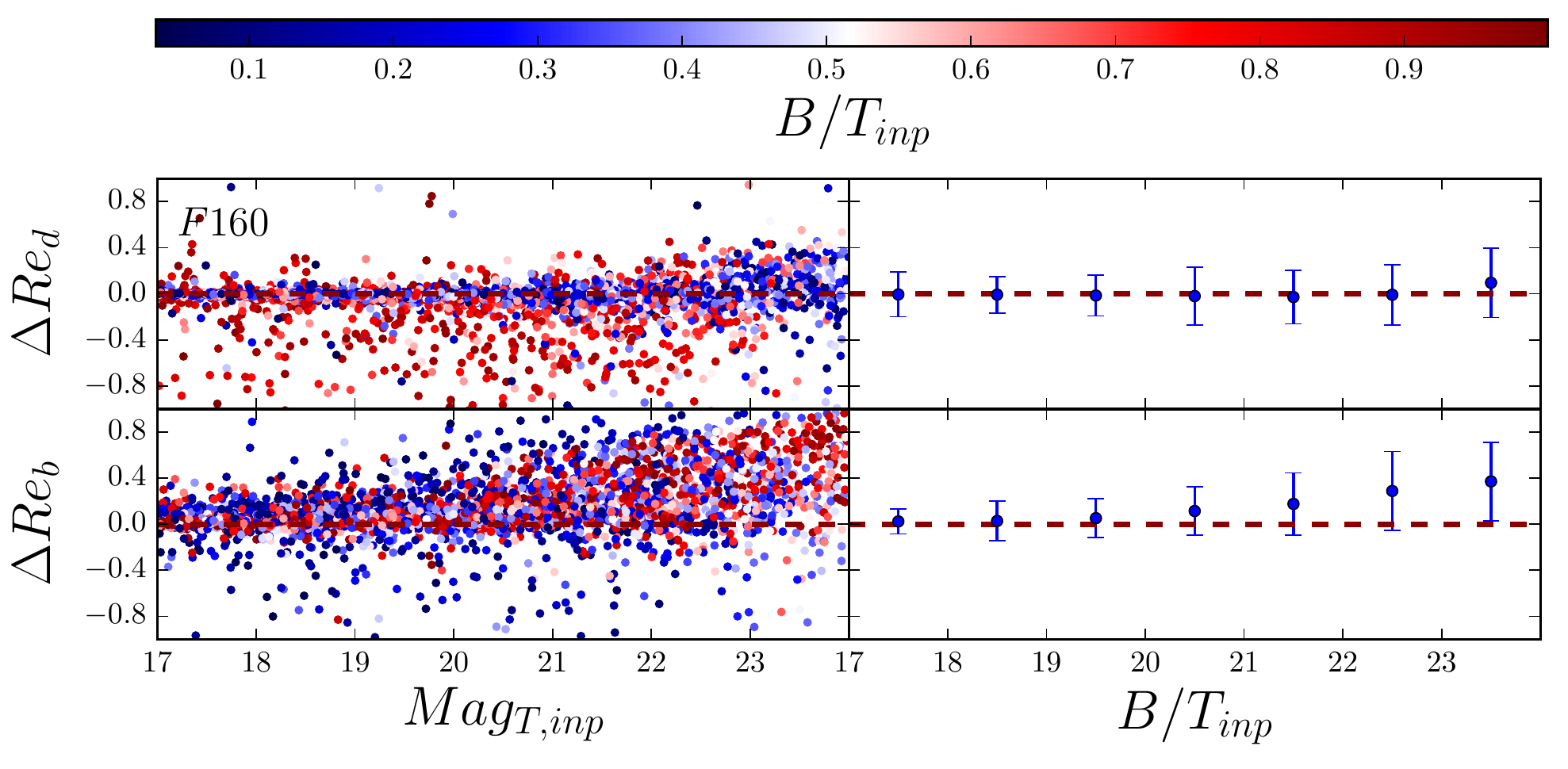} \\ 

\end{array}$
\caption{Comparison between simulated and recovered properties of bulges and disks, namely magnitude and half-light radii in two filters: F850 and F160. All panels show in the y-axis the relative difference between input and output values: $\Delta Mag=Mag_{in}-Mag_{out}$, $\Delta Re=\frac{R_{e,in}-R_{e,out}}{R_{e,in}}$. The results are plotted against $B/T$ (total magnitude) and color coded with total magnitude ($B/T$) in the left (right) panels. In each plot we show individual galaxies in the left panels all galaxies up to $mag\leq24$ are shown here in order to test the magnitude limit of the method ) and binned median values and scatters on the right panels (median and scatter are computed only for $mag\leq23$).} 
\label{fig:b_t_color}
\end{center}
\end{figure*}

A complete visualization of all the parameter space is complex given the number of parameters. We decided to highlight the dependence of magnitude and size of bulges and disks on galaxy morphology (quantified through $B/T$), magnitude and redshift. Indeed figure \ref{fig:b_t_color} compares input and output magnitudes and sizes for two different filters: F850 and F160W. The accuracy of the photometry and stellar masses is discussed in section~\ref{sec:SEDs}. As expected, we observe that the errors in the structural parameters of the disk (bulge) increase towards high (low) $B/T$ values. When one of the two components dominates over the other it becomes more difficult to quantify both components properly.  
 For fainter galaxies ($F160>23-23.5$) the bulge magnitude start deviating from the zero bias line. This suggests that we have reached the $S/N$ limit below which results become biased (as pointed out by \citealp{vanderWel2012} for 1 component fits). We apply a magnitude cut at $F160W=23$ to keep a zero bias and a scatter lower than $30\%$. Only galaxies brighter than this magnitude limit will be included in the final catalog released with this work. 
 Similar behavior is observed for the sizes. However, globally the errors remain within $\sim10-20\%$.
 We emphasize that the simulations performed are not representative of the \emph{real} evolution of galaxy SEDs since we are using a unique template. The effect of the redshift on the measurements is explored in section \ref{sec:errorbars} and figure \ref{fig:random_errs2}.

 \subsection{Individual errors}
 \label{sec:errorbars}
The above procedure has provided a statistical analysis of the systematic uncertainties which are on average close to zero, meaning that the method is intrinsically unbiased. In this section we estimate individual uncertainties (random and systematic) for all the galaxies in the catalog. These are particularly important to derive stellar population parameters through SED fitting.
Although \textsc{Galfit} provides error bars, they are known to be underestimated and not very indicative of the true error (see e.g. \citealp{haeussler2007}). This is generally explained because \textsc{Galfit} assumes a gaussian noise distribution when computing the errors (which is not a good approximation for HST drizzled images) and also does not consider the effects of companions. In \textsc{GalfitM}, the situation is even more dramatic since the fits in all bands are not independent and the constraints in the different parameters are therefore coupled. The \textsc{galfit} manual states that in such a case, the estimation of errors is not reliable. We thus use an alternative approach to estimate the errors. 

The relation between the measured value of a given parameter and the true value is approximated by:
$$Measured = True+ b \pm \sigma $$

The bias (b) or systematic uncertainties is expected to be close to zero as seen from the simulations and therefore the error budget is dominated by the random uncertainties ($\sigma$).
We adopt a similar method as the one described in  \cite{vanderWel2012}. The method main assumption is that galaxies with similar structural properties share similar errors.

We first use the analytic simulations of section 5.2 to estimate the systematic uncertainties. Since the ground truth is known and the same model is  used to generate galaxies and to fit them, the differences between the input and output in this idealized case can be interpreted as the intrinsic systematic errors.
We thus define for every galaxy in the catalog and the simulation the following vector $\mathbf{p}$ :
\begin{ceqn}
\begin{gather*}
\mathbf{p}=\bigg(\frac{m}{\sigma_m},\frac{log(Re_b)}{\sigma_{log(Re_b)}},\frac{log(Re_d)}{\sigma_{log(Re_d)}},\frac{n_b}{log(n_b)},\frac{BT}{log(BT)}\bigg)
\end{gather*}
\end{ceqn}

In the above equation, $m$ designates the apparent magnitude in a given filter, $n$ is the \Sersic index, $Re_d$ and $Re_b$ are the effective radii of the disk and the bulge component and $B/T$ is the ratio between the flux of the bulge and the total flux. Each value is normalized by the dispersion to have similar variation ranges for all parameters. 
We then compute the euclidian distance of each real galaxy to all simulated objects and select the 30 closest objects. The bias for that galaxy is estimated as the 3-sigma clipped median of the difference between the input values of the simulations and the ones recovered by \GalfitM.

To estimate the random errors, we cannot rely on the simulations since they do not capture all the complexity of real data.  Since we have fitted all galaxies with different settings for each type of model (\Sersic + Exponential,\ Sersic) the differences on the resulting fits between the two settings can be used to estimate random uncertainties.  By doing this we assume that both settings have similar systematic biases. Therefore,  for each galaxy in the catalog and every band we compute the following two vectors $\mathbf{p^{1C}}$ $\mathbf{p^{2C}}$ for 1-component and 2-component respectively:
\begin{ceqn}
 \begin{equation*}
   \mathbf{p^{1C}}=\bigg(\frac{m}{\sigma_m},\frac{log(n)}{\sigma_{log(n)}},\frac{log(R_e)}{\sigma_{log(R_e)}},f_{sph,disk,irr} \bigg)
\end{equation*}

\begin{equation*}   
\begin{split}
  \mathbf{p^{2C}}=\bigg(\frac{m}{\sigma_m},\frac{log(n_b)}{\sigma_{log(n_b)}},\frac{log(Re_d)}{\sigma_{log(Re_d)}},\frac{log(Re_b)}{\sigma_{log(Re_b)}}, \\
  \frac{BT}{log(BT)},f_{sph,disk,irr}\bigg)
  \end{split}
\end{equation*}
\end{ceqn}

The meaning of the  different symbols are the same of the previous equations. $f_{sph}$, $f_{disk}$ and $f_{irr}$ are the probabilities that the galaxy looks like a spheroid, disk or irregular respectively \citep{Huertas2016}. 

\begin{figure*}
\begin{center}
$\begin{array}{c c}
 \includegraphics[width=0.5\textwidth]{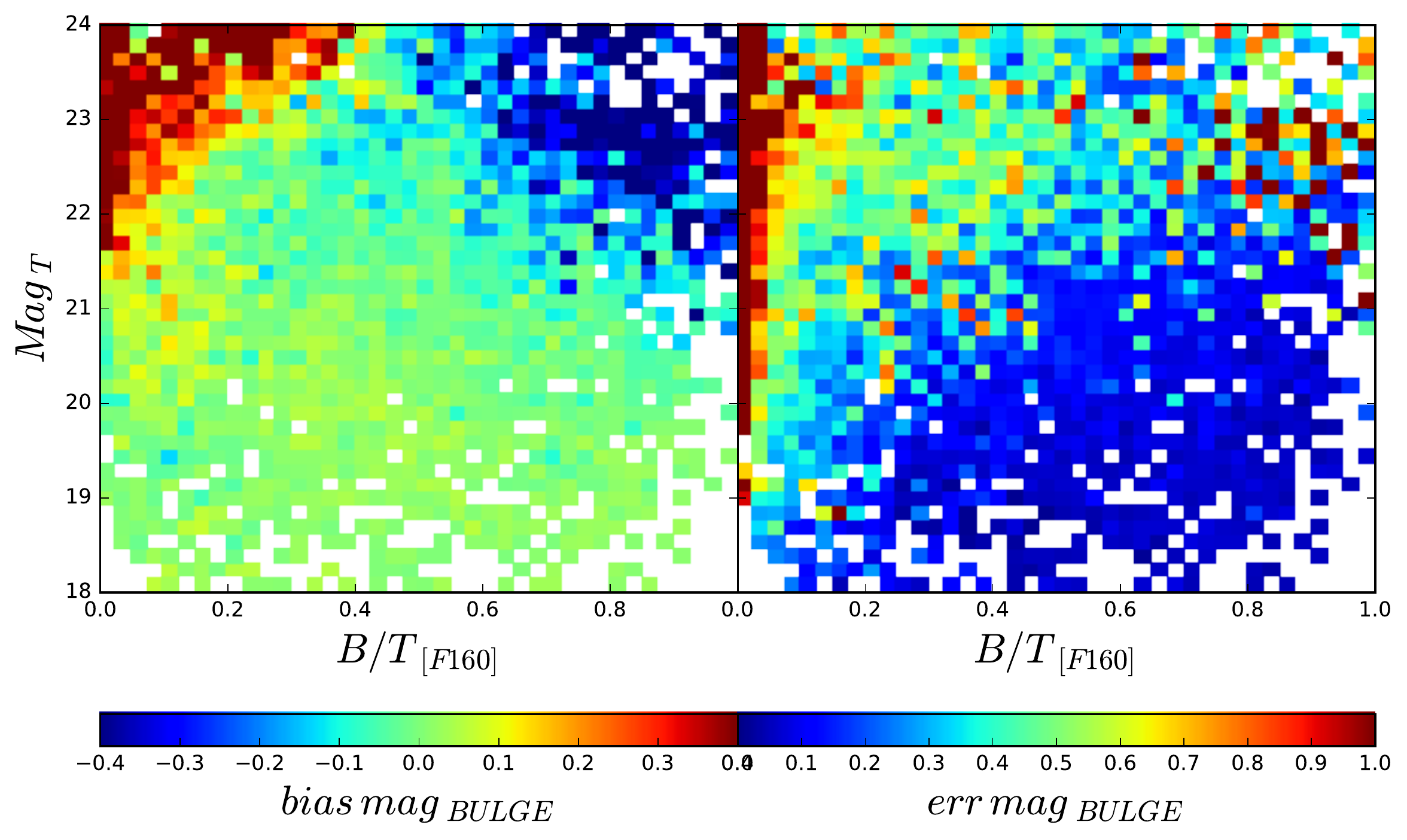} &
\includegraphics[width=0.5\textwidth]{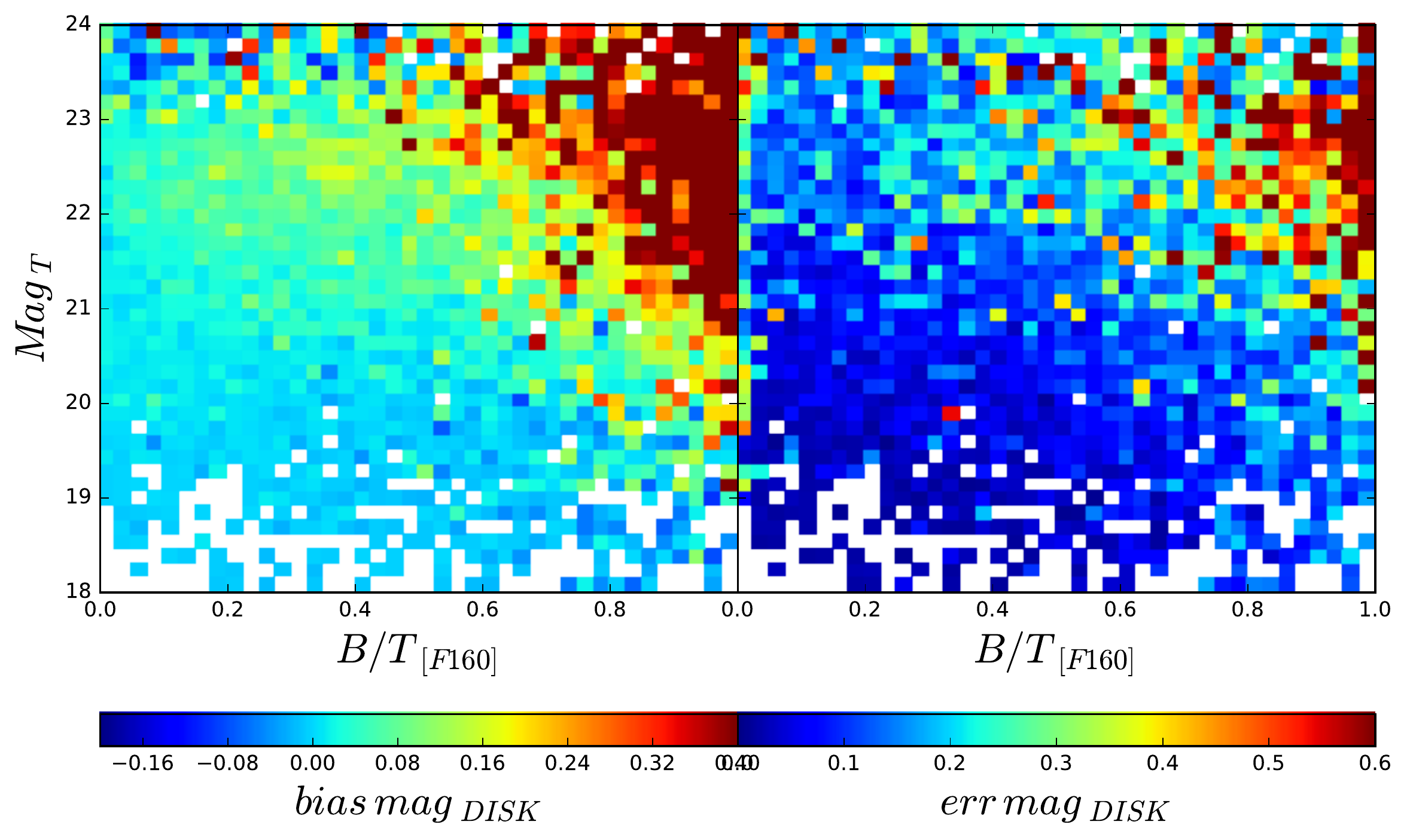} \\
 \includegraphics[width=0.5\textwidth]{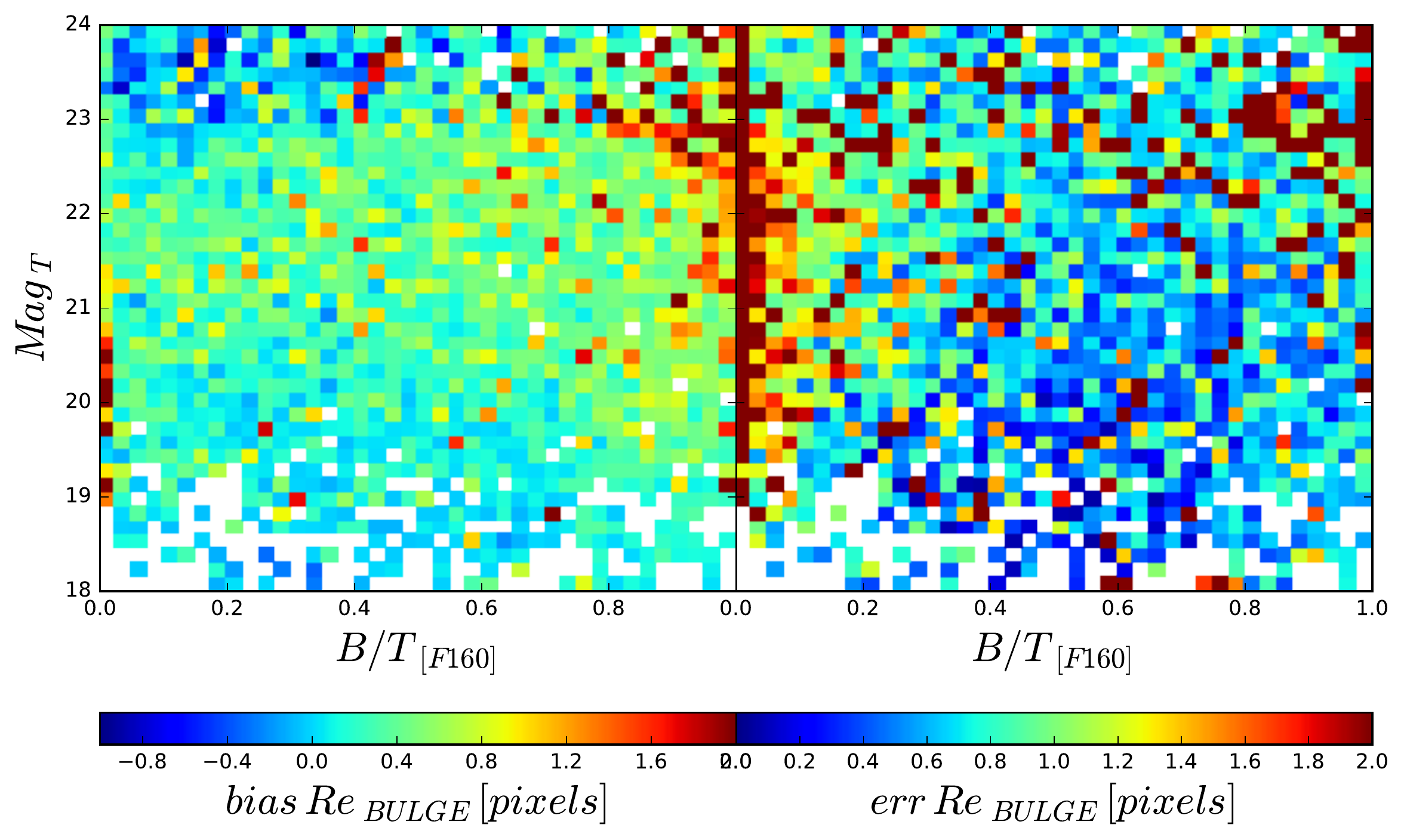} &
\includegraphics[width=0.5\textwidth]{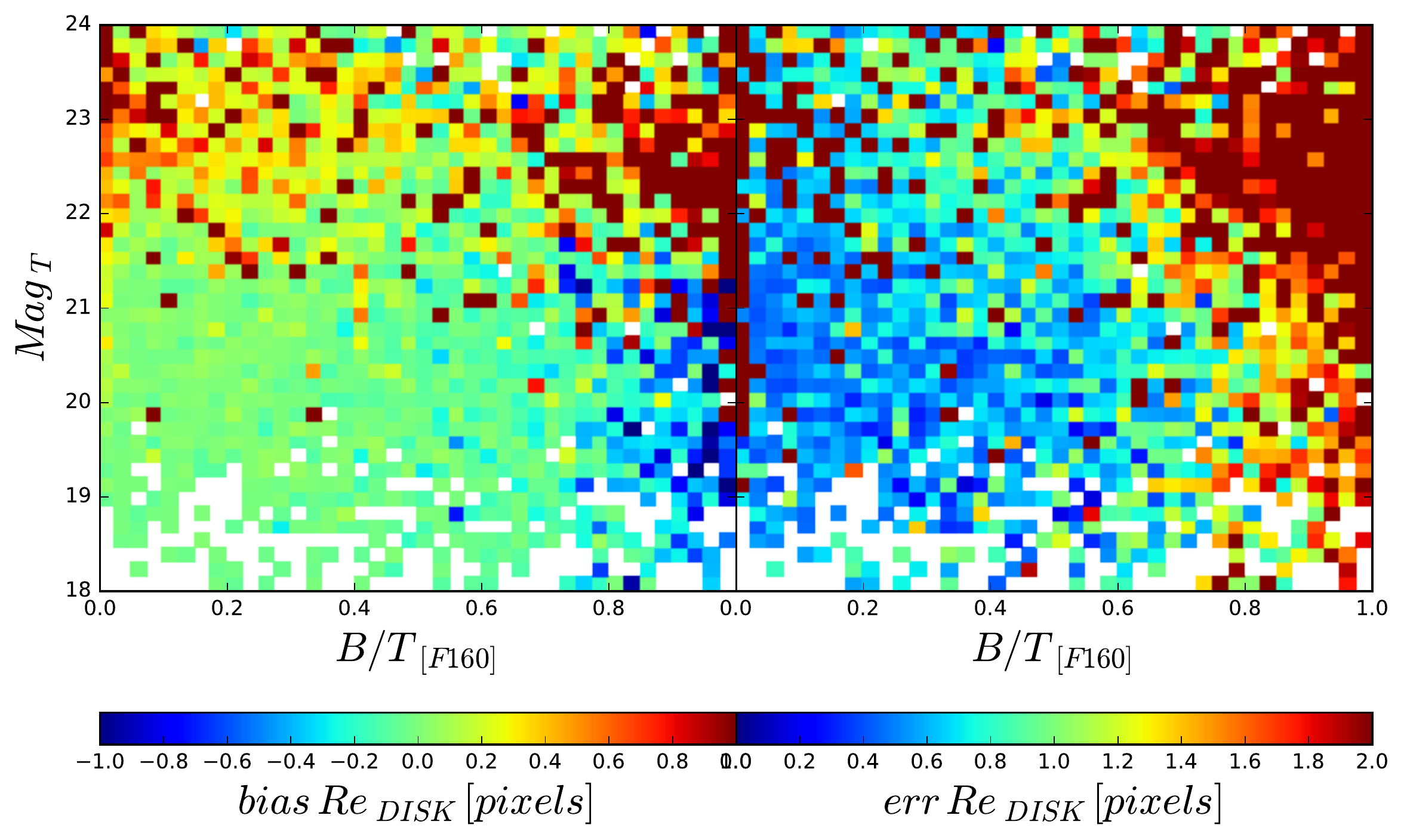} \\
 \includegraphics[width=0.5\textwidth]{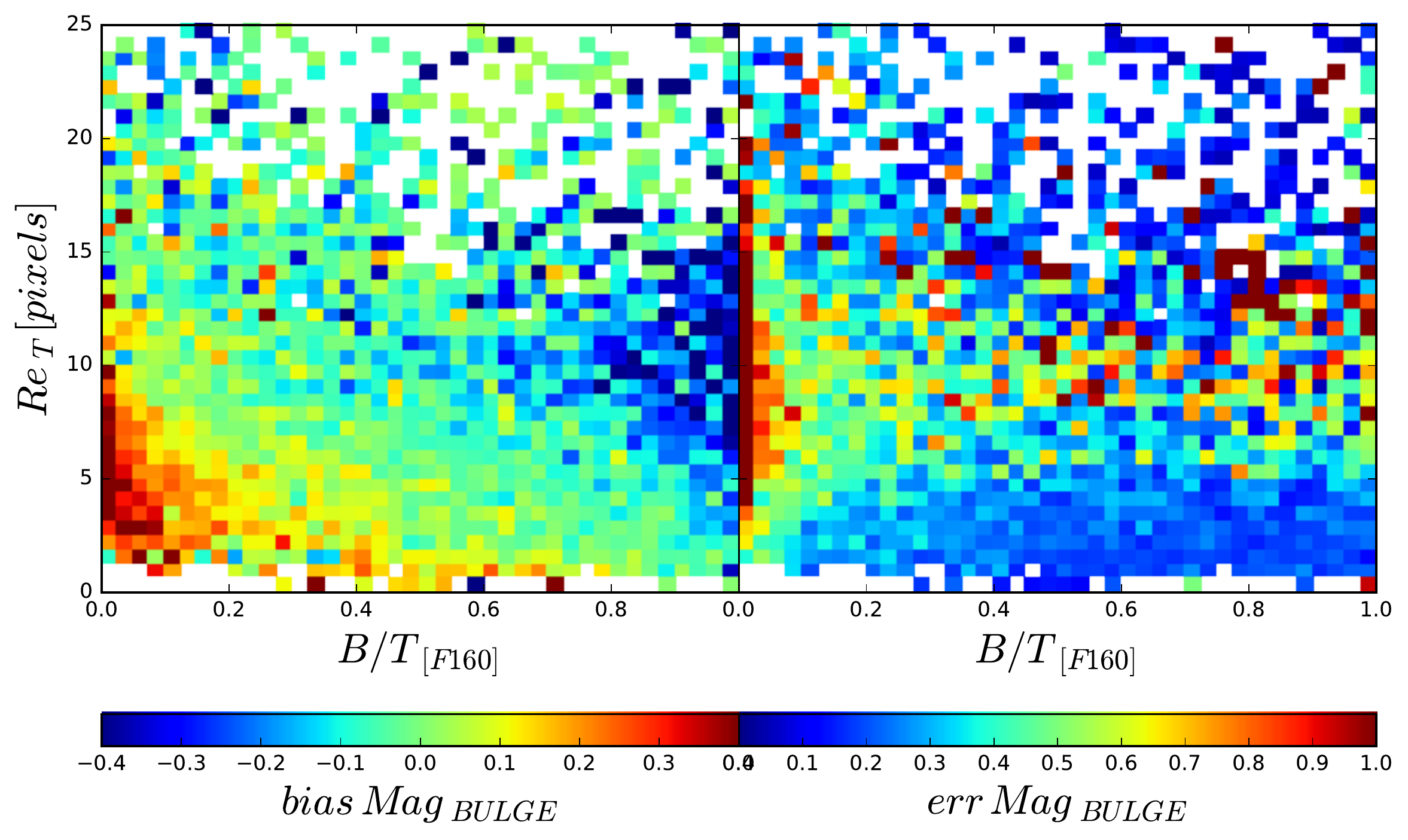} &
\includegraphics[width=0.5\textwidth]{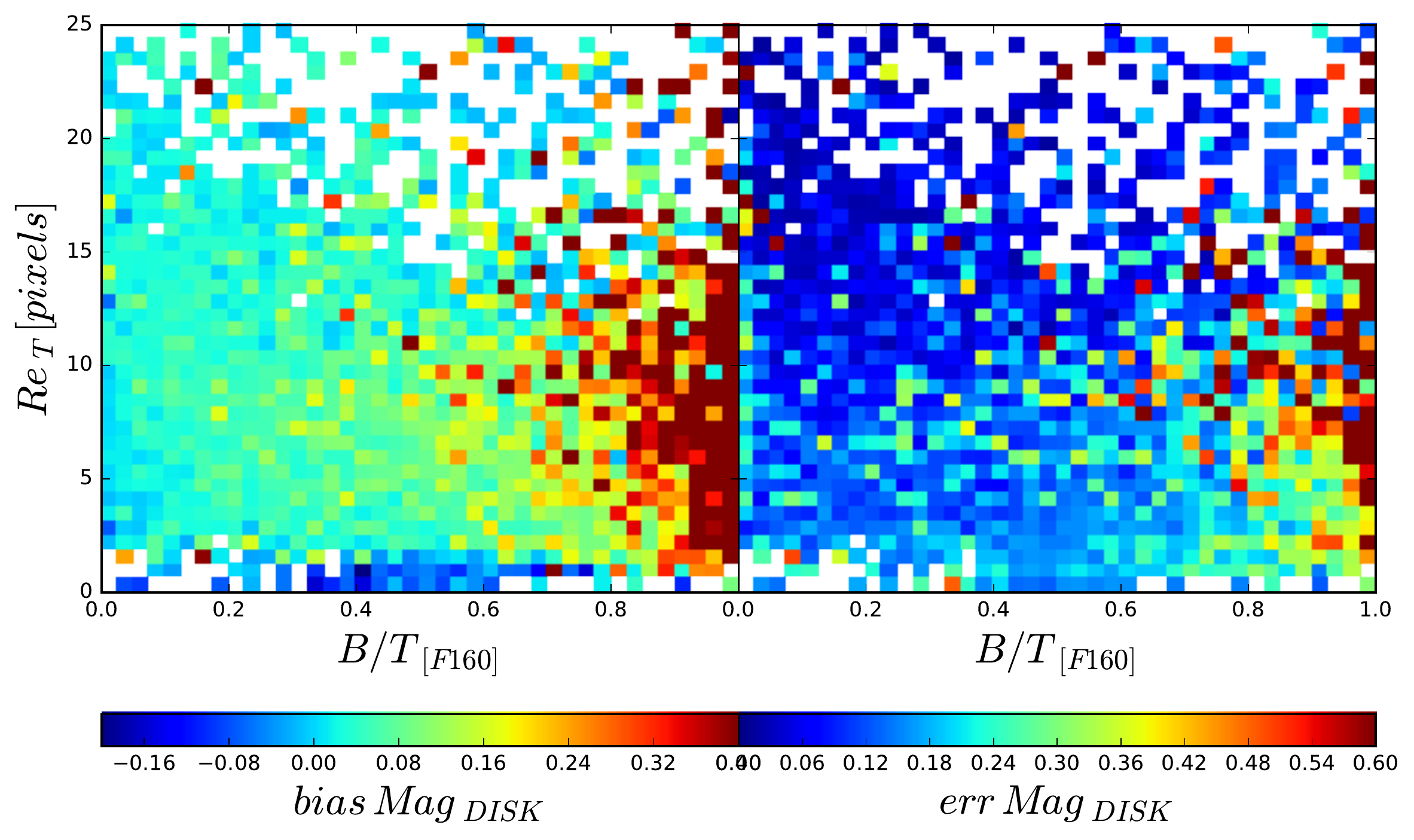} \\
 \includegraphics[width=0.5\textwidth]{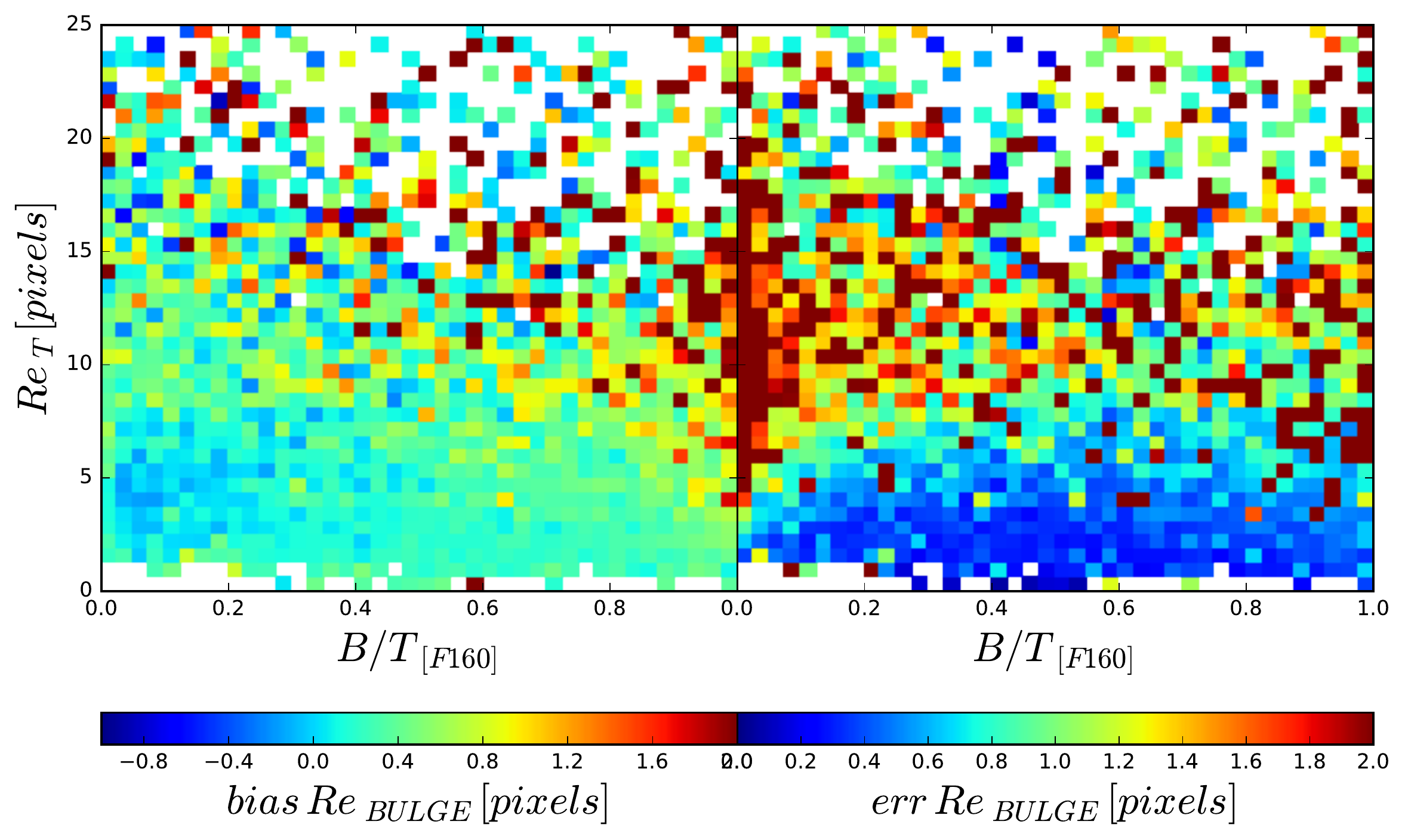} &
\includegraphics[width=0.5\textwidth]{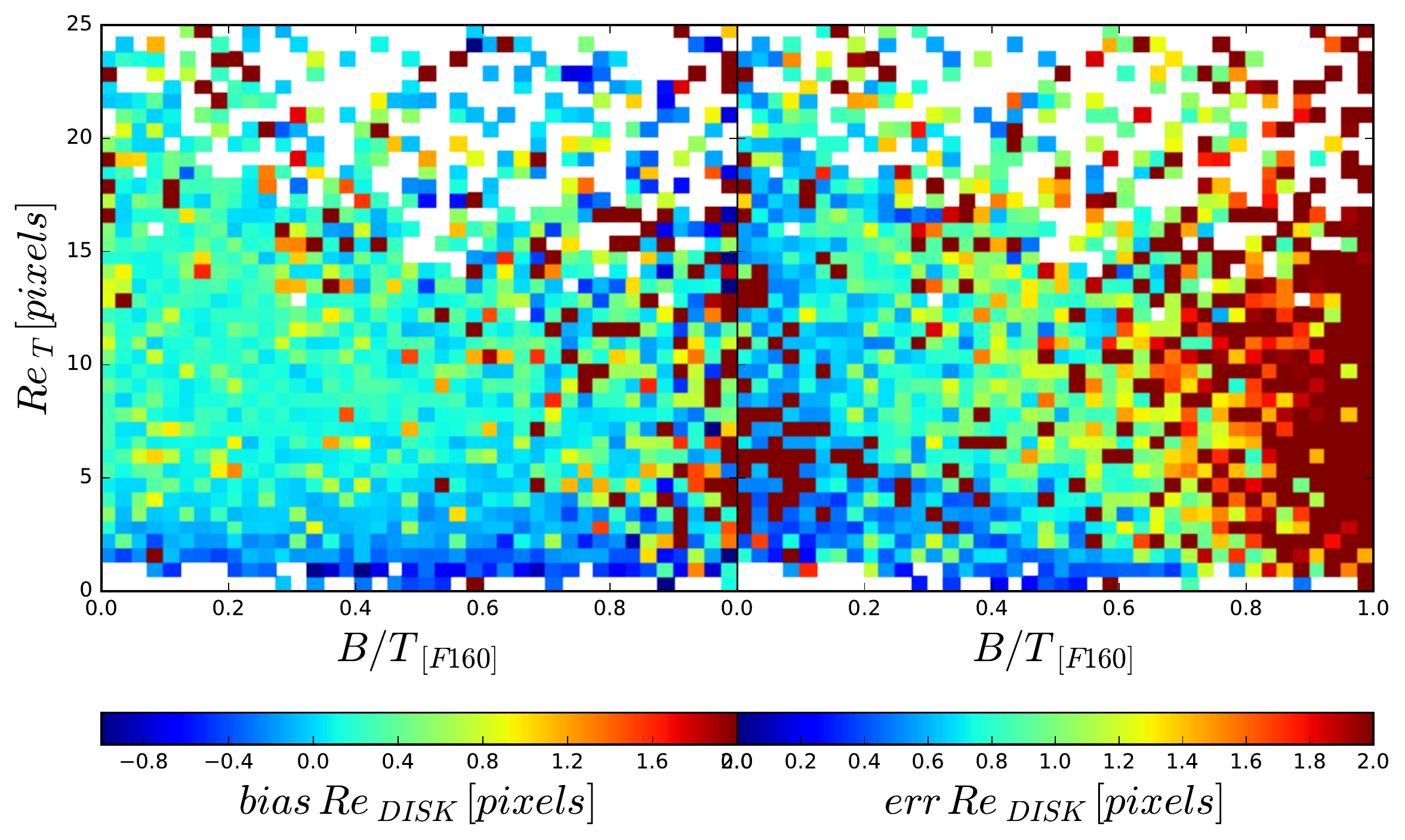} \\

\end{array}$
\caption{Systematic and random uncertainties on size and magnitude of bulges and disks shown in color code in the $Mag$ - $B/T$ plane as well as in the $Re$-$B/T$. The general trend is that errors mostly depend on $B/T$. They increase for the bulge (disk) component at low (large)  $B/T$. All panels show galaxies with $mag_{F160}<23$ except the ones in the top. }

\label{fig:random_errs1} 
\end{center}
\end{figure*}

\begin{figure*}
\begin{center}
$\begin{array}{c c}
 \includegraphics[width=0.5\textwidth]{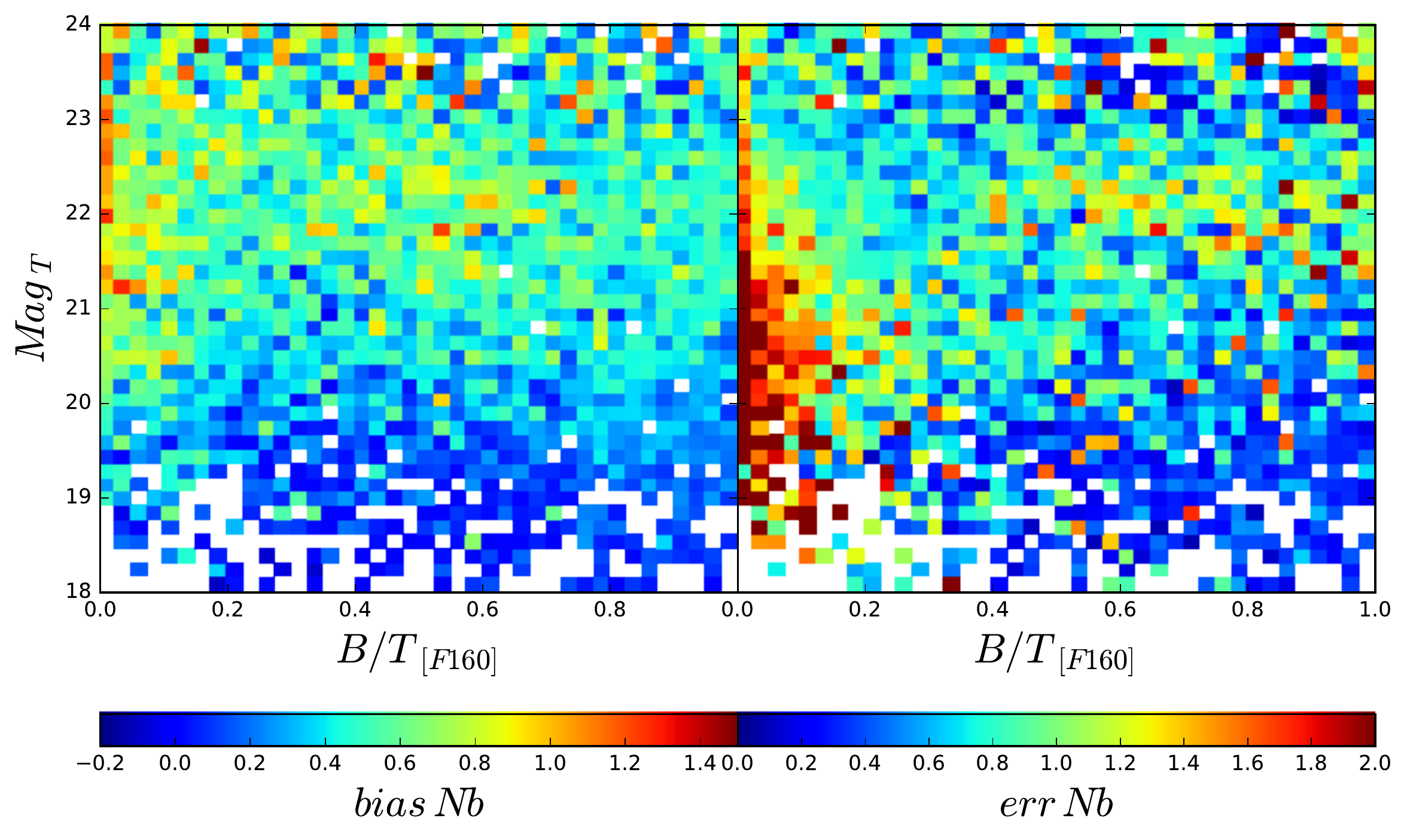} &
\includegraphics[width=0.5\textwidth]{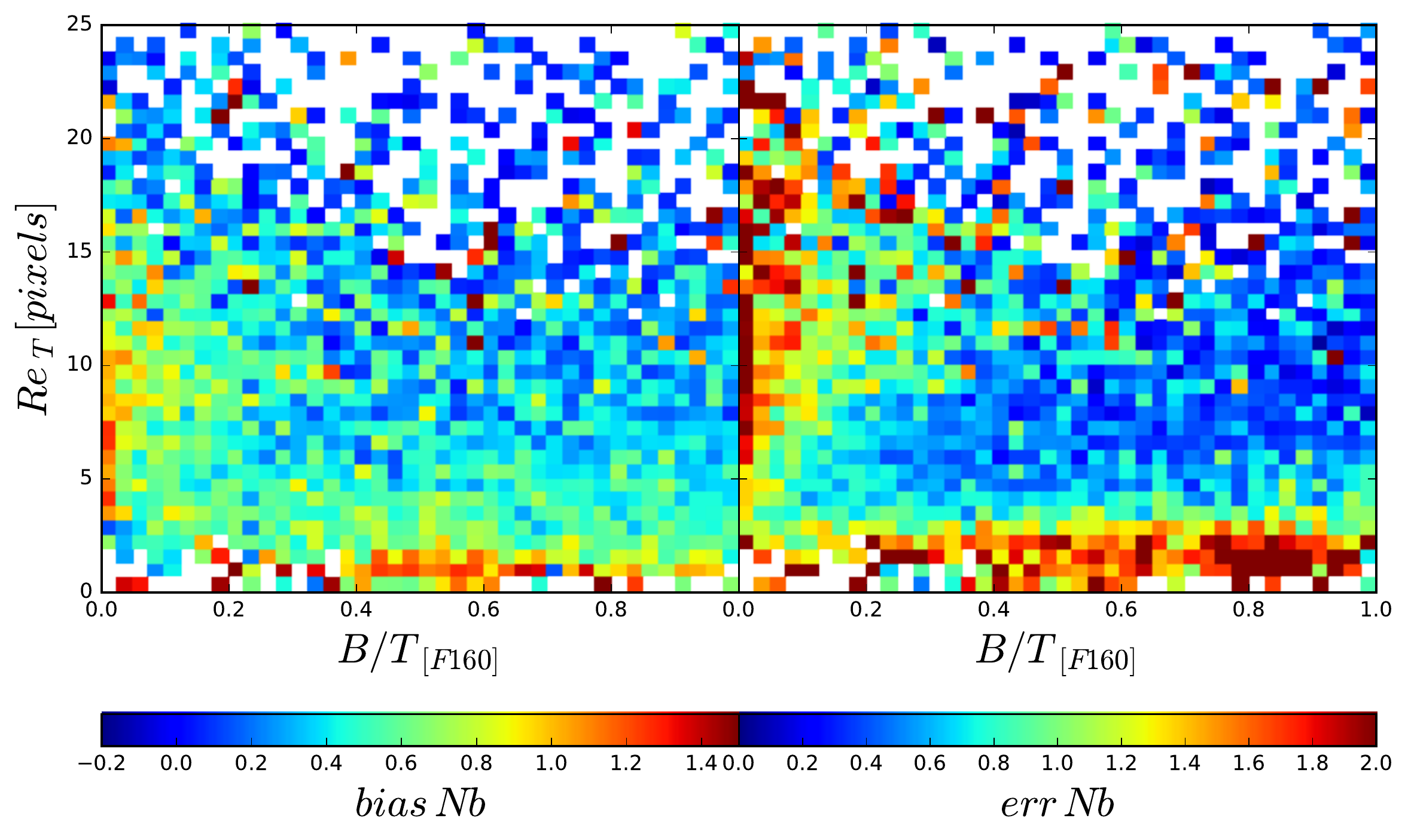} \\

\includegraphics[width=0.5\textwidth]{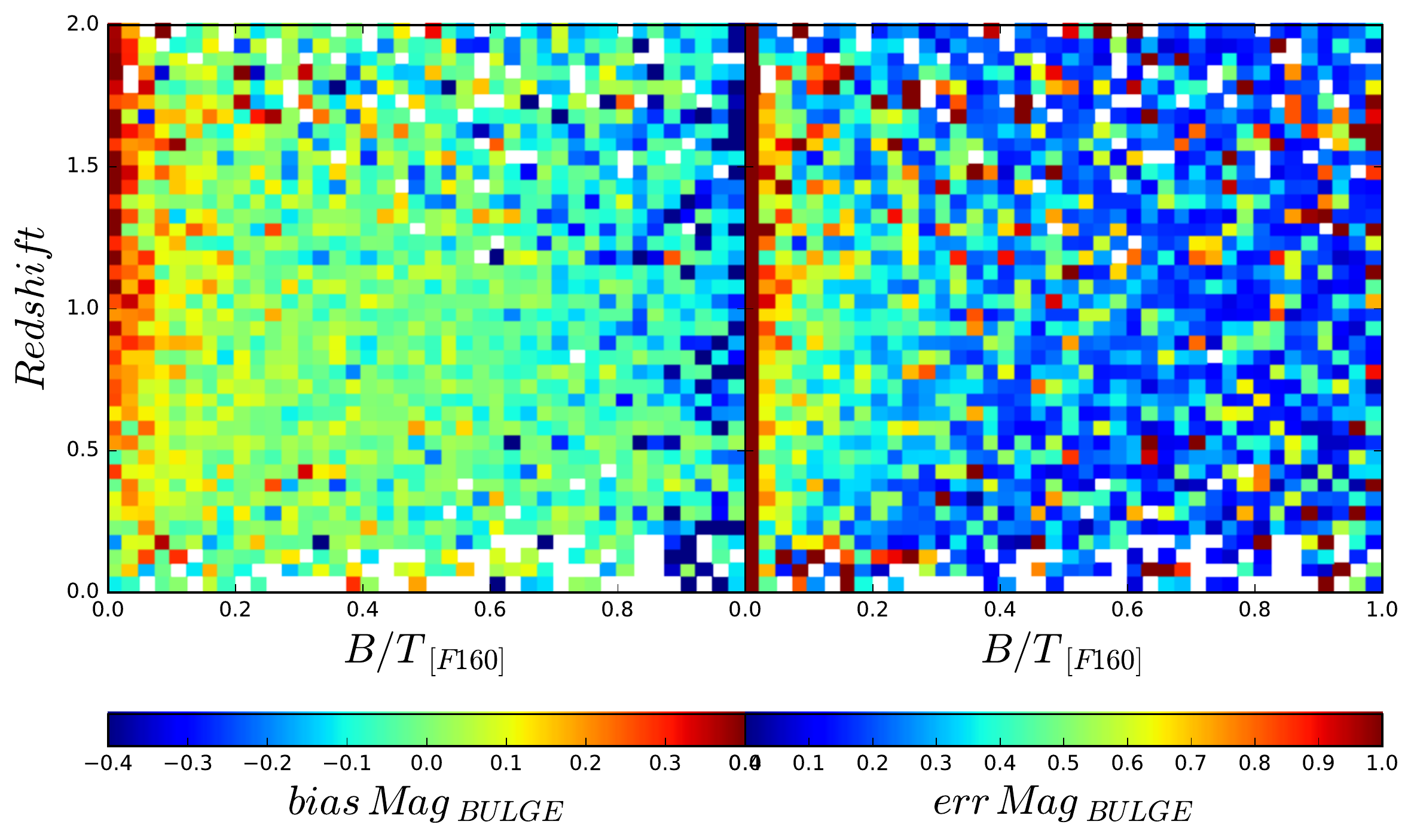} &
\includegraphics[width=0.5\textwidth]{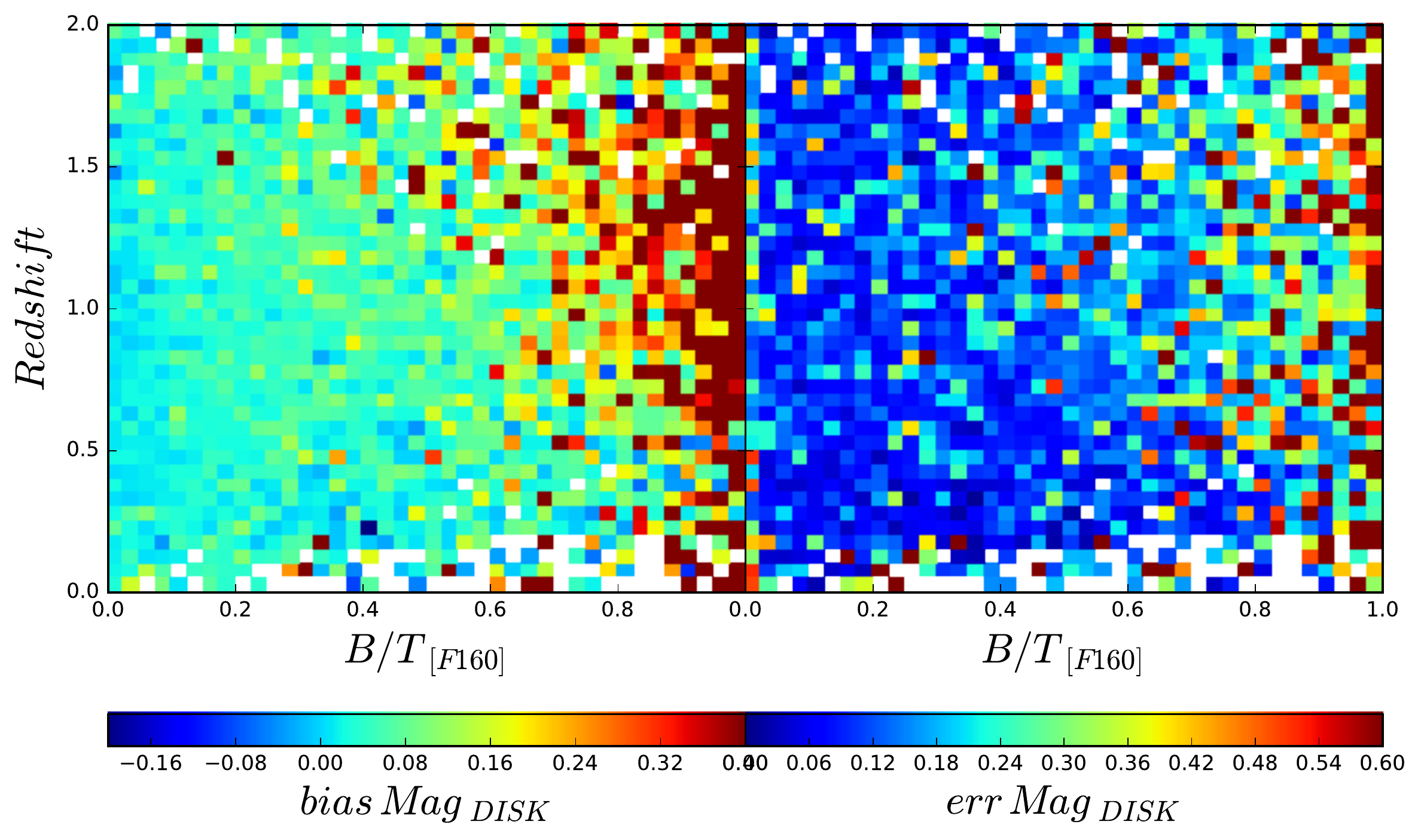} \\

\includegraphics[width=0.5\textwidth]{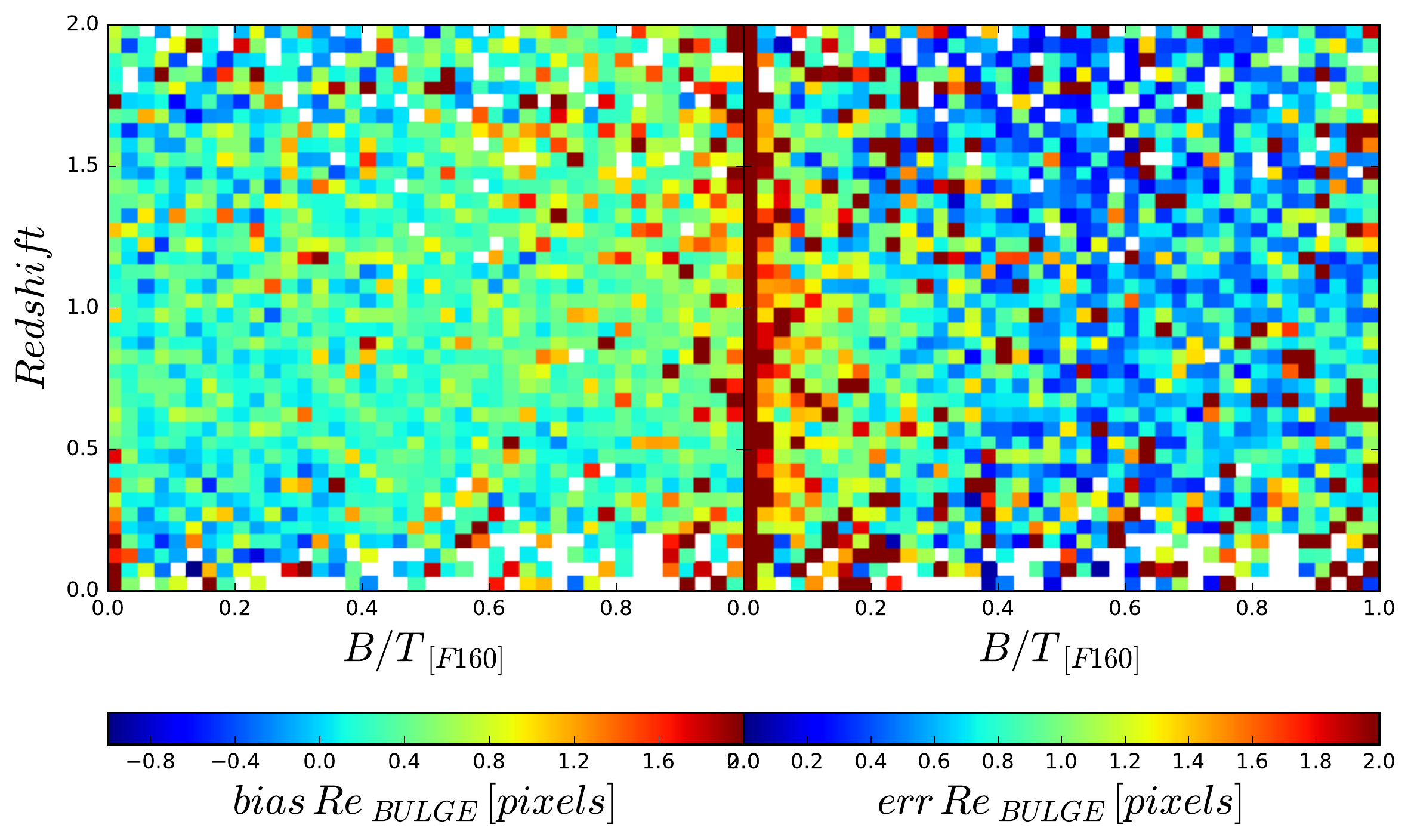} &
\includegraphics[width=0.5\textwidth]{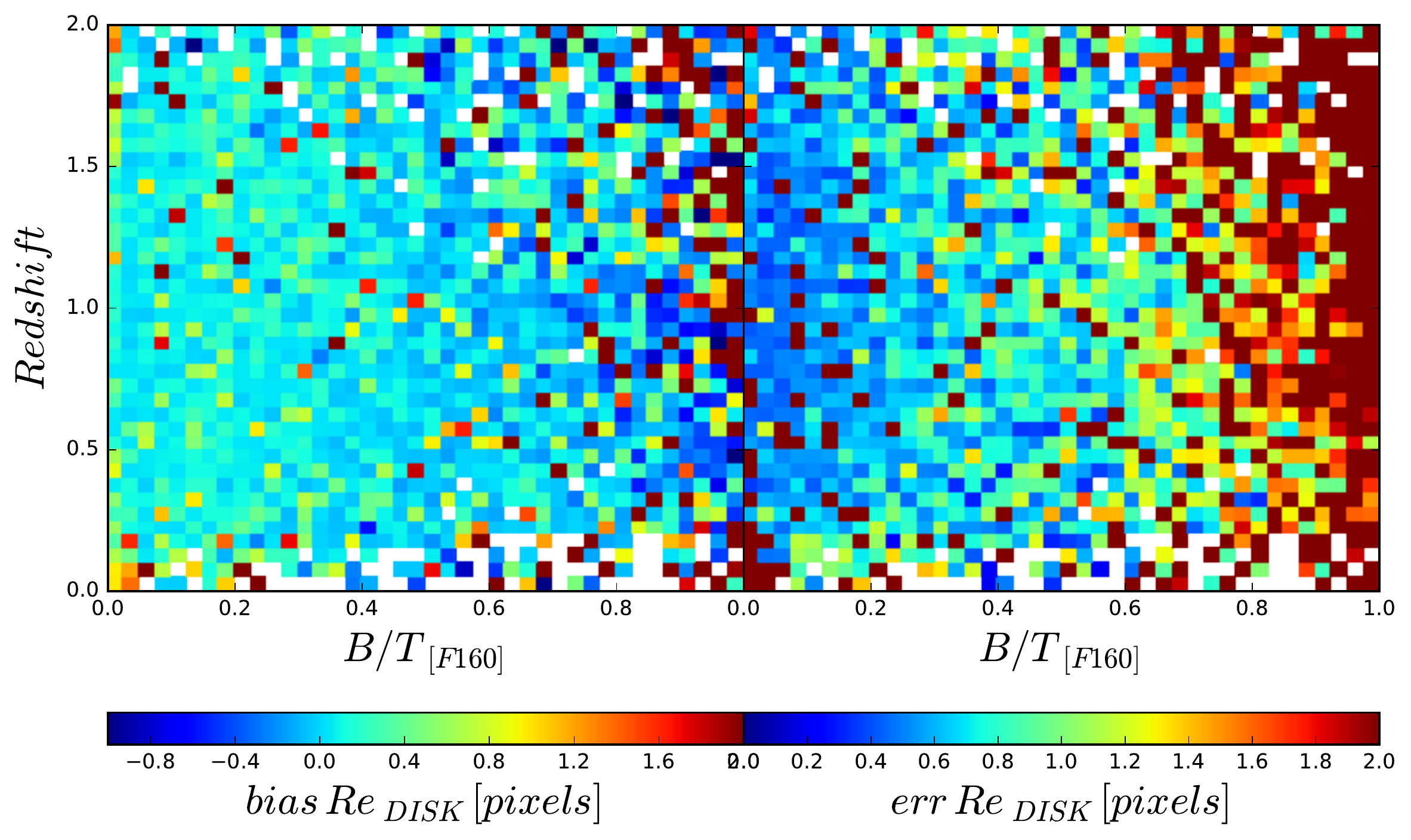} \\
\end{array}$ 
\caption{Systematic and random uncertainties on size, magnitude and \Sersic index  of bulges and disks shown in color code in the $Mag$ - $B/T$ plane as well as in the $Re$-$B/T$ and in $Redshift$-$B/T$ plane. The general trend is that errors mostly depend on $B/T$. They increase for the bulge (disk) component at low (large)  $B/T$, while they do not show strong correlation with the redshift.  } 
\label{fig:random_errs2} 

 \end{center}
\end{figure*}

Results for systematic and random uncertainties are shown in figures \ref{fig:random_errs1} and \ref{fig:random_errs2}. The systematic errors are close to zero over most of the parameter space. The measurements are essentially dominated by random uncertainties. The typical random uncertainty for the magnitude of the bulge and disk depends mostly on  the total magnitude and the bulge-to-total ratio as already assessed by the simulations from the previous section. Indeed the random errors increase for galaxies magnitude larger then 23 as it can be see in the bottom panels of figure  \ref{fig:random_errs1}. Moreover, for objects with $B/T<0.2$, the error on the bulge magnitude increases to $0.6-0.7$ magnitudes and the error on the disk magnitude rises to $\sim0.5$ for $B/T>0.8$. This confirms that the properties of embedded components are estimated accurately only if the dominant component represents less than $80\%$ of the light. There is little or no dependence of the magnitude errors on other parameters such as the size or the \Sersic index of the bulge. Regarding the errors on sizes of both components, the average error for the bulge is $\sim20\%$ and $\sim10\%$ for the disks, with again, a dependence on $B/T$.
Finally, the two panels in the bottom side of figure \ref{fig:random_errs2} explore the dependence of size and magnitude errors with the redshift. We do not observe strong gradients or correlations. This is due to the fact that the majority of our galaxies are concentrated between 0.5<z<1.6 where the angular scale do not drastically change, thus the uncertainties on the size are not affected by this effect. In addition to that the absence of correlation is related on the magnitude selection. Indeed, by selecting bright galaxies (H<23), we are also selecting the largest ones. These two effects dominate the final trends observed.

\begin{figure}
\begin{center}
$\begin{array}{c}
\includegraphics[width=0.5\textwidth]{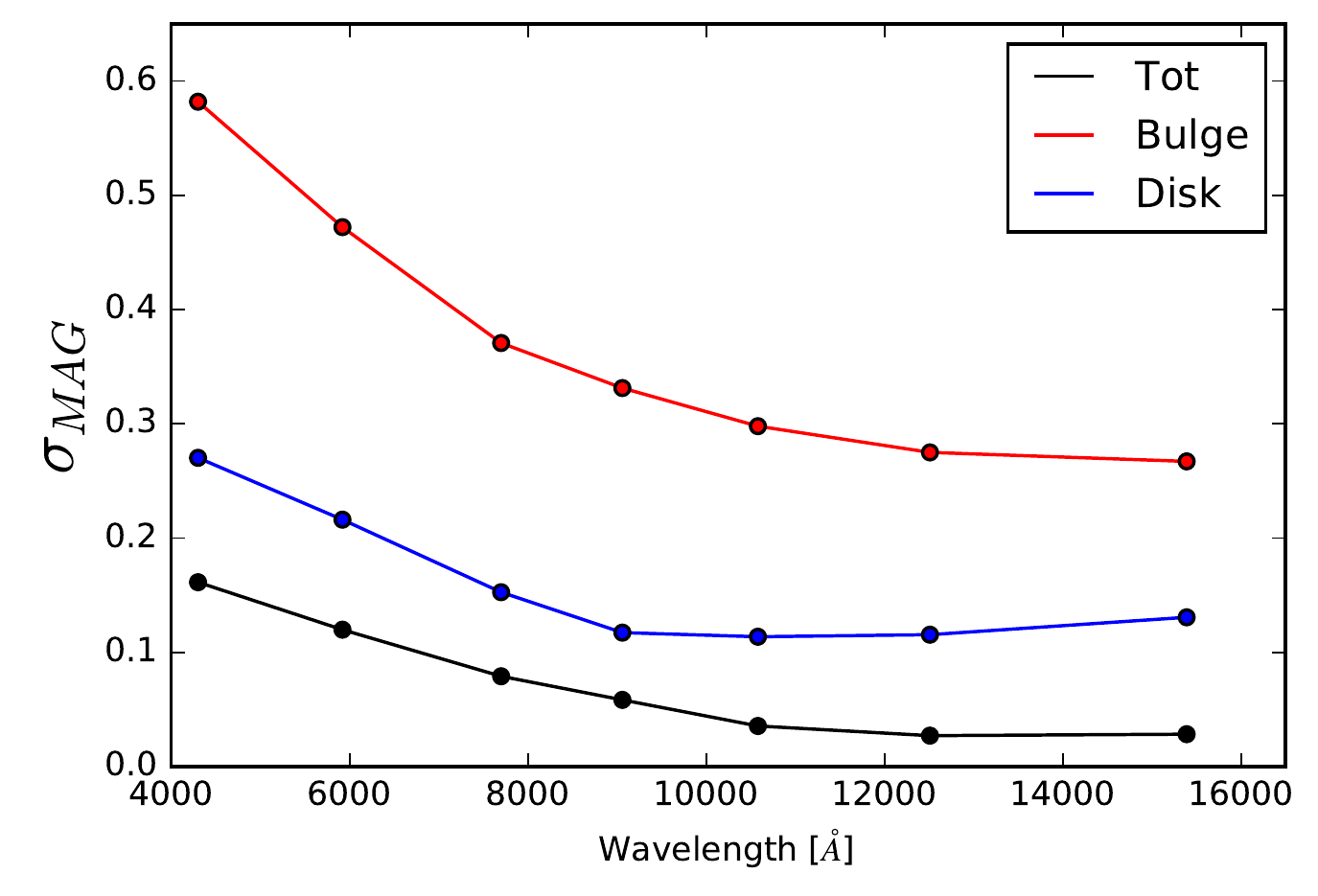} \\ 
\includegraphics[width=0.5\textwidth]{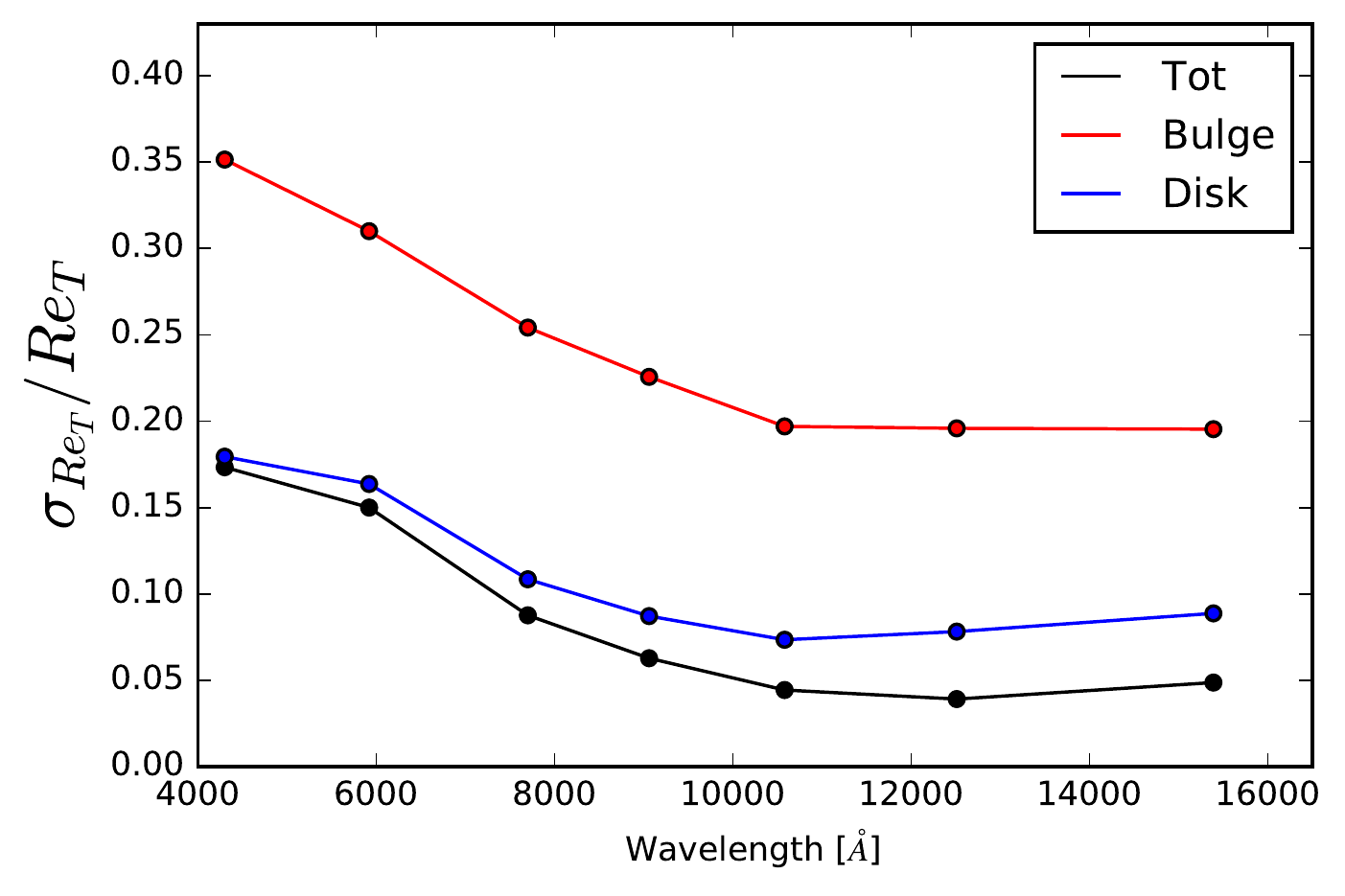} \\ 
\end{array}$
\caption{Average estimated random uncertainties for bulges (red lines), disks (blue lines) and the full galaxy (black lines) as a function of wavelength. Errors are larger in the bluer bands, especially for the bulge component as expected.   } 
\label{fig:random_errs_wl}

\end{center}
\end{figure}

 Figure~\ref{fig:random_errs_wl} shows the error dependence on wavelength. Both errors on magnitude and size are larger in shorter wavelengths. This is somehow expected since bluer bands are shallower. It is also easy to explain that bulges are more severely affected since they are expected to be redder and therefore fainter in the bluer wavelengths.
 
  \subsection{Comparison with the literature}

\begin{figure}
\begin{center}
$\begin{array}{c}
\includegraphics[width=0.5\textwidth]{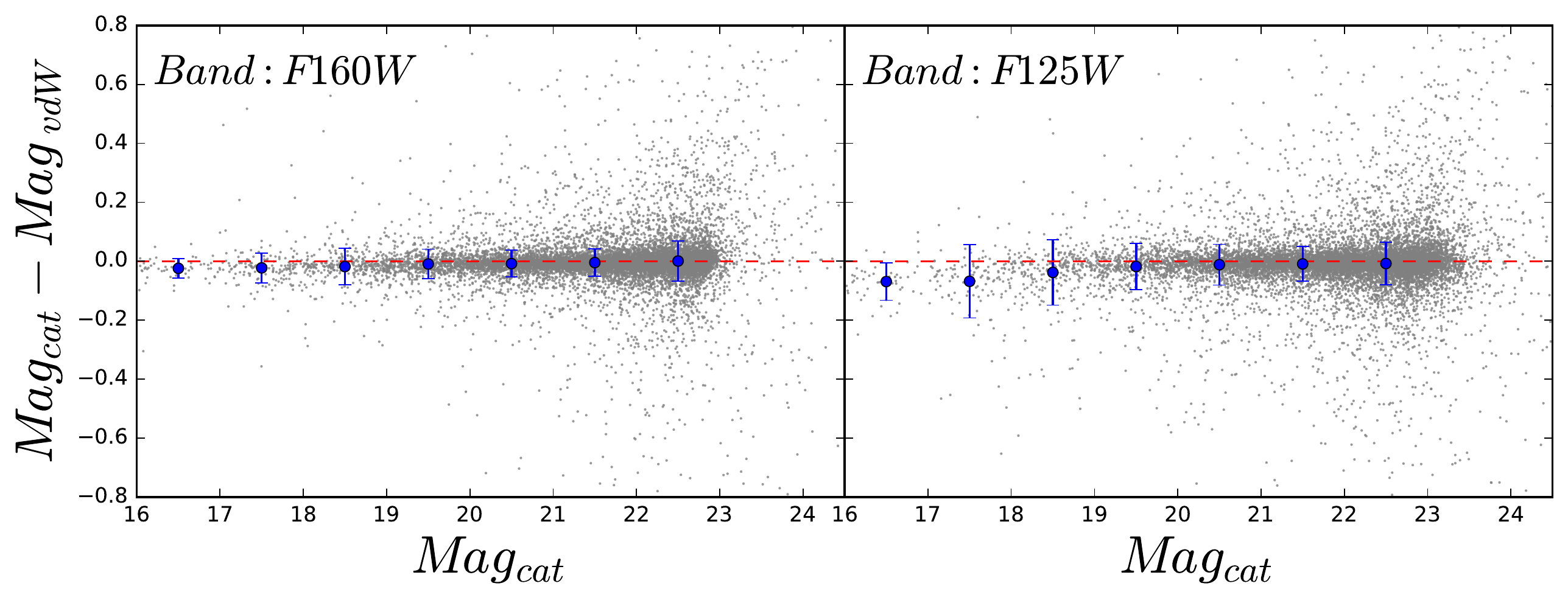} \\  \includegraphics[width=0.5\textwidth]{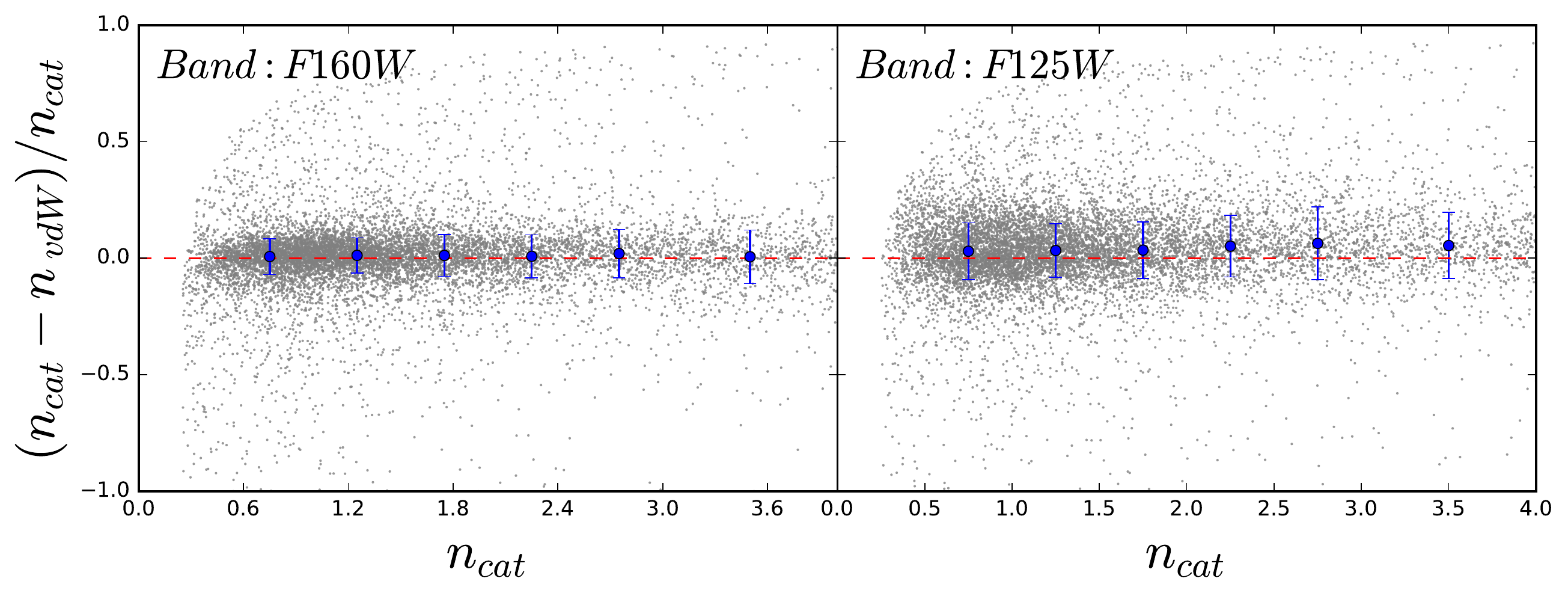} \\
\includegraphics[width=0.5\textwidth]{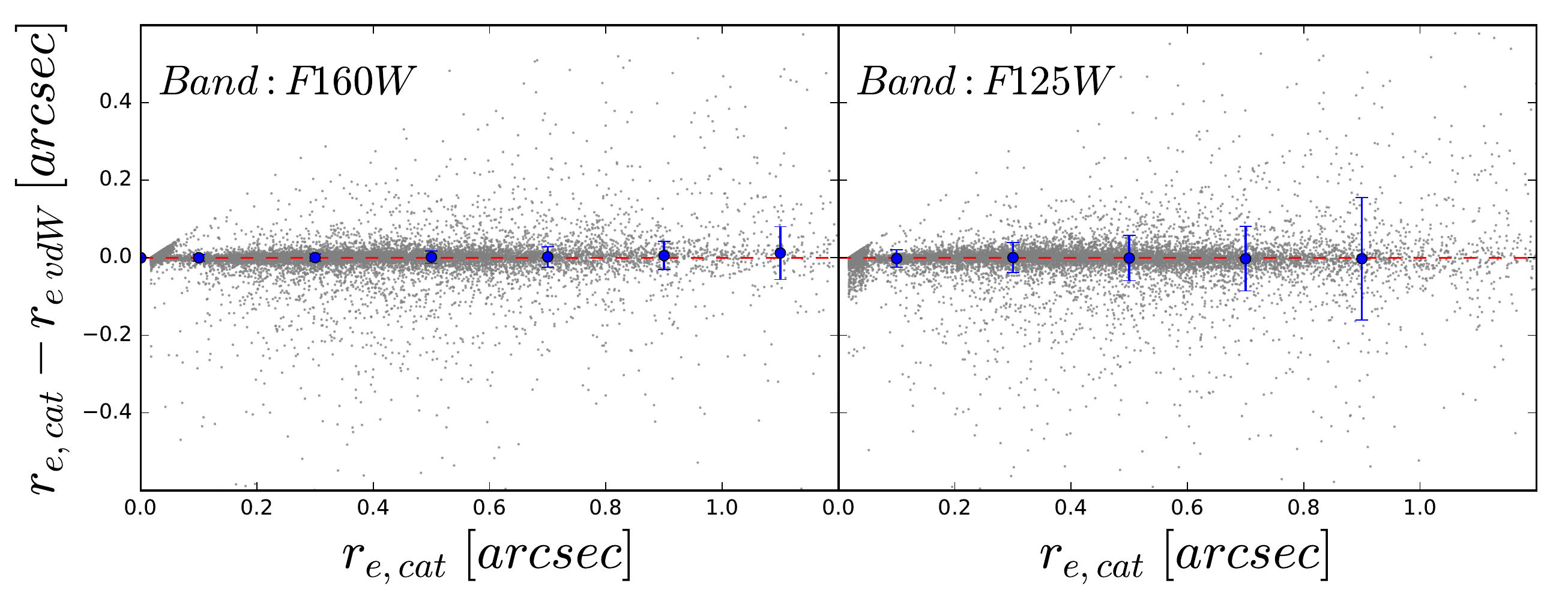} \\
\end{array}$
\caption{Comparison between our 1-component fits and the ones from van der Wel et al. (2012) in two filters. Results for the F160W and F125W filters are shown in the left and right panel respectively. Gray points in all panels show the difference between our measurements and the published ones for individual galaxies. Blue larger points are the median difference values and the error bars show the $3-\sigma$ clipped scatter. Top panels: magnitude difference. Middle panels: \Sersic index. Bottom panels: half light radii.}
\label{fig:match_mag}
\end{center}
\end{figure}

Finally, we also perform a comparison with the literature. Namely, \citealp{vanderWel2012} did a 1-component \Sersic fit to all galaxies in CANDELS down to $H=24.5$ in two NIR filters (F125 and F160) independently. As described in the previous sections, the method used in this work differs from \citealp{vanderWel2012} in the sense that all bands are fitted simultaneously. Although this technique is intended to benefit from a better $S/N$, one needs to make sure that no systematics are introduced in the process. At least for bright objects, similar results should be obtained in both works.
Figure \ref{fig:match_mag} compares the magnitude, \Sersic index and half light radii from our catalog with the ones from \cite{vanderWel2012}. There is reasonable agreement with a systematic difference compatible with zero and a scatter on the order of $\sim10\%$ increasing at fainter magnitudes as expected. The scatter is on the order of the error reported in the measurement as discussed by \cite{vanderWel2012}. This result confirms that our procedure works as expected at least for a 1 component fit and that no systematic biases are introduced by using all filters jointly.

\section{Stellar population properties of bulges and disks}
\label{sec:SEDs}

Using the methods described in the previous sections, we obtain 4-7 point SEDs for bulges and disks in our sample together with a measure of uncertainties in the derived fluxes in each band (see section~\ref{sec:errors}). We then perform a standard SED fitting with the FAST code \citep{Kriek2009} and derive stellar masses, SFRs, ages and metallicities for all bulges and disks. Additionally, using the best-fit models we also derive rest-frame U,V,J colors for bulges and disks. The input models are grids of Bruzal $\&$ Charlot (2003) models that assume a Chabrier (2003) IMF and Calzetti et al. (2000) extinction law. For the star formation history (SFH) we applied an exponentially declining model, with tau in the range of [8:10]. 

\begin{figure*}
$\begin{array}{c c}
\includegraphics[width=0.45\textwidth]{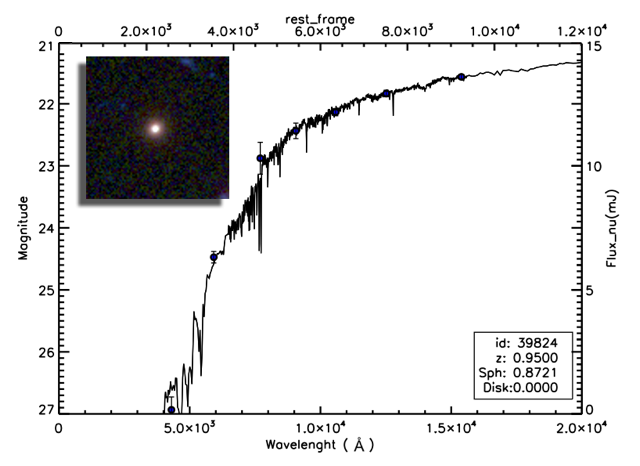} &
\includegraphics[width=0.45\textwidth]{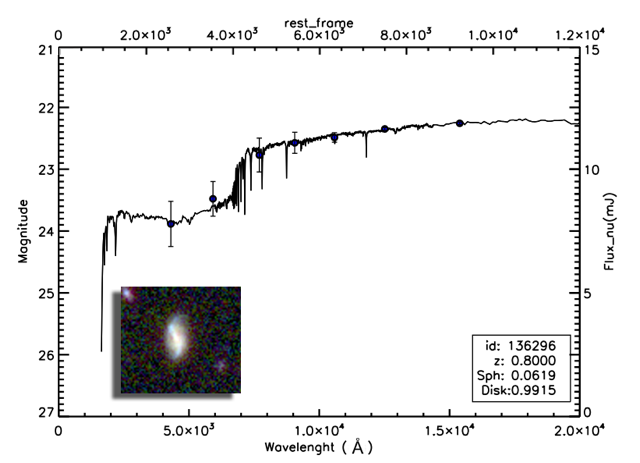} \\
\includegraphics[width=0.45\textwidth]{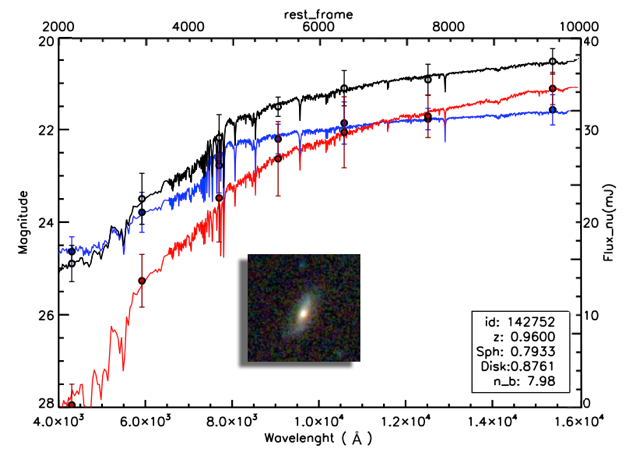} &
\includegraphics[width=0.45\textwidth]{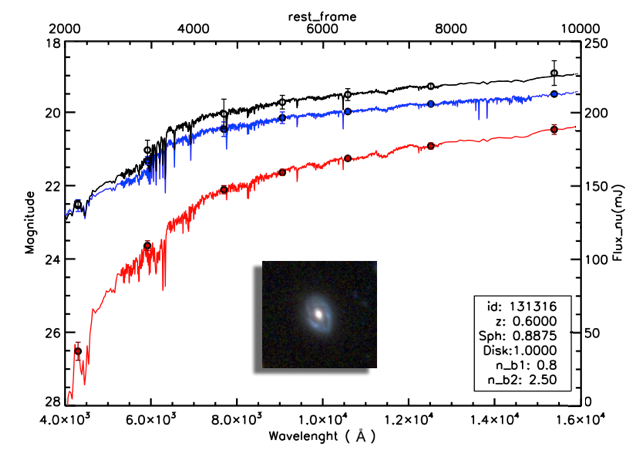}
\end{array}$
\caption{Examples of different morphologies, classified as explained in section \ref{sec:mod_sel}. Top: pure bulge galaxy on the left and a pure disk one on the right. For both just a single profile is required. Bottom: Bulge+Exponential on the left and a Pseudo-Bulge+Exponential on the right. The color code is the same than for figure \ref{fig:simulation}, Results from single \Sersic fits are shown in black, red and blue are for the disk and the bulge components respectively. The points show the measured flux in the 7 bands. The solid line are the best fit models obtained with FAST.}
\label{fig:examples_sed}
\end{figure*}

Figure \ref{fig:examples_sed} shows some examples of the best-fit templates for galaxies with one or two components according to the CNN based classification. More examples can be checked online in the public release of the catalog\footnote{\url{https://lerma.obspm.fr/huertas/form_CANDELS}}.

\begin{figure*}
\begin{center}
$\begin{array}{c c}
\includegraphics[width=0.4\textwidth]{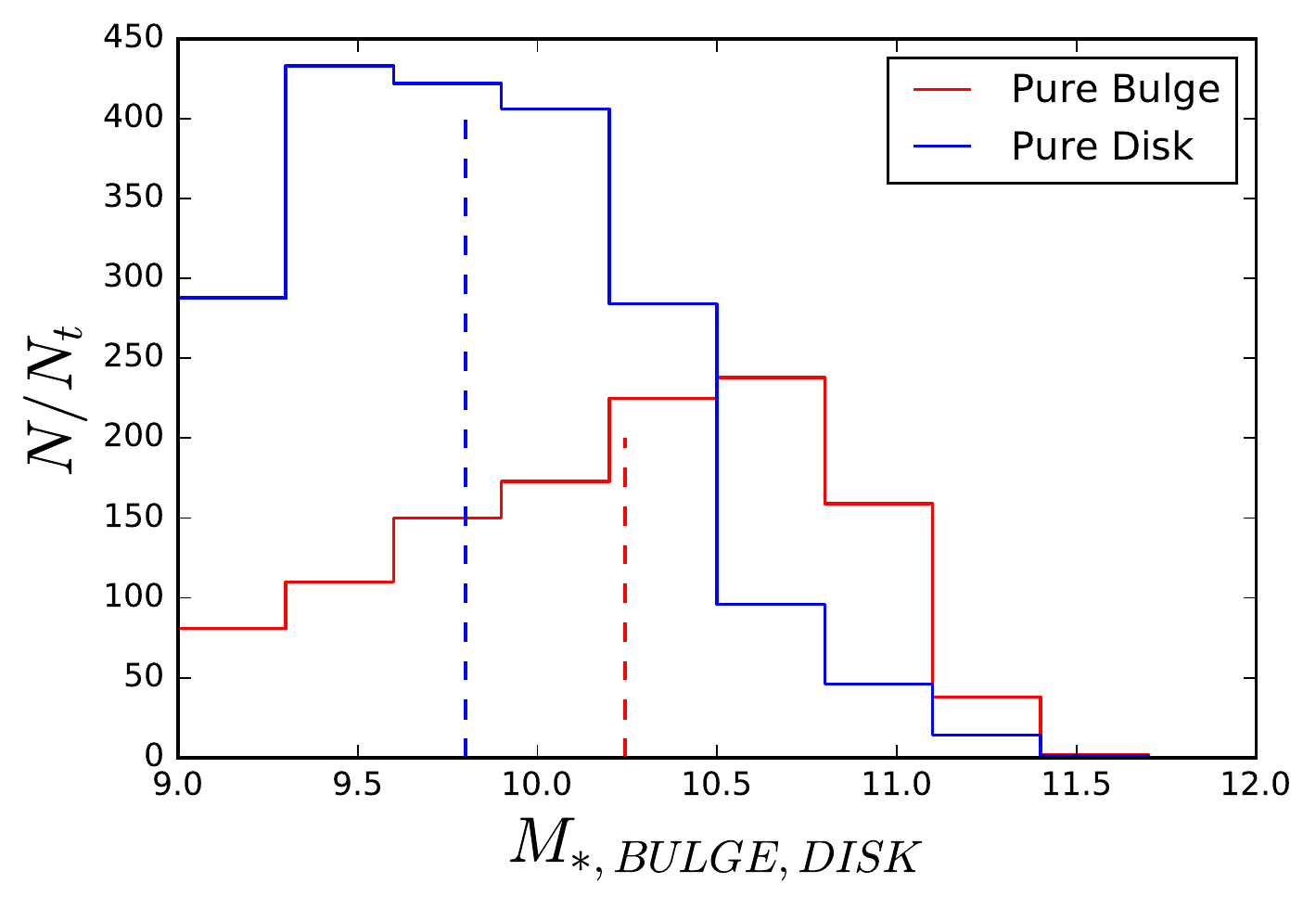} &
 \includegraphics[width=0.4\textwidth]{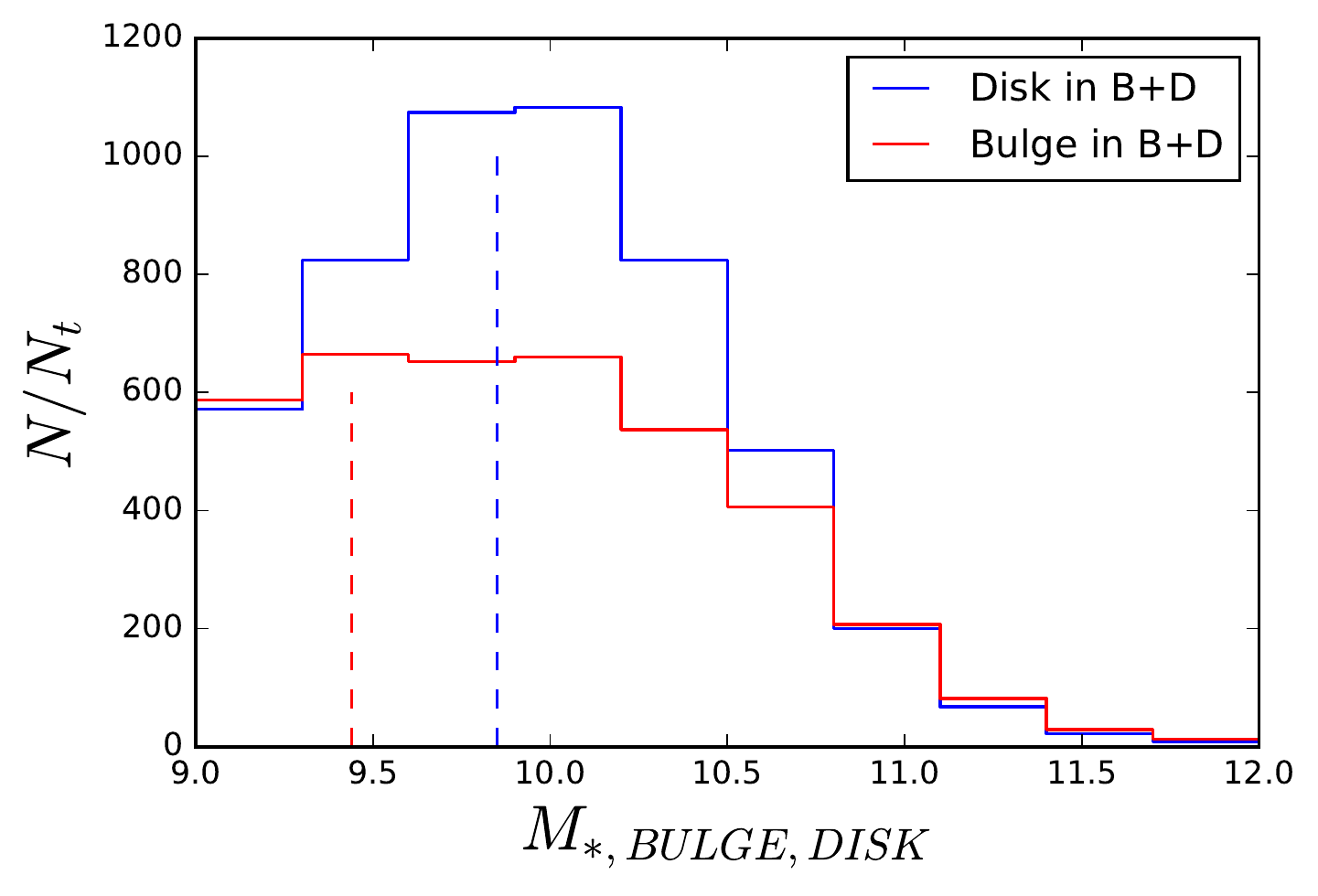}
\end{array}$
\caption{Stellar mass distribution obtained through SED fitting. Panels show mass ranges for bulges and disks in different systems, classified as explained in section \ref{sec:mod_sel}. The left panel shows the mass trend for pure bulges and pure disk, while the right panel mass distribution of bulges and disks in double component galaxies. }
\label{fig:mass_hist}
\end{center}
\end{figure*}

The mass distributions obtained for bulges and disks are shown in the two panels of figure \ref{fig:mass_hist}.

Random uncertainties in the stellar population parameters are estimated by performing MonteCarlo simulations for each galaxy. Assuming the errors on the fluxes as computed in section~\ref{sec:errors}, we generate $500$ SED realizations for each galaxy that we fit with FAST. The uncertainty on each parameter is derived as the minimum and maximum value within the $68\%$ of realizations with the smallest reduced $\chi^2$ values. Rest-frame colors are also computed for each realization and the uncertainty is computed in an analogous way.

Using the simulations described in~\ref{sec:sims}, it is also possible to estimate a global statistical uncertainty on the stellar populations properties. Since the main quantity that will be used in forthcoming works are $B/T$ ratios, we focus here on the stellar masses of both components. We notice that uncertainties in other parameters might be large given the reduced wavelength coverage and therefore should be used with caution. 

\begin{figure*}
\centering
\includegraphics[width=1.\textwidth]{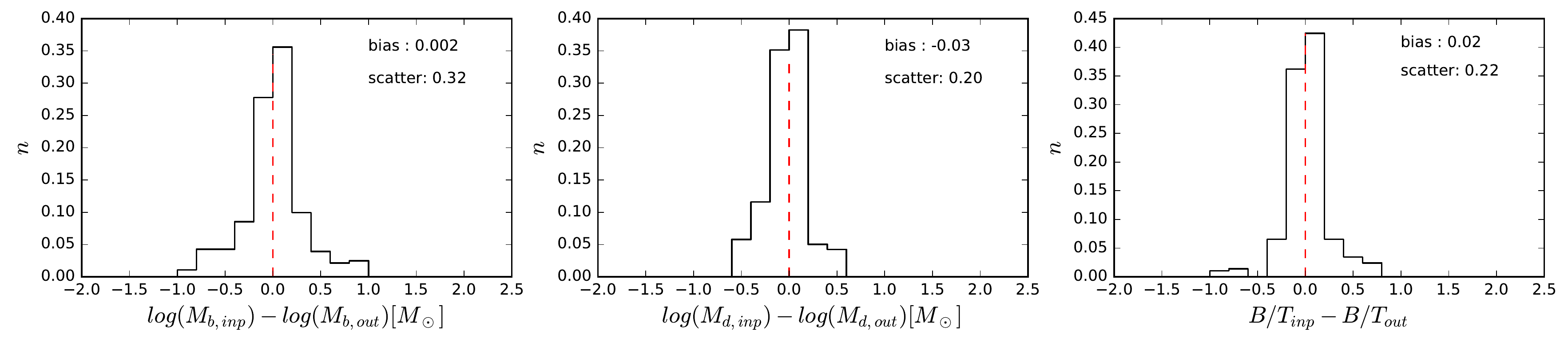} 
\caption{ Distribution of the difference between the stellar masses estimated using the simulated photometry and the ones obtained with the recovered fluxes by \textsc{MEGAMORPH} (see text for details). Left panel: bulge mass, middle panel: disk mass. The right panel shows the error distribution of the stellar mass bulge-to-total ratio.}
\label{fig:mass_bd}
\end{figure*}

\begin{figure*}
\centering
\includegraphics[width=\textwidth]{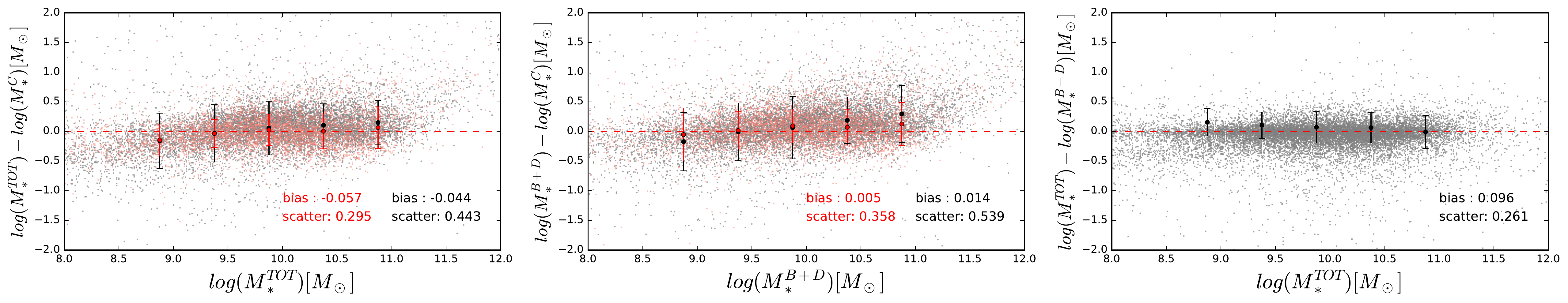}
\caption{Uncertainties on stellar mass-estimates. The left panel shows the comparison between the stellar mass estimated in CANDELS ($M_*^C$) and our estimates based on fits to the photometry from a 2D \Sersic fit ($M_*^{TOT}$). The middle panel shows the same comparison but using the stellar mass estimated by adding the masses of the disk and the bulge components ($M_*^{B+D}$). The red sample are galaxies covered by all 7 filters. Gray points show galaxies whose SED was fitted using four bands. The bias and the scatter are estimated independently for the two samples.  Finally, the right panel shows the comparison between the total stellar mass obtained from fitting the photometry estimated with a Single \Sersic and the stellar mass obtained by adding the stellar masses of the bulge and the disk.} 
\label{fig:mass_check}
\end{figure*}

 We first run FAST on the simulated SEDs of both components and then perform a second run on the recovered fluxes from \textsc{GalfitM} on the same simulated galaxies.  Figure \ref{fig:mass_bd} shows the comparison of both estimates. We find that the bias is close to zero and the dispersion is on the order of $\sim0.2-0.3$ dex, which is the typical error expected for SED based stellar masses. This results in an unbiased estimate of the stellar-mass bulge-to-total ratio with a typical scatter of $\sim0.2$. Notice that this does not mean that the \emph{true} stellar mass is recovered. It indicates however that \textsc{GalfitM} properly recovers the fluxes of the disk and bulge components without introducing additional systematics as already demonstrated for the sizes in section~\ref{sec:sims}. Another additional check is shown in the right panel of figure~\ref{fig:mass_check}, which compares the stellar mass obtained by fitting the photometry of a 1-component fit to the stellar mass obtained by adding the masses of the bulge and disk components. The two estimates agree within a $\sim0.2$ dex uncertainty as expected. \\

Finally, in order to have an independent estimate of the reliability of the stellar masses derived with only 4-7 filters, we compare our values with the stellar masses estimated in CANDELS which cover a larger spectral range. This is shown in the two first panels of figure~\ref{fig:mass_check}. Our estimates are unbiased, but with a scatter of $\sim0.4$. This scatter should be a combination of model dependence (we did not use the same stellar populations models) and spectral sampling. As a matter of fact, when the sample is divided between galaxies for which we have a spectral sampling of 4 points or less and the others, the results are still unbiased but the scatter increases from $\sim0.3$ with $4+$ filters to $\sim0.4$ with less than 4 filters.

\section{Final catalog}

The catalog release is made in three different tables. The same unique identifier is used to link all tables. 

\begin{itemize}

\item \emph{Gold catalog:} Contains a table with our best selected configuration for each galaxy.  For each galaxy we provide the structural parameters in all bands derived from what we think is the best setup (see Table~\ref{tbl:table1}) according to the CNN best class. The selection essentially follows the flow chart of figure~\ref{fig:flow_chart}. For each parameter we provide an errorbar computed with the method described in section~\ref{sec:errorbars}. Additionally, we provide an estimate of the stellar mass bulge-to-total ratio as well as rest-frame colors for bulges and disks in the two GOODS fields. For standard use, we recommend this catalog.

\item An additional table with \emph{all} structural measurements in all bands and setups described in this work is also provided along with the CNN outputs. This table should be used to explore different selections.

\item A third table with the derived stellar population properties for bulges and disks resulting from the SED fitting as well as the uncertainties. This table allows one to compute bulge to total stellar mass ratios. 

\end{itemize}

\section{Summary and conclusions}

This work presents a catalog of multi-wavelength bulge-disk decompositions of $\sim17.600$ galaxies in the five CANDELS fields. The dataset is mass complete down to $\sim5\times10^{10} M_{\odot}$  at $z\sim2$. 

Each galaxy is fitted with a 1-component \Sersic model and a 2 component \Sersic (bulge) + exponential (disk) model and with three different setups each in which we modify the wavelength dependence of the size and \Sersic index of the bulge component. We used the \textsc{galfitM/galapagos2} code from the \textsc{megamorph} project, allowing US to simultaneously fit images at different wavelengths.  One key new ingredient of this work is that we introduced a new method to address the systematic uncertainties arising from the use of a wrong model to describe the surface brightness profile of galaxies. The technique, based on convolutional neural networks, provides a quantitative measurement  of how well a given profile (1 or 2 components) describes a given galaxy. We show that our proposed method can distinguish between different profiles with a $\sim80-90\%$ accuracy. It also allows us to reduce the contamination from unphysical components at the $\sim10\%$ level and study a clean sample of bulges and disks. 

Through extensive simulations, we show that the structural parameters derived for bulges and disks are globally unbiased. We develop a method based on the comparison of results from different runs to provide individual error bars for each structural parameter. We show that the typical error for the bulge (disk) magnitudes are $\sim0.2$ ($\sim0.1$) mags. For galaxies with $B/T<0.2$ ($B/T>0.8$) the error on the bulge (disk) magnitudes increases to $\sim0.5$ mags. Sizes are estimated within $\sim10-20\%$ uncertainty.

The derived spectral energy distributions of both components with realistic errors are then fitted with stellar population models to estimate stellar masses and bulge-to-total mass ratios. We show that the statistical uncertainties are on the order of $\sim20\%$. We also provide rest-frame colors and SFRs which will be analyzed in forthcoming woks.

The catalog including all derived quantities is made public with the present work. 

\section*{Acknowledgments}

The present work has been funded by a French Doctoral Contract. The authors are also grateful to Google for the unrestricted gift given to the University of Santa Cruz to carry out the project: "Deep-Learning for Galaxies".

\bibliographystyle{mnras}
\bibliography{catalog}  

\end{document}